\documentclass[aps,prb,twocolumn,floatfix]{revtex4-2}
\usepackage{graphicx}
\usepackage{multirow}
\usepackage{color}
\usepackage{amsmath}
\usepackage{amsfonts}
\usepackage{amssymb}
\usepackage{graphicx}
\graphicspath{{Figures/}}
\usepackage{epstopdf}
\usepackage{empheq}
\usepackage{float}
\usepackage{cases} 
\usepackage{mathtools}
\usepackage{dcolumn}
\usepackage{bbold}
\usepackage{xfrac}
\usepackage{bm}
\usepackage{siunitx}

\usepackage[dvipsnames]{xcolor}
\usepackage[colorlinks]{hyperref}
\hypersetup{
colorlinks=True,
linkbordercolor=White,
linkcolor=BrickRed,
citecolor=Blue,	 
}

\DeclarePairedDelimiter\bra{\langle}{\rvert}
\DeclarePairedDelimiter\ket{\lvert}{\rangle}

\renewcommand{\vec}[1]{\boldsymbol{#1}}

\newcommand\hl[1]{{#1}}

\begin{document}

\title{Dressed basis sets for the modeling of exchange interactions in double quantum dots}

\author{Mauricio J. Rodr\'iguez}
\affiliation{Univ. Grenoble Alpes, CEA, IRIG-MEM-L\_Sim, Grenoble, France.}%
\author{Esteban A. Rodr\'iguez-Mena}
\affiliation{Univ. Grenoble Alpes, CEA, IRIG-MEM-L\_Sim, Grenoble, France.}%
\author{Ahmad Fouad Kalo}
\affiliation{Univ. Grenoble Alpes, CEA, IRIG-MEM-L\_Sim, Grenoble, France.}%
\author{Yann-Michel Niquet}
\email{yniquet@cea.fr}
\affiliation{Univ. Grenoble Alpes, CEA, IRIG-MEM-L\_Sim, Grenoble, France.}%

\date{\today}

\begin{abstract}
We discuss the microscopic modeling of exchange interactions between double semiconductor quantum dots used as spin qubits. Starting from a reference full configuration interaction (CI) calculation for the two-particle wave functions, we build a reduced basis set of dressed states that can describe the ground-state singlets and triplets over the whole operational range with as few as one hundred basis functions (as compared to a few thousands for the full CI). This enables fast explorations of the exchange interactions landscape as well as efficient time-dependent simulations. We apply this methodology to a double hole quantum dot in germanium, and discuss the physics of exchange interactions in this system. We show that the net exchange splitting results from a complex interplay between inter-dot tunneling, Coulomb exchange and correlations. We analyze, moreover, the effects of confinement, strains and Rashba interactions on the anisotropic exchange and singlet-triplet mixings at finite magnetic field. We finally illustrate the relevance of this methodology for time-dependent calculations on a singlet-triplet qubit.
\end{abstract}

\maketitle

\section{Introduction}

Spin qubits in semiconductor quantum dots are attracting increasing interest for quantum computing and simulation owing to their favorable ratio between coherence and manipulation times and their prospects for large scale integration \cite{Loss98,Stano22,Burkard2023Review,Fang2023Review}. Silicon and germanium materials can, in particular, be isotopically purified in order to get rid of the nuclear spins that spoil the coherence of the electron or hole spins in the quantum dots \cite{Itoh14,Cvitkovich24}. Coherent and high-fidelity single- and two-qubit gates have thus been demonstrated in silicon electron spin qubits \cite{Yoneda18,Philips22,Mills22,Xue22,Noiri22,Huang24,Steinacker24}, and, more recently, in silicon and germanium hole spin qubits \cite{Maurand16,Watzinger18,Hendrickx20b,Hendrickx20,Froning21,Hendrickx21,Camenzind22,wang2022ultrafast,Lawrie23,Valentin25}. The latter can be efficiently manipulated electrically thanks to the strong intrinsic spin-orbit coupling (SOC) in the valence band of semiconductor materials \cite{Winkler03}. Although more sensitive to electrical noise, they exhibit operational sweet spots with long coherence times comparable to electron spin qubits driven with the help of micro-magnets \cite{Wang21,Piot22,Hendrickx2024,Mauro24,Bassi24}. Germanium heterostructures \cite{Sammak19,Scappucci20} have, in particular, made outstanding progress recently, with the demonstration of charge and spin control, spin entanglement, and spin shuttling in 4 to 16 quantum dots devices \cite{Wang2023,Borsoi24,van_riggelen_shuttling_2024,Wang2024}. Considerable progress has also been made in the theoretical understanding of SOC in hole spin qubit devices. The role played by Rashba and Dresselhaus spin-orbit interactions \cite{Winkler03,Rashba03,Golovach06}, and by $g$-tensor modulations \cite{Kato03,Crippa18} in the manipulation, coherence and relaxation of hole spins has been clarified in many relevant situations (strong one- and two-dimensional confinement, anharmonic and non-separable potentials, inhomogeneous strains...) \cite{Kloeffel11,Kloeffel13,Marcellina17,Venitucci18,Li20,Terrazos21,Michal21,Bosco21b,martinez2022hole,Sarkar23,Abadillo2023,Rodriguez2023,Wang24b,Stano25,Mauro25}. 

The exchange coupling mediated by tunnel and Coulomb interactions between quantum dots is the driving mechanism for most two-qubit gates \cite{Loss98,Burkard99}. It is also the basis for singlet-triplet \cite{Petta05,Jock18,Takeda20,jirovec_Freqs_ST-Ge,Jirovec23,Saezmollejo24,rooney_strains_ST-Ge_ST_-,Liles24,Zhang25,Tsoukalas2025} and ``exchange-only'' qubits \cite{DiVincenzo00,Laird10,Medford13,Eng15,Weinstein23,Acuna24} that can be manipulated with baseband signals. Spin-orbit coupling also reshapes the interactions between quantum dots, by enabling spin-flip tunneling and thus giving rise to anisotropic singlet-triplet mixings and exchange corrections \cite{harvey-collard_spin-orbit_ST-Si,froning_strong_SO_2021,Geyer24}. It is, therefore, essential to understand both the challenges and opportunities posed by SOC for quantum protocols \cite{Bosco24}. So far, the understanding of these interactions has mostly relied on effective Hamiltonians \cite{Stepanenko12,Hetenyi20,mutter_all-electrical_ST-Ge,Sen23} that catch the essential physics but hardly capture the complexity of real quantum devices. As a matter of fact, the description of Coulomb interactions in realistic geometries requires numerically intensive techniques such as configuration interaction (CI) \cite{Sherrill99,Rontani06,ElectroConfinedCI(EM)}. The CI method basically expands the many-body wave functions in a basis of Slater determinants built on a subset of $N$ single-particle wave functions. The size of the CI basis set thus scales as $N^2$ (for two particles), so that CI is usually unsuitable for intensive tasks such as time-dependent simulations. Yet the CI method has provided many insights in the physics of Coulomb interactions in single and multiple quantum dots \cite{Cimente07,Baruffa10,lownoisexchGCI,timedepCI,Barnes11,CtrlPhaseGCI,SpinRelaxCI,Nielsen13,ExCplDonCI(EM),Deng20,Chan2021,IntHolesCI(KP),MicroSweetSpotCI,4eDDQDCI,Dodson22,Shehata23}, highlighting, for example, the role of Wigner localization on the many-body spectrum and wave functions \cite{Bryant87,Reimann00,Ellenberg06,Kalliakos08,Singha10,Abadillo21,Corrigan21,e-eCI(TB),ValleytronicCI(EM),Yannouleas22}.

In this work, we introduce a methodology for the efficient modeling of Coulomb interactions in double quantum dot (DQD) devices. Starting from reference CI calculations, we build a reduced basis set of ``dressed'' states accounting for the main Coulomb correlations. This basis set can describe the lowest singlet and triplet states relevant for exchange-driven operations over the whole operational gate voltages range with as few as a hundred basis functions (whereas the original CI basis set contains $>4500$ Slater determinants). This allows for fast explorations of the device physics and for efficient time-dependent simulations. We illustrate this framework with hole DQDs in germanium heterostructures. %We discuss, in particular, the physics of the exchange interaction in this system, and highlight the effects of dot asymmetry and strains on the exchange energy and mixing between singlet and triplet states.  

We first review the methodology in section \ref{sec:methodology}. We describe the CI calculations and dressed basis set and assess convergence. We next investigate the physics of exchange interactions in germanium DQDs in section \ref{sec:physics}. We show that the net exchange interaction is ruled by a complex interplay between inter-dot tunneling, Coulomb exchange and Coulomb correlations. We then discuss the effects of a finite magnetic field in section \ref{sec:magnetic}. We analyze, in particular the role of confinement, strains and Rashba interactions on the anisotropic exchange and singlet-triplet mixings. Finally, we illustrate in section \ref{sec:tdsimus} the suitability of the dressed basis set for time-dependent calculations on a singlet-triplet qubit.

\section{Device and methodology}
\label{sec:methodology}

\subsection{Device}

\begin{figure}[t]
\centering
\includegraphics[width=.9\linewidth]{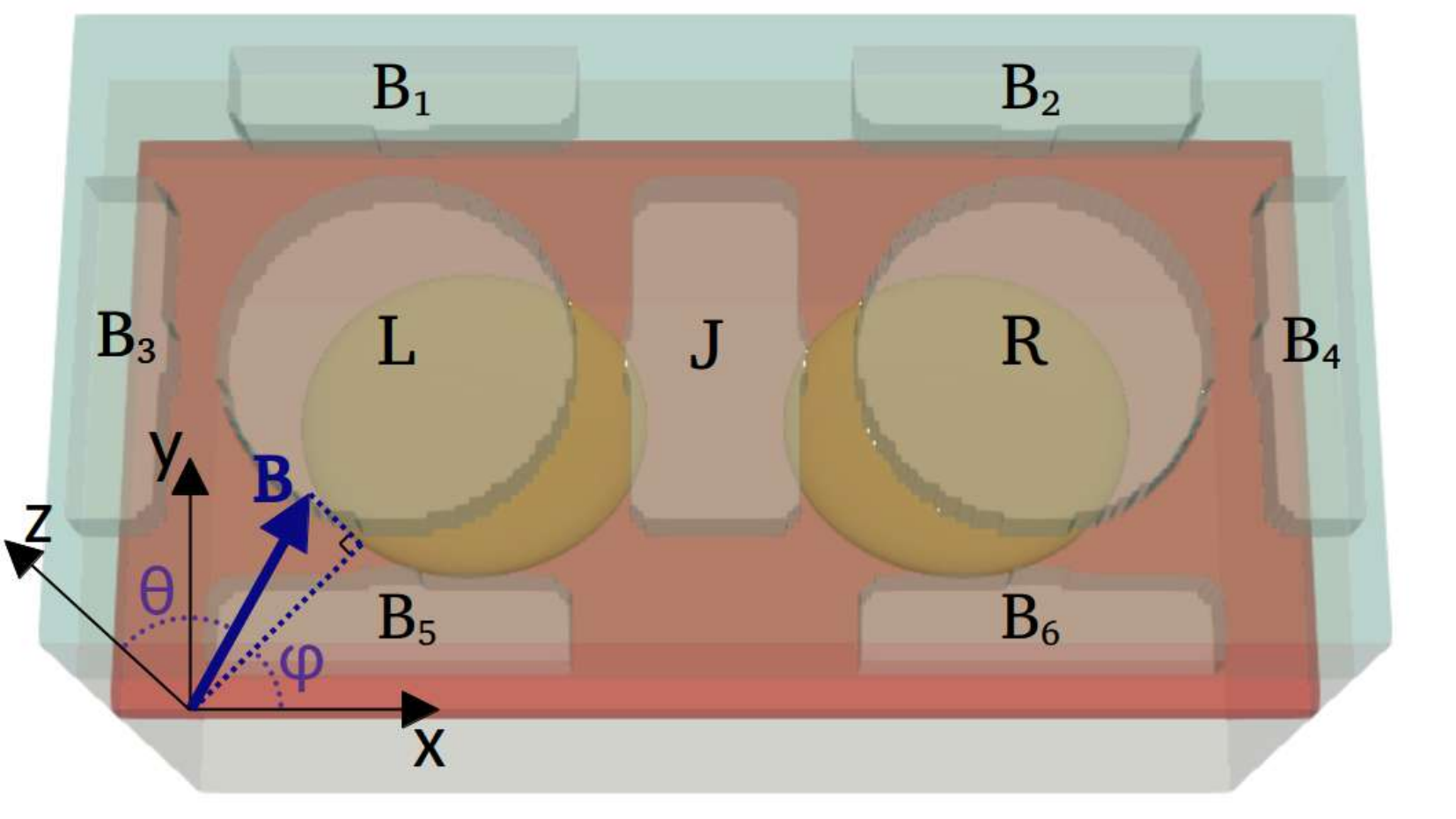}
\caption{A double unit cell of the 2D array of spin qubit devices. The heterostructure is a 16-nm-thick Ge quantum well (red) grown on a thick Ge$_{0.8}$Si$_{0.2}$ buffer capped with a 50-nm-thick Ge$_{0.8}$Si$_{0.2}$ barrier (blue). The dots are shaped by plunger (L, R) and barrier (J, B$_i$) gates (gray). The yellow contour is the isodensity surface that encloses 90\% of the  hole charge in the ground $(1,1)$ state at bias $V_\mathrm{L}=V_\mathrm{R}=-40$\,mV and $V_\mathrm{J}=-15$\,mV (B gates grounded). The orientation of the magnetic field $\vec{B}$ is characterized by the angles $\theta$ and $\varphi$ in the crystallographic axes set $x=[100]$, $y=[010]$ and $z=[001]$.}
\label{fig:device}
\end{figure}

As an illustration, we consider the prototypical two-dimensional (2D) array of quantum dots shown in Fig.~\ref{fig:device}. The holes are confined in a 16-nm-thick Ge well grown on a Ge$_{0.8}$Si$_{0.2}$ buffer and capped with a 50-nm-thick Ge$_{0.8}$Si$_{0.2}$ barrier. The quantum dots are shaped by 100-nm-diameter circular plunger gates interleaved with rectangular exchange gates. The 20-nm-thick aluminium gates are embedded in Al$_2$O$_3$ and are meant to be connected with vias to a routing metal level above. The side of the unit cell of the array (the distance between the centers of nearest neighbor plunger gates) is $a_\mathrm{2D}=180$\,nm.

We assume residual biaxial in-plane strains $\varepsilon_{xx}=\varepsilon_{yy}=\varepsilon_\mathrm{buf}=0.26\%$ in the Ge$_{0.8}$Si$_{0.2}$ buffer \cite{Sammak19}. The compressive biaxial strains in the Ge well are, therefore, $\varepsilon_{xx}=\varepsilon_{yy}=\varepsilon_\parallel=-0.61\%$ and $\varepsilon_{zz}=\varepsilon_\perp=+0.45\%$. We may, additionally, account for the inhomogeneous strains brought by the differential thermal contraction of materials upon cool-down. When relevant, the latter are computed with a finite-elements discretization of the continuum elasticity equations \cite{Abadillo2023}. All material parameters are borrowed from Ref.~\onlinecite{Abadillo2023}.

\subsection{Single particle states}

\begin{figure}[t]
\centering
\includegraphics[width=.9\linewidth]{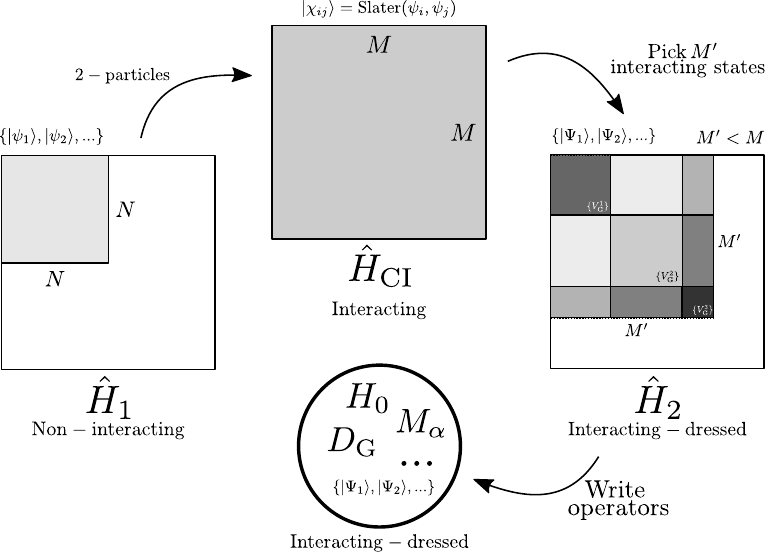}
\caption{Outline of the methodology. The $N$ lowest eigenstates of the finite-differences Luttinger-Kohn Hamiltonian are used as a basis set for fast one-particle calculations and for CI calculations. Then the lowest eigenstates of the CI Hamiltonian, sampled at different gate voltages $\{V_\mathrm{G}^k\}$ and orthogonalized against each other, are used as a ``dressed'' basis set for two-particle calculations over the whole operational gate voltages range.}
\label{fig:methodology}
\end{figure}

The whole methodology is outlined in Fig.~\ref{fig:methodology}. We first compute the electrostatic potential $V_t(\vec{r})$ in the empty device. The latter fulfills Poisson's equation
\begin{equation}
\boldsymbol{\nabla}_{\vec{r}}\cdot\kappa(\vec{r})\boldsymbol{\nabla}_{\vec{r}}V_t(\vec{r})=0\,,
\end{equation}
with $\kappa(\vec{r})=\kappa_0\kappa_\mathrm{r}(\vec{r})$, $\kappa_0$ the vacuum permittivity and $\kappa_\mathrm{r}(\vec{r})$ the dielectric constant at point $\vec{r}$. This equation is solved iteratively on a finite-volumes cartesian mesh. The bias voltages $V_\mathrm{G}$ \hl{on the gates $\mathrm{G}$ in the set $\{\mathrm{L},\mathrm{R},\mathrm{J},\mathrm{B}_i\}$} are used as boundary conditions for $V_t(\vec{r})$. They differ from the electro-chemical potentials applied on these gates by rigid shifts that depend on the ionization potentials of the materials of the gate stack and heterostructure. We next solve the Luttinger-Kohn (LK) model \cite{Luttinger56,Winkler03,KP09} for the hole wave functions $\psi_n$ in this potential \hl{(see Appendix \ref{app:LKH})}. The physical spin components $\sigma\in\{\uparrow,\downarrow\}$ of the $\psi_n$'s are therefore expanded as
\begin{equation}
\psi_n(\vec{r},\sigma)=\sum_{\nu\in\left\{-\tfrac{3}{2},-\tfrac{1}{2},\tfrac{1}{2},\tfrac{3}{2}\right\}}\varphi_n^\nu(\vec{r})u_\nu(\vec{r},\sigma)\,,
\end{equation}
where $u_\nu$ is the Bloch function with angular momentum $j_z=\nu$ along $z=[001]$ ($\nu=\pm 3/2$ for HH and $\nu=\pm 1/2$ for LH components), and $\varphi_n^\nu$ is an envelope function. The set of differential equations fulfilled by the $\varphi_n^\nu$'s is solved on the same mesh as the potential using finite differences (FDs). To reduce computational cost, we apply hard-wall boundary conditions ($\varphi_n^\nu=0$) $8$ nm below and $16$ nm above the Ge well. The typical mesh step in the Ge well is $\delta_\parallel=2$\,nm along $x=[100]$ and $y=[010]$, and $\delta_\perp=5$\,\AA\ along $z$ \footnote{The mesh is inhomogeneous along $z$ outside the well}. We consider holes with positive dispersion throughout this work.

The FD mesh of the device of Fig.~\ref{fig:device} contains $1.04$ million points, thus $4.14$ million complex degrees of freedom. The lowest-lying eigenstates of the (sparse) FD Hamiltonian are computed with an iterative Jacobi-Davidson method \cite{Sleijpen00,Templates00}. Given the size of the problem, FDs are, however, hardly suitable for fast explorations of the device properties around a given bias point. For that purpose, we build an effective, low-energy Hamiltonian in a basis set of $N$ single-particle states $\ket{\psi_n}$. We first expand the LK Hamiltonian $H_\mathrm{LK}$ in powers of the gate potentials $V_\mathrm{G}$ and magnetic field $\vec{B}$:
\begin{align}
H_1&=H_0+\sum_\mathrm{G} e\delta V_\mathrm{G} D_\mathrm{G} \nonumber \\
&+\sum_{\alpha\in\{x,y,z\}}B_\alpha M_\alpha+\frac{1}{2}\sum_{\alpha,\beta\in\{x,y,z\}}B_\alpha B_\beta N_{\alpha\beta}\,.
\end{align}
Here $H_0=H_\mathrm{LK}(\{V_\mathrm{G}^0\},\vec{B}=\vec{0})$ is the LK Hamiltonian at a reference bias point $\{V_\mathrm{G}^0\}$, $D_\mathrm{G}(\vec{r})=\partial V_t(\vec{r})/\partial V_\mathrm{G}$ is the derivative of the total potential $V_t(\vec{r})$ with respect to gate voltage $V_\mathrm{G}$, $M_\alpha=\partial H_\mathrm{LK}/\partial B_\alpha$ and $N_{\alpha\beta}=\partial^2 H_\mathrm{LK}/\partial B_\alpha\partial B_\beta$ are the first and second derivatives of $H_\mathrm{LK}$ with respect to the components $B_\alpha$ of the magnetic field, and $\delta V_\mathrm{G}=V_\mathrm{G}-V_\mathrm{G}^0$. This expansion is formally exact if the electrostatics is linear \footnote{This is the case as long as screening by remote hole gases is not included in, e.g., the Thomas-Fermi approximation. Even in that case, the response of the hole gases can be linearized with respect to small variations of the potentials, which turns Poisson's equation into a linear Helmholtz equation.} as the LK Hamiltonian is quadratic in the magnetic field. $D_\mathrm{G}(\vec{r})$ is then nothing else than the potential created when applying $1$\,V on gate G while grounding all other gates. The operators $M_\alpha$ and $N_{\alpha\beta}$ are computed from numerical differences between LK Hamiltonians at small magnetic fields.

We next project $H_1$ in the subspace spanned by the $N$ chosen single-particle wave functions $\psi_n$ and introduce the $N\times N$ matrix $\hat{H}_1=P^\dagger H_1 P$ with elements $[\hat{H}_1]_{ij}=\bra{\psi_i}H_1\ket{\psi_j}$ (the columns of $P$ being the $\psi_n$'s on the FD mesh). The easiest choice for the $\psi_n$'s is the $N$ lowest-lying hole states at reference bias $\{V_\mathrm{G}^0\}$, in which case the matrix $\hat{H}_0=P^\dagger H_0 P$ is diagonal. The simplest models for DQDs are typically built in such a $N=4$ basis set (the lowest two bonding and two anti-bonding states of the DQD) \cite{Burkard99,Burkard2023Review}. The low-energy eigenstates of $\hat{H}_1$ are essentially exact at the reference bias point, and are the better the larger $N$ for finite $\delta V_\mathrm{G}$'s. We will discuss convergence with respect to $N$ in section \ref{ssec:cvg}. The accurate description of far detuned $\delta V_\mathrm{G}$'s may however call for huge basis sets. To address a wider range of $\delta V_\mathrm{G}$'s, we may also merge the wave functions $\psi_n$ from $N_\mathrm{b}$ different $\{V_\mathrm{G}^k\}$'s in this range. The low-lying eigenstates of $\hat{H}_1$ are then exact at each $\{V_\mathrm{G}^k\}$ and get interpolated in between. The $\ket{\psi_n}$ from different $\{V_\mathrm{G}^k\}$'s must be orthogonalized against each other and possibly discarded if the residual norm after orthogonalization is too small (which means that they do not bring significant information). The matrices $\hat{H}_0$, $\hat{D}_\mathrm{G}=P^\dagger D_\mathrm{G} P$, $\hat{M}_\alpha=P^\dagger M_\alpha P$ and $\hat{N}_{\alpha\beta}=P^\dagger N_{\alpha\beta} P$ are pre-computed once for all to speed up the calculation of $\hat{H}_1$ for arbitrary $\{\delta V_\mathrm{G}\}$ and $\vec{B}$.

\subsection{Two particle states}
\label{subsec:twop}

To compute the two-particle states, we add the Coulomb interaction, $W$, and diagonalize the Hamiltonian \cite{Sherrill99,Rontani06,Abadillo21,ElectroConfinedCI(EM)}
\begin{equation}
H_\mathrm{int}=\sum_{i,j}[\hat{H}_1]_{ij}c^\dagger_i c_j+\frac{1}{2}\sum_{i,j,k,l}W_{kjil}c_i^\dagger c_j^\dagger c_k c_l
\label{eq:Hint}
\end{equation}
in the basis set of the $M=N(N-1)/2$ Slater determinants $\ket{\chi_{ij}}\equiv c_i^\dagger c_j^\dagger\ket{0}$ ($i<j$). Here $c_i^\dagger$ and $c_i$ are the creation and annihilation operators for state $\ket{\psi_i}$, $\ket{0}$ is the vacuum state, and the Coulomb integrals $W_{ijkl}$ read
\begin{align}
W_{ijkl}=e^2\sum_{\sigma,\sigma^\prime}\int d^3\vec{r}\int d^3\vec{r}^\prime\,&\psi_i(\vec{r},\sigma)\psi_j^*(\vec{r},\sigma)W(\vec{r},\vec{r}^\prime) \nonumber \\
\times&\psi_k^*(\vec{r}^\prime,\sigma^\prime)\psi_l(\vec{r}^\prime,\sigma^\prime)\,,
\label{eq:Wijkl}
\end{align}
where $W(\vec{r},\vec{r}^\prime)$ is the potential created at $\vec{r}$ by a unit test charge at $\vec{r}^\prime$. We may introduce the joint charge density
\begin{align}
\rho_{kl}(\vec{r})&=e\sum_\sigma\psi_k^*(\vec{r},\sigma)\psi_l(\vec{r},\sigma) \nonumber \\
&=e\sum_{\mu,\nu,\sigma}\varphi_k^{\mu*}(\vec{r})\varphi_l^\nu(\vec{r})u_\mu^*(\vec{r},\sigma)u_\nu(\vec{r},\sigma)\,
\label{eq:rhokl}
\end{align}
and write $W_{ijkl}=\bra{\rho_{ij}}W\ket{\rho_{kl}}$. The calculation of the $W_{ijkl}$'s in a realistic dielectric environment is demanding. The screening is, in particular, different at long and short ranges \cite{Vinsome71,Richardson71}. The physics of the DQD is, nonetheless, dominated by the long-range tail of the Coulomb interaction (since the holes repel each other and hardly come close together in principle). Therefore, we discard the $\mu\ne\nu$ terms of Eq.~\eqref{eq:rhokl} that only give rise to short-range multipolar contributions to $W_{ijkl}$ \cite{BandCoulCI(KP)}. Indeed, the Bloch functions are orthogonal and normalized so that the product $u_\mu^*u_\nu$ averages to $\delta_{\mu\nu}$ over any unit cell of the diamond lattice \footnote{Moreover, the low-energy states of this system are almost pure HH states, and $u_{+3/2}^*(\vec{r},\sigma) u_{-3/2}(\vec{r},\sigma)=0$ since the two Bloch functions have different physical spins.}. We thus introduce the ``macroscopic'' charge density locally averaged over such a unit cell
\begin{equation}
\bar{\rho}_{kl}(\vec{r})=e\sum_\nu\varphi_k^{\nu*}(\vec{r})\varphi_l^\nu(\vec{r})\,,
\end{equation}
substitute $\bar{\rho}_{kl}(\vec{r})$ for $\rho_{kl}(\vec{r})$ and use, accordingly, the macroscopic screened Coulomb interaction $\bar{W}(\vec{r},\vec{r}^\prime)$ in Eq.~\eqref{eq:Wijkl}. The latter satisfies Poisson's equation:
\begin{equation}
\boldsymbol{\nabla}_{\vec{r}}\cdot\kappa(\vec{r})
\boldsymbol{\nabla}_{\vec{r}}\bar{W}(\vec{r},\vec{r}^\prime)=-\delta(\vec{r}-\vec{r}^\prime)\,.
\end{equation}
Therefore,
\begin{equation}
W_{ijkl}=\int d^3\vec{r}\,\bar{\rho}_{ij}^*(\vec{r})\bar{V}_{kl}(\vec{r})
\end{equation}
where
\begin{equation}
\bar{V}_{kl}(\vec{r})=\int d^3\vec{r}^\prime\,\bar{W}(\vec{r},\vec{r}^\prime)\bar{\rho}_{kl}(\vec{r}^\prime)
\end{equation}
is the solution of Poisson's equation for the complex density $\bar{\rho}_{kl}(\vec{r})$:
\begin{equation}
\boldsymbol{\nabla}_{\vec{r}}\cdot\kappa(\vec{r})
\boldsymbol{\nabla}_{\vec{r}}\bar{V}_{kl}(\vec{r})=-\bar{\rho}_{kl}(\vec{r})\,.
\end{equation}
The $N(N+1)/2$ potentials $\bar{V}_{kl}(\vec{r})$ are computed with the same Poisson solver as the total potential $V_t(\vec{r})$ (with all gates grounded). Although numerically expensive, the procedure can be efficiently parallelized, the calculations for different $(k,l)$ pairs being independent. 

The ``configuration interaction'' (CI) Hamiltonian matrix $\hat{H}_\mathrm{CI}$ in the basis set $\ket{\chi_{ij}}$ follows from straightforward commutations in Eq.~\eqref{eq:Hint} \cite{Sherrill99,Rontani06}. For a single-particle basis set size $N\approx 100$, the dimension of $\hat{H}_\mathrm{CI}$, $M=N(N-1)/2\approx 5000$, is still inappropriate for, e.g., fast explorations of the stability diagram or time-dependent simulations. To describe the low-energy physics we are interested in, we may, therefore, want to build a reduced interacting Hamiltonian matrix $\hat{H}_2$ in a subspace of $M^\prime<M$ two-particle states $\ket{\Psi_n}$ already dressed by Coulomb interactions. As in the single-particle case, the chosen $\Psi_n$'s may be the lowest-lying eigenstates of $\hat{H}_\mathrm{CI}$ at the reference bias $\{V_\mathrm{G}^0\}$, or a combination of such eigenstates at different $\{V_\mathrm{G}^k\}$'s (orthogonalized against each other). In that case, the low-energy eigenstates of $\hat{H}_2$ are ``exact'' at each $\{V_\mathrm{G}^k\}$ (as compared to the original $\hat{H}_\mathrm{CI}$). The $M^\prime\times M^\prime$ matrices of $H_0$, the $D_\mathrm{G}$'s and $M_\alpha$'s, and the Coulomb interaction can be computed once for all (first in the single-particle basis set $\ket{\psi_n}$, then in the CI basis set $\ket{\chi_{ij}}$ and finally in the dressed basis set $\ket{\Psi_n}$) and combined to construct $\hat{H}_2$ for arbitrary magnetic fields and $\delta V_\mathrm{G}$'s. \hl{A similar basis of many-body wave functions was also used for time-dependent calculations in Ref.~\cite{Castelano2018}.}

As demonstrated below, we can achieve an accurate description of the DQD physics over the operational gate voltages range with as low as $N=96$ and $M^\prime=112$ states. We give additional details about the implementation in Appendix \ref{app:details}. 

\subsection{Convergence}
\label{ssec:cvg}

We illustrate the above methodology on a symmetric DQD shaped by bias voltages $V_\mathrm{L}=V_\mathrm{R}=-40$\,mV with all B gates grounded. \hl{We do not take into account the inhomogeneous strains imposed by the differential thermal contraction of materials in this section.}

We first discuss the single-particle tunnel coupling $|t|$ between the two dots as a function of $V_\mathrm{J}$, as extracted from the gap $\Delta=2|t|$ between the bonding ground-state and anti-bonding first excited state of the DQD. The tunnel coupling $t$ spans orders of magnitudes and is exponentially dependent on the barrier height and width. It provides, therefore, a very stringent test of the convergence in the single-particle basis set. It is, moreover, a key ingredient of the many-body exchange interaction between the two quantum dots.

\begin{figure}[t]
\centering
\includegraphics[width=.9\linewidth]{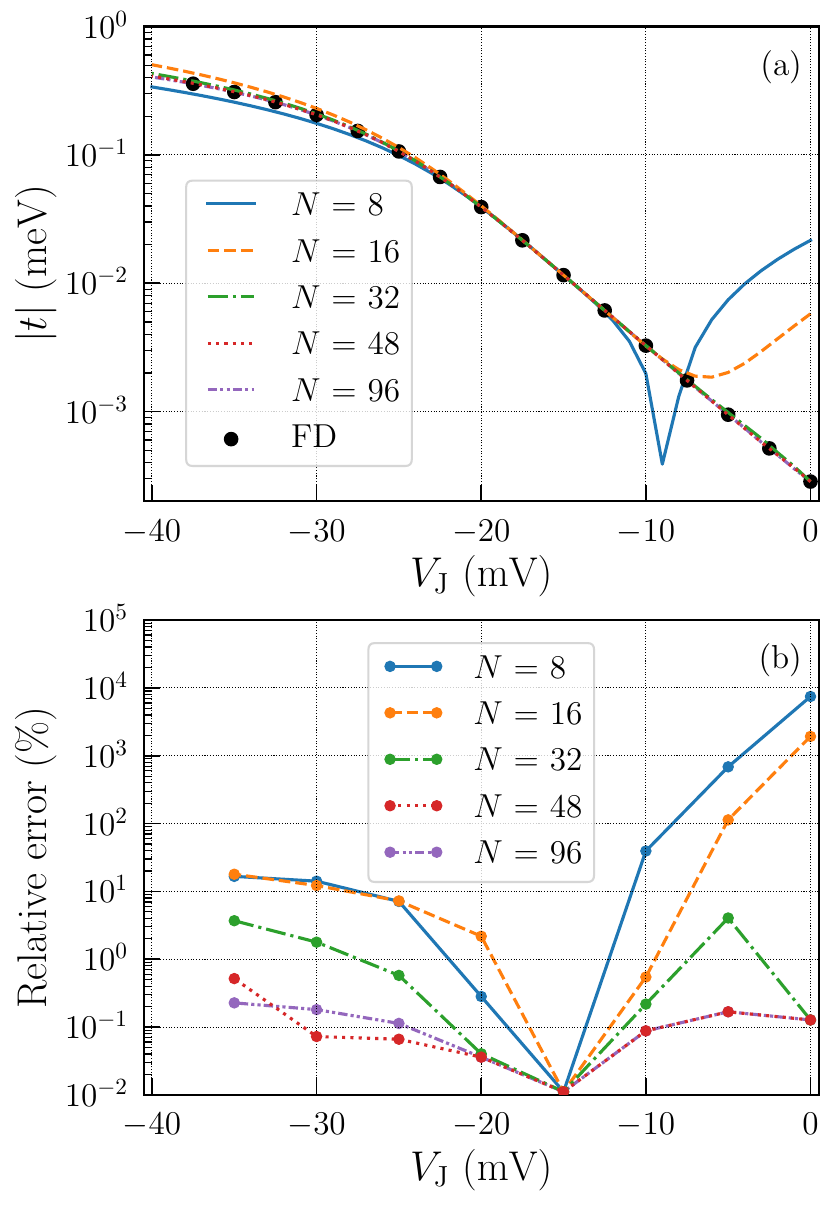}
\caption{(a) The tunnel coupling $|t|$ as a function of $V_\mathrm{J}$, calculated on the FD mesh and in finite basis sets of $N=8$ to $96$ lowest-lying eigenstates $\ket{\psi_n(V_\mathrm{J}^0=-15\,\mathrm{mV})}$. (b) The error in the finite basis sets (with respect to the FD solution). The B gates are grounded and $V_\mathrm{L}=V_\mathrm{R}=-40$\,mV.}
\label{fig:tVJ}
\end{figure}

The ``exact'' $|t|$, calculated on the FD mesh, is compared in Fig.~\ref{fig:tVJ} to the values computed in finite basis sets of $N=8$ to $96$ low-lying eigenstates $\ket{\psi_n}$ at $V_\mathrm{J}^0=-15$\,mV. The error is, by design, minimal at $V_\mathrm{J}=V_\mathrm{J}^0$ (where it shall actually be zero in the absence of numerical round-off errors \footnote{All single-particle states $\ket{\psi_n}$ and energies $E_n=\bra{\psi_n}H_0\ket{\psi_n}$ are converged so that $|H_0\ket{\psi_n}-E_n\ket{\psi_n}|<10^{-10}$ Hartree.}). It remains smaller than $0.5\%$ over the whole investigated $V_\mathrm{J}$ range for $N=96$ (while $|t|$ spans three orders of magnitude). A similar agreement can be reached for other values of $V_\mathrm{J}^0$ in this range, although the error usually grows when $|V_\mathrm{J}-V_\mathrm{J}^0|$ increases. It is, therefore, best to select $V_\mathrm{J}^0$ (and other $V_\mathrm{G}^0$'s) near the middle of the targeted bias range. We did not achieve significant improvements by mixing $\psi_n$'s from different $V_\mathrm{J}^0$'s in this particular system.

\begin{figure}[t]
\centering
\includegraphics[width=.9\linewidth]{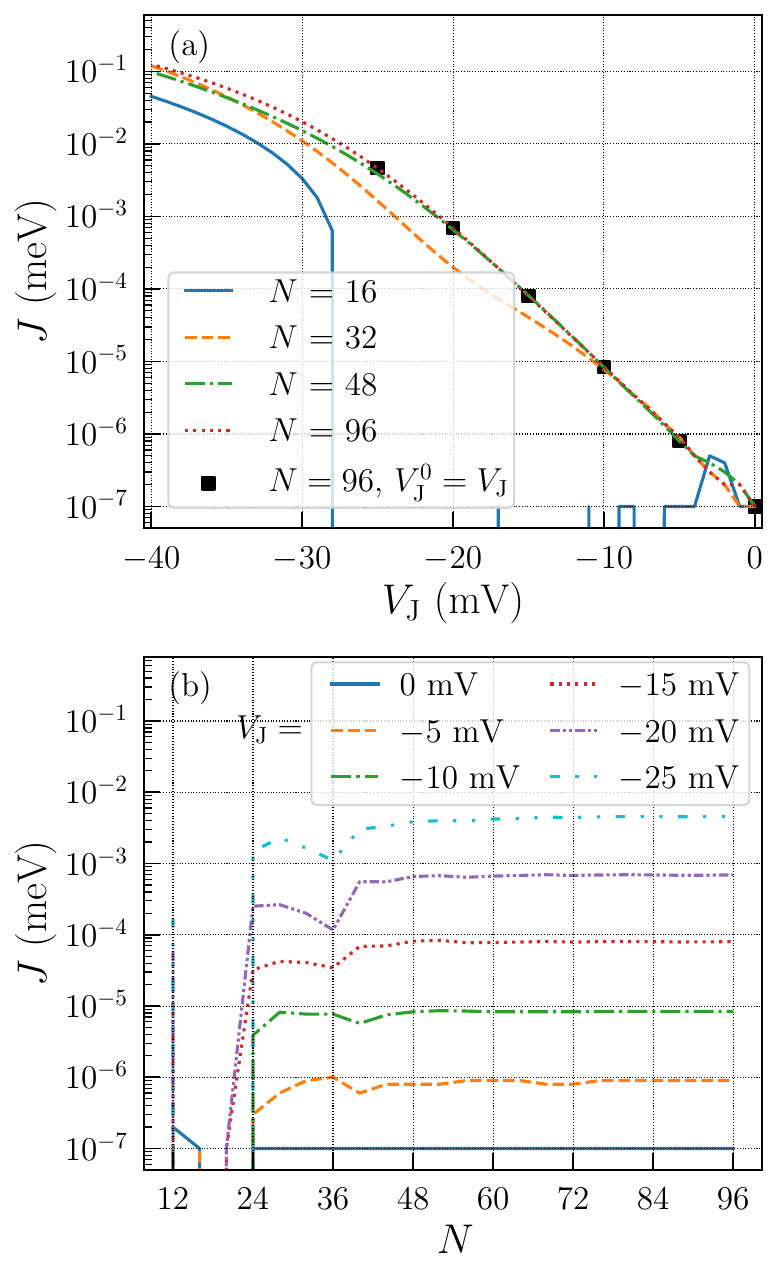}
\caption{(a) Exchange energy $J$ as a function of $V_\mathrm{J}$, for different sizes of the single-particle basis set $N$ and CI basis set $M=N(N-1)/2$. The $N$ single-particle wave functions are either computed at $V_\mathrm{J}^0=V_\mathrm{J}$ (dots), or borrowed from a unique $V_\mathrm{J}^0=-15$\,mV (lines). (b) Exchange energy $J$ at $V_\mathrm{J}=0$, $-5$, $-10$, $-15$, $-20$, $-25$\,mV as a function of the number of single-particle states $N$ calculated at  $V_\mathrm{J}^0=-15$\,mV.}
\label{fig:JVJ}
\end{figure}

We next assess the convergence of the exchange energy at this symmetric operation point. At zero magnetic field, the two-particle states split into singlets and triplets. The exchange energy is defined as $J=E_T-E_S$, with $E_S$ and $E_T$ the ground-state singlet and triplet energies. $J$ is plotted as a function of $V_\mathrm{J}$ in Fig.~\ref{fig:JVJ}, for different sizes of the single-particle basis set $N$ and CI basis set $M=N(N-1)/2$. The black $J(V_\mathrm{J})$ dots are computed with the single-particle states $\ket{\psi_n(V_\mathrm{J})}$ at the same $V_\mathrm{J}$, while the lines are computed with a unique set of 96 $\ket{\psi_n(V_\mathrm{J}^0)}$ at $V_\mathrm{J}^0=-15$\,mV. The exchange energy is found negative over a large range of gate voltages for $N=16$ (see section \ref{sec:physics}), so that data are missing in the log scale plot. Although there is no ``exact'' reference for this quantity, we achieve good relative convergence for $N=96$ ($M=4560$) in the whole range of exchange gate voltage $-20\le V_\mathrm{J}\le -5$\,mV and plunger gate detuning $|\delta V_\mathrm{d}|\le 15$\,mV (the latter being defined as $\delta V_\mathrm{d}=\delta V_\mathrm{L}-\delta V_\mathrm{R}$, with $\delta V_\mathrm{R}=-\delta V_\mathrm{L}$ in symmetric dots). In this range, the median relative variation of $E_T-E_S$ from $N=48$ to $N=96$ is $2.65\%$.

\begin{figure}[t]
\centering
\includegraphics[width=.9\linewidth]{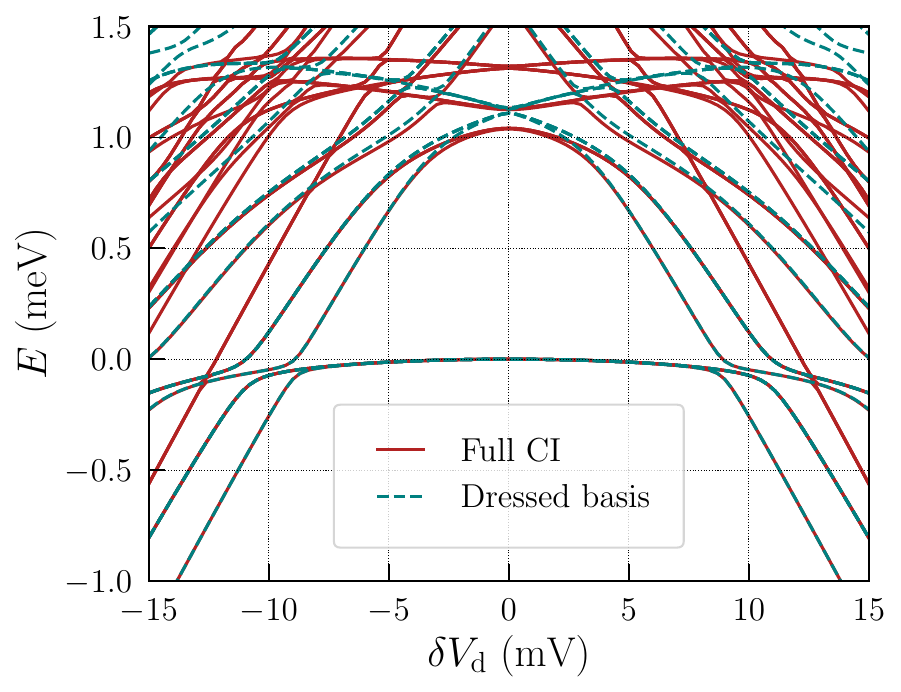}
\caption{The two-hole spectrum as a function of the detuning voltage $\delta V_\mathrm{d}=2\delta V_\mathrm{L}=-2\delta V_\mathrm{R}$ with respect to the bias point $V_\mathrm{L}=V_\mathrm{R}=-40$\,mV, $V_\mathrm{J}=-15$ mV, computed in the original CI basis set ($M=4560$) and in the dressed basis set ($M^\prime=112)$.}
\label{fig:cmpspectrum}
\end{figure}

\begin{figure}[t]
\centering
\includegraphics[width=.9\linewidth]{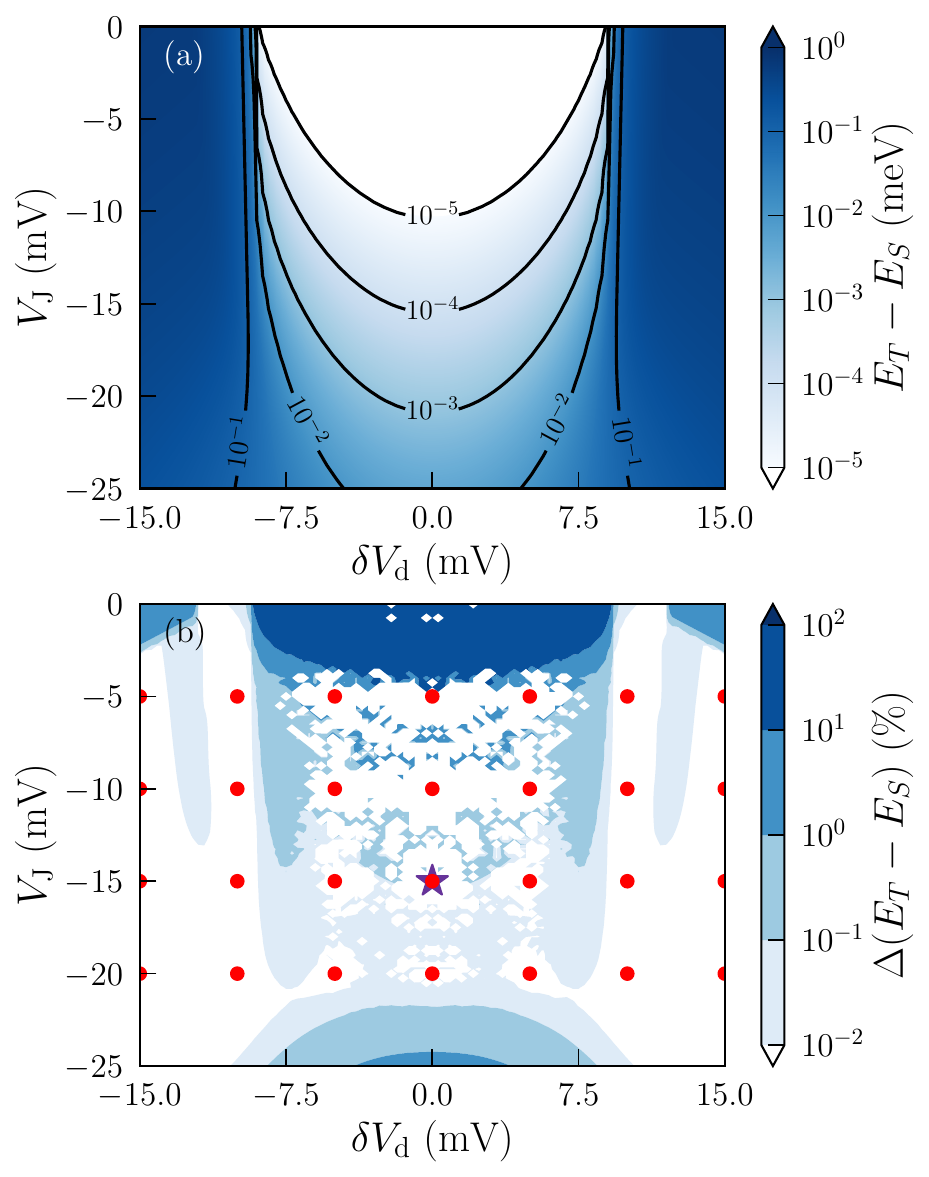}
\caption{(a) The difference $\Delta_{ST}=E_T-E_S$ between the ground triplet and singlet state energies in the $(\delta V_\mathrm{d},V_\mathrm{J})$ plane, calculated in the dressed $M^\prime=112$ basis set. $\Delta_{ST}$ is the exchange energy $J$ in the $(1,1)$ charge sector $|\delta V_\mathrm{d}|\lesssim 8$\,meV. (b) The difference between $\Delta_{ST}$ computed in the full CI ($M=4560$) and in the dressed ($M^\prime=112$) basis sets as a function of $V_\mathrm{J}$ and $V_\mathrm{d}$. The dressed basis set is constructed by picking the ground singlet and triplet states at the 28 red points (including the star). The full CI basis set is built from 96 single-particle states at the red star.}
\label{fig:JVdVJ}
\end{figure}

Finally, we address the accuracy of the dressed two-particle basis sets. We build a basis set $\ket{\Psi_n}$ comprising the lowest singlet and triplet eigenstates on a grid of J gate voltages $V_\mathrm{J}=-5$, $-10$, $-15$, and $-20$\,mV and plunger gate detunings $\delta V_\mathrm{d}=-15$, $-10$, $-5$, $0$, $5$, $10$, and $15$\,mV. These $M^\prime=112$ $\Psi_n$'s are computed in the basis of $M=4560$ Slater determinants built from the $96$ single-particle wave functions at $V_\mathrm{J}^0=-15$\,mV and $\delta V_\mathrm{d}^0=0$. As an illustration, the two-hole spectrum in the original $M=4560$ and dressed $M^\prime=112$ basis sets are plotted as a function of $\delta V_\mathrm{d}$ in Fig.~\ref{fig:cmpspectrum} ($V_\mathrm{J}=-15$\,meV). The mostly horizontal lines are $(1,1)$ charge states of the DQD, whereas the oblique lines are $(2,0)$ and $(0,2)$ states that anti-cross the former owing to inter-dot tunneling (see section \ref{sec:physics}). The dressed basis set reproduces very well the low-lying states over the whole range of detunings. Remarkably, there are excited states missing in the dressed basis set. These are $(0,2)$ and $(2,0)$ states that do not mix with the ground $(1,1)$ states for symmetry reasons. They can not, therefore, be captured by sampling the ground singlet and triplet states in the $(\delta V_\mathrm{d},V_\mathrm{J})$ plane. However, they hardly take part in exchange-driven operations in that plane \footnote{They can be slightly recoupled to $S(1,1)$ by the magnetic field (that breaks symmetries), but make negligible contributions.}. The difference $\Delta_{ST}=E_T-E_S$ between the ground triplet and singlet state energies (which is the exchange $J=\Delta_{ST}$ in the $(1,1)$ sector $|\delta V_\mathrm{d}|\lesssim 8$\,mV), as well as the relative difference between $\Delta_{ST}$ in the $M=4560$ and $M^\prime=112$ basis sets are plotted in the $(\delta V_\mathrm{d},V_\mathrm{J})$ plane in Fig.~\ref{fig:JVdVJ}. As expected, this difference is $\approx 0$ for the points of the sampling grid, and remains typically $<1$\% in-between. The dressed basis set thus matches the original CI calculation in the range $-20<V_\mathrm{J}<-5$\,mV and $-15<\delta V_\mathrm{d}<15$\,mV suitable for the modeling of two qubit operations, yet at a much lower numerical cost.

\section{Physics of the exchange interaction}
\label{sec:physics}

\hl{In this section, we discuss the physics of the exchange interaction in Ge/GeSi quantum dots. We first introduce the minimal CI Hamiltonian for exchange (the Hund-Mulliken model). We then discuss the shortcomings of this minimal model by comparisons with CI calculations for large $M$'s. We show, in particular, that it misses relevant excitations as well as dynamical correlations. They are actually captured by the dressed basis set at a reasonable cost.}

In the simplest picture, the exchange splitting between the $(1,1)$ states results from the competition between tunneling and Coulomb interactions, described by the following (Hund-Mulliken) Hamiltonian in the subspace of the $\{T_0(1,1),S(1,1),S(0,2),S(2,0)\}$ ground-states \cite{Loss98,Burkard99,Burkard2023Review}:
\begin{equation}
H_\mathrm{eff}=
\begin{bmatrix}
-J_c & 0 & 0 & 0 \\
 0 & 0 & \tau_S & \tau_S \\
0 & \tau_S & U-\varepsilon & 0 \\
0 & \tau_S & 0 & U+\varepsilon
\end{bmatrix}\,.
\label{eq:Hexchange}
\end{equation}
Here $\tau_S=\sqrt{2}t$, $\varepsilon$ is the detuning energy with respect to the symmetric operation point and $U$ is the charging energy in the $(2,0)$ and $(0,2)$ states (assumed equal for simplicity) \footnote{It is more precisely the difference between the charging energy in the $(0,2)/(2,0)$ and $(1,1)$ states.}. On the one hand, tunneling from $(1,1)$ to $(0,2)$ and $(2,0)$ states delocalizes the holes and lowers the total energy of the system. The ground $(0,2)$ and $(2,0)$ states are however singlets, so that tunneling (which does not break time-reversal symmetry) gets blocked in the $T(1,1)$ states. This selectively pushes down $S(1,1)$ with respect to the triplet states (all degenerate with the uncoupled $T_0(1,1)$). On the other hand, tunneling to $S(0,2)$ and $S(2,0)$ costs the intra-dot charging energy $U$. Treating tunneling as a perturbation in Eq.~\eqref{eq:Hexchange} yields to second-order:
\begin{equation}
J\approx\frac{2U\tau_S^2}{U^2-\varepsilon^2}-J_c\,.
\label{eq:J}
\end{equation}
$J_c$ is a pure Coulomb exchange correction to $J$. Indeed, the Coulomb interaction favors triplets, whose wave functions are anti-symmetric with respect to the exchange of positions (but symmetric with respect to the exchange of spins, in contrast to singlets). Holes in a $T(1,1)$ state can not, therefore, be at the same position and repel each other less than in the $S(1,1)$ state. $J_c$ also decreases as the barrier is closed and the holes get further separated in the $(1,1)$ state, and is often neglected with respect to $2\tau_S^2/U$. In a plot of the singlet/triplet energies with respect to $\varepsilon$, the exchange then emerges from the anti-crossing (admixture) between the $S(1,1)$ and $S(2,0)/S(0,2)$ states that bends the former down.

\begin{figure}[t]
\centering
\includegraphics[width=.9\linewidth]{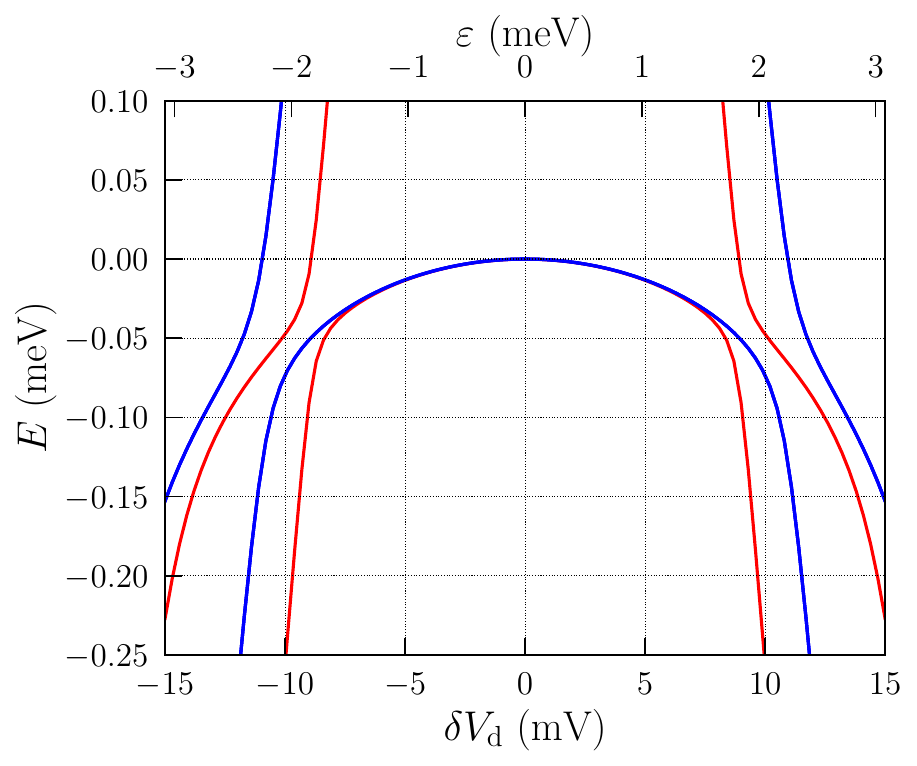}
\caption{The two-hole spectrum of the DQD as a function of the detuning voltage $\delta V_\mathrm{d}$ (close-up of Fig.~\ref{fig:cmpspectrum}). The singlet states are colored in red and the triplet states in blue. The calculations are performed in the dressed two-particle basis set with size $M^\prime=112$. The top scale is the detuning energy $\varepsilon$.}
\label{fig:spectrum}
\end{figure}

We discuss below the quantitative relevance of this model in Ge/SiGe quantum dots, and address, in particular, the complex role played by Coulomb correlations. 

The low-energy spectrum of the DQD is plotted in Fig.~\ref{fig:spectrum} as a function of the detuning voltage $\delta V_\mathrm{d}=2\delta V_\mathrm{L}=-2\delta V_\mathrm{R}$ with respect to the bias point $V_\mathrm{L}=V_\mathrm{R}=-40$\,mV ($V_\mathrm{J}=-15$ mV). This close-up of Fig.~\ref{fig:cmpspectrum} highlights the singlets in red and the triplets in blue. \hl{The anti-crossings between the singlet states are clearly visible on this figure; however, Eq.~\eqref{eq:Hexchange} obviously misses the nearby anti-crossings between the $T(1,1)$ and $T(0,2)/T(2,0)$ states, whose effects on the exchange energy will be addressed later.}

We focus on the ground-state singlets to start with. The $S(1,1)$ state anti-crosses the $S(0,2)$ and $S(2,0)$ states at $\delta V_\mathrm{d}=\pm\delta V_\mathrm{d}^*=\pm8.92$\,mV. To convert $\delta V_\mathrm{d}$'s into relevant energies, we introduce the detuning lever arm $\alpha_\mathrm{d}=\partial\varepsilon/\partial V_\mathrm{d}$. From the expectation values of the $D_\mathrm{L}$ and $D_\mathrm{R}$ operators, we estimate $\alpha_\mathrm{d}=0.205$. We can next assess the charging energy $U\approx \alpha_\mathrm{d}\delta V_\mathrm{d}^*=1.83$\,meV. This is in good agreement with a direct CI calculation $U=1.97$\,meV at $V_\mathrm{L}=V_\mathrm{J}=0$ (i.e., with the left dot emptied) \footnote{This net charging energy is computed as $U=U_{02}-U_{11}$, where $U_{02}=2.32$\,meV is the charging energy in the $(0,2)$ state (beyond the anti-crossing) and $U_{11}=0.346$\,meV is the charging in the $(1,1)$ state (at the symmetric operation point). The former is the difference $U_{02}=E[S(0,2)]-E[S^*(0,2)]$ between the energies $E[S(0,2)]$ and $E[S^*(0,2)]=2E_0$ of the interacting and non-interacting singlets, respectively, with $E_0$ the single-particle ground-state energy. Likewise, $U_{11}=E[S(1,1)]-E[S^*(1,1)]=E[S(1,1)]-2E_b$, with $E_b$ the bonding ground-state energy at zero detuning.}. The anti-crossing gaps at $\delta V_\mathrm{d}=\pm\delta V_\mathrm{d}^*$ are $\Delta_S=2|\tau_S|=80$\,$\mu$eV. Therefore, $|\tau_S|=40$\,$\mu$eV is significantly larger than the value $|\tau_S^\mathrm{sp}|=\sqrt{2}|t|$ expected from the single-particle tunneling $|t|=11.6$\,$\mu$eV extracted at the $(1,0)/(0,1)$ anti-crossing at zero detuning. The shape of the barrier is indeed different at finite detuning, and, as discussed below, $\tau_S$ is altered by Coulomb interactions.

We proceed with a more detailed assessment of Eq.~\eqref{eq:Hexchange}. \hl{For that purpose, we construct the CI basis set from the ground bonding and anti-bonding single-particle states of the LK Hamiltonian at the symmetric operation point. The $M=6$ Slater determinants built from these $N=4$ bonding and anti-bonding states are actually linear combinations of the $\{T_\alpha(1,1),S(1,1),S(0,2),S(2,0)\}$ states acted upon by the Hund-Mulliken Hamiltonian (they span the same subspace).} We define $J_0$ as the exchange energy in this CI basis set. Once we have identified the singlet/triplet eigenstates of ${\hat H}_\mathrm{CI}$ and computed $J_0$, we diagonalize the detuning operator $D_\mathrm{d}=D_\mathrm{L}-D_\mathrm{R}$ in the singlet subspace in order to split the pure $S(1,1)$ state (with eigenvalue 0) from the pure $S(0,2)$ and $S(2,0)$ states (with eigenvalues $\pm d$). We finally rotate ${\hat H}_\mathrm{CI}$ in the basis set ${\cal B}=\{T_\alpha(1,1),S(1,1),S(0,2),S(2,0)\}$ of the pure singlets and triplets and map it to Eq.~\eqref{eq:Hexchange}. We extract that way $J_c$, $\tau_S$, and $U$. 

\begin{figure}[t]
\centering
\includegraphics[width=.9\linewidth]{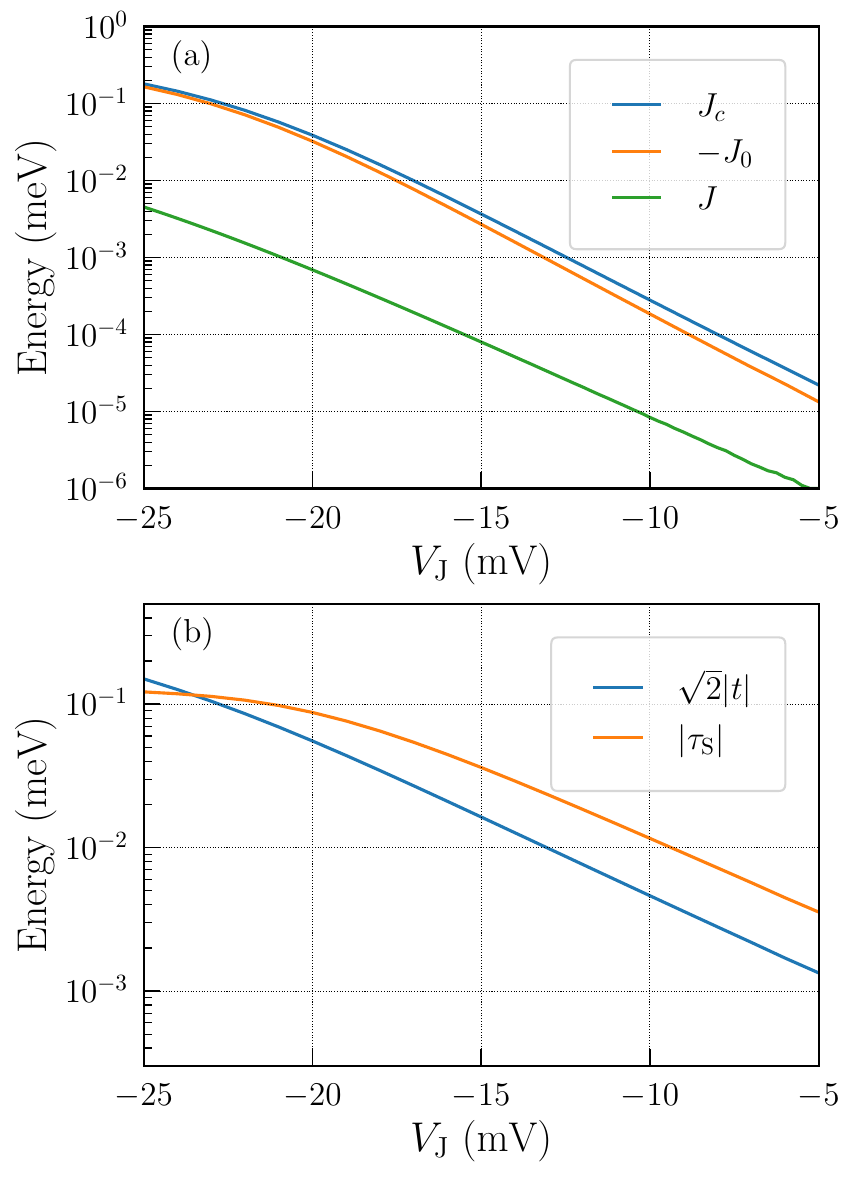}
\caption{(a) Exchange energies $J$ and $-J_0$, and Coulomb exchange correction $J_c$ as a function of $V_\mathrm{J}$ at the symmetric operation point $\delta V_\mathrm{d}=0$. (b) Tunnel couplings $|\tau_S|$ and $\sqrt{2}|t|$ as a function of $V_\mathrm{J}$ at $\delta V_\mathrm{d}=0$.}
\label{fig:Jandt}
\end{figure}

\hl{Fig.~\ref{fig:Jandt}} actually shows that this minimal basis set fails to achieve even qualitative agreement with the converged CI \cite{Calderon06,Pedersen07}. Its deficiencies nonetheless highlight some key features of the exchange interaction. It turns out that the triplet is the ground-state in the basis set ${\cal B}$, at variance with the theory of Ref. \cite{Lieb62} (which only holds, however, in complete basis sets). $J_c$ is indeed larger than the ``kinetic'' exchange $2\tau_S^2/U$, so that $J_0$ is negative. We plot for comparison $J_c$ and $-J_0$ as a function of $V_\mathrm{J}$ in Fig.~\ref{fig:Jandt}a, as well as the value of $J$ from the CI calculation in the dressed $M^\prime=112$ basis set. The net exchange interaction is thus ruled by strong ``dynamical'' Coulomb correlations outside the quasi-degenerate ground bonding/anti-bonding subspace.

As discussed previously, triplets are favored over singlets by Coulomb interactions owing to the formation of an ``exchange hole'' in the pair correlation function of the two particles: due to Pauli exclusion principle, two holes in a triplet state can not sit at the same position, while two holes in a singlet can, which strengthens their interaction. The large $J_c$ in Fig.~\ref{fig:Jandt}a originates in the barrier where the two holes leak in the $(1,1)$ states. This is why $J_c$ decreases exponentially with $V_\mathrm{J}$ as does $t$. The minimal basis set ${\cal B}$ does not, however, allow for correlations between the motions of the two particles (beyond those imposed by anti-symmetry). Yet the two holes practically tend to avoid each other owing to their mutual repulsion. This digs an additional ``Coulomb hole'' \cite{Hedin65} in the pair correlation function of both singlets and triplets that reduces the advantage of the latter \hl{(the singlet being the expected ground-state at zero magnetic field \cite{Lieb62})}. Such dynamical correlations can only be captured in an extended basis set of Slater determinants.

Moreover, the effective tunneling strength between the two dots is significantly affected by Coulomb correlations. The effect of Coulomb interactions on the coupling between the $(1,1)$ and $(2,0)/(0,2)$ states is already sizeable in the minimal basis set ${\cal B}$. We compare in Fig.~\ref{fig:Jandt}b the value of $\tau_S$ extracted in basis set ${\cal B}$ with the expected $\sqrt{2}t$ from Fig.~\ref{fig:tVJ}. They differ significantly over the whole $V_\mathrm{J}$ range. Indeed, the Coulomb interaction has non zero matrix elements between the $S(1,1)$ and $S(0,2)/S(2,0)$ states, and even between the $S(0,2)$ and $S(2,0)$ states themselves (although they are all much smaller than the charging energy $U$). The Coulomb interaction thus plays a key role in shaping the net interactions between the $(1,1)$ and $(0,2)/(2,0)$ charge sectors.

Another quantitative flaw of Eq.~\eqref{eq:Hexchange} is that it misses the curvature of the triplet states \cite{White18,Abadillo21}. Indeed, the triplet $T(1,1)$ state anti-crosses $T(2,0)$ and $T(0,2)$ states at higher detuning and thus also bends down, which reduces the net exchange energy but is not accounted for by Eq.~\eqref{eq:Hexchange}. This correction is controlled by the gap $\Delta_{ST}$ between the $S(0,2)$ and $T(0,2)$ states (or between the $S(2,0)$ and $T(2,0)$ states), and by the triplet tunneling strength $\tau_T$. The Coulomb interaction has naturally a strong impact on the $(0,2)$ states, and leads, in particular, to a renormalization of $\Delta_{ST}$. In a naive, weakly interacting picture, $\Delta_{ST}^0=\varepsilon_1-\varepsilon_0$ is the difference between the first excited ($\varepsilon_1$) and ground ($\varepsilon_0$) single-particle energy levels of the right quantum dot (as a triplet can not be built from a single level). As discussed above, Coulomb interactions however favor triplets over singlets, which lowers the actual splitting $\Delta_{ST}<\Delta_{ST}^0$ \cite{Merkt91}. This reduction is significant in the present device. A CI calculation with the left dot empty at $V_\mathrm{L}=0$\,V and $V_\mathrm{J}=0$\,V indeed yields $\Delta_{ST}=0.47$\,meV while $\Delta_{ST}^0=1.32$\,meV. This estimate of $\Delta_{ST}$ is roughly consistent with the splitting between the singlet and triplet anti-crossings on Fig.~\ref{fig:spectrum}, $\Delta V_\mathrm{d}=2$\,mV or $\Delta\varepsilon=0.41$\,meV. The renormalization of $\Delta_{ST}$ is a key ingredient of the physics of the exchange interaction, as a smaller $\Delta_{ST}$ increases the curvature of the triplet states in the $(1,1)$ sector and lowers the exchange energy. We remind that $\Delta_{ST}$ may collapse in elongated quantum dots due to Wigner localization effects \cite{Bryant87,Reimann00,Ellenberg06,Kalliakos08,Singha10,Abadillo21,Corrigan21,e-eCI(TB),ValleytronicCI(EM),Yannouleas22}: Coulomb correlations localize the two holes one on each side of the dot, so that the density looks alike in the $S(0,2)$ and $T(0,2)$ states (which have the fingerprints of an effective $(1,1)$ state built in the right dot). $\Delta_{ST}$ then decreases almost exponentially with the aspect ratio of the dot. Nevertheless, the effects of Coulomb correlations scale with the effective mass, which makes germanium more resilient to Wigner localization than, e.g., silicon. The gap $2\tau_T$ is also visibly larger at the $T(1,1)/T(0,2)$ than at the $S(1,1)/S(0,2)$ anti-crossing on Fig.~\ref{fig:spectrum}, which further enhances triplet curvature. This results from the different symmetry and extension of the states, and from an overall reduction of the effective barrier height when increasing detuning.

The model of Eq.~\eqref{eq:Hexchange} thus provides a qualitative, yet not a quantitative picture of the exchange interactions as given by the numerical simulations. Of course, the $\{T_\alpha(1,1),S(1,1),S(0,2),S(2,0)\}$ subspace may be enriched with $T(0,2)$ and $T(2,0)$ states \cite{White18} and thought as a basis of ``dressed'' states already mixing thousands of Slater determinants and accounting for the main correlations. This is actually the rationale of the methodology introduced in Section \ref{subsec:twop}. However, the present calculations show that higher excited singlet and triplet states must also be included in this dressed basis set to achieve quantitative accuracy over the whole operational gate voltages range. As a trade-off between accuracy and efficiency, the couplings with these higher excited states may be folded in the original subspace with, e.g., a Schrieffer-Wolff transformation, leading to an effective Hamiltonian similar to Eq.~\eqref{eq:Hexchange} but with bias-dependent parameters. The size of the dressed basis constructed in this work ($M^\prime=112$) is, nonetheless, small enough to address intensive calculations such as time-dependent simulations. 

\section{Effects of the magnetic field}
\label{sec:magnetic}

In this section, we illustrate the effects of the magnetic field on the two-particle states. 

In the absence of SOC, the magnetic field simply splits the threefold degenerate triplet into $T_0$, $T_+$ and $T_-$ states. Vertical confinement and SOC however split the $p$-like Bloch functions into HH, LH and split-off subbands with different angular momenta and $g$-factors. The admixture of HH and LH components by lateral confinement and strains modulates the net $g$-factors that can be different in the two dots \cite{Michal21,martinez2022hole,Abadillo2023,Stano25}. Additionally, SOC gives rise to Rashba- and Dresselhaus-like interactions that couple the pseudo-spin and orbital motion (envelopes) of the hole \cite{Marcellina17,Terrazos21,Bosco21b,Abadillo2023,Rodriguez2023}. This mixes the split singlet and triplet states and gives rise to a complex pattern of anti-crossings as a function of magnetic field strength and orientation \cite{jirovec_Freqs_ST-Ge,Jirovec23,rooney_strains_ST-Ge_ST_-,Saezmollejo24}. 

The effects of SOC on coupled quantum dots can be qualitatively understood in the minimal $\{S(2,0),S(0,2),S(1,1),T_0(1,1),T_+(1,1),T_-(1,1)\}$ subspace. For that purpose, we introduce the ground-state orbitals $\ket{L\uparrow}$ and $\ket{L\downarrow}$ of the left dot, as well as the ground-state orbitals $\ket{R\uparrow}$ and $\ket{R\downarrow}$ of the right dot. The pseudo-spin index $\uparrow/\downarrow$ labels the Kramers-degenerate left and right orbitals at $B=0$, and may be chosen differently in the two dots. We next assume the following one-particle Hamiltonian in the $\{\ket{L\uparrow},\ket{L\downarrow},\ket{R\uparrow},\ket{R\downarrow}\}$ basis set \cite{Sen23}
\begin{equation}
\hat{H}_1=\begin{bmatrix}
H_L & T \\
T^\dagger & H_R
\end{bmatrix}\,,
\label{eq:H}
\end{equation}
with $H_L$ and $H_R$ the $2\times 2$ sub-blocks of the left and right dots:
\begin{subequations}
\begin{align}
H_L&=\frac{1}{2}(\varepsilon I_2+\mu_B\vec{\sigma}\cdot g_L\vec{B}) \\
H_R&=\frac{1}{2}(-\varepsilon I_2+\mu_B\vec{\sigma}\cdot g_R\vec{B})\,.
\label{eq:HLR}
\end{align}
\end{subequations}
Here $I_2$ is the $2\times 2$ identity matrix, $\vec{\sigma}$ is the vector of Pauli matrices, and $g_L$, $g_R$ are the $g$-matrices of the left and right dots, respectively \cite{Abragam1970,Venitucci18}. The inter-dot tunneling block $T$ takes the following form owing to time-reversal symmetry at $B=0$:
\begin{equation}
T=\begin{bmatrix}
t_1 & t_2 \\
-t_2^* & t_1^*
\end{bmatrix}\,.
\label{eq:T}
\end{equation}
We emphasize that $\{\ket{L\uparrow},\ket{L\downarrow}\}$ and $\{\ket{R\uparrow},\ket{R\downarrow}\}$ are both defined up to a unitary transformation (i.e., choice of pseudo-spin). Therefore, the set of matrices $g_L$, $g_R$ and $T$ is not unique \cite{Chiboratu08}. Since such unitary transformations preserve the determinant of $T$,
\begin{equation}
t=\sqrt{|t_1|^2+|t_2|^2}    
\end{equation}
is, nonetheless, independent on the choice of pseudo-spins. Practically, the pseudo-spins can always be chosen so that $T=tI_2$ is diagonal \cite{Geyer24}. In such a ``spin-orbit frame'' \footnote{The spin-orbit frame is not uniquely defined either, as a same unitary transform on both dots leaves $T$ invariant. A particular spin-orbit frame can always be built starting from given $g_L$, $g_R$ matrices and tunneling block $T$ parametrized as $T=t(\cos\theta_\mathrm{so}I_2-i\sin\theta_\mathrm{so}\vec{n}_\mathrm{so}\cdot\vec{\sigma})$, with $\vec{n}_\mathrm{so}$ and $\theta_\mathrm{so}$ the so-called spin-orbit vector and spin-orbit angle. Indeed, the transformations $U_L=\exp(-i\theta_\mathrm{so}\vec{n}_\mathrm{so}\cdot\vec{\sigma}/2)$ in the $\{\ket{L\uparrow},\ket{L\downarrow}\}$ subspace and $U_R=\exp(i\theta_\mathrm{so}\vec{n}_\mathrm{so}\cdot\vec{\sigma}/2)$ in the $\{\ket{R\uparrow},\ket{R\downarrow}\}$ subspace diagonalize the $T$ block $\tilde{T}=U_L^\dagger TU_R=tI_2$. The $g$-matrices then become $\tilde{g}_L=R(\vec{n}_\mathrm{so},-\theta_\mathrm{so})g_L$ and $\tilde{g}_R=R(\vec{n}_\mathrm{so},\theta_\mathrm{so})g_R$ with $R(\vec{u},\theta)$ the matrix of the real-space rotation of angle $\theta$ around the unit vector $\vec{u}$ \cite{Geyer24,Saezmollejo24}.}, the effects of SOC are fully embedded in the $g$-matrices $\tilde{g}_L$ and $\tilde{g}_R$ (whose analysis may, however, be intricate). Alternatively, $\ket{\uparrow}$ and $\ket{\downarrow}$ may be chosen, respectively, as the states with largest $\ket{+\tfrac{3}{2}}$ and $\ket{-\tfrac{3}{2}}$ components in each dot. In that case, the effects of SOC are distributed between the $g$-matrices and $T$ block, which are, however, more easily analyzed against perturbation theories of pure HH states \cite{martinez2022hole,Abadillo2023}. The observables are independent on this basis choice, including the effective $g$-factors of the left dot, $g_L^*=|g_L\vec{b}|$, and right dot, $g_R^*=|g_R\vec{b}|$, with $\vec{b}=\vec{B}/B$ the unit vector along the magnetic field.

\begin{figure}[t]
\centering
\includegraphics[width=.9\linewidth]{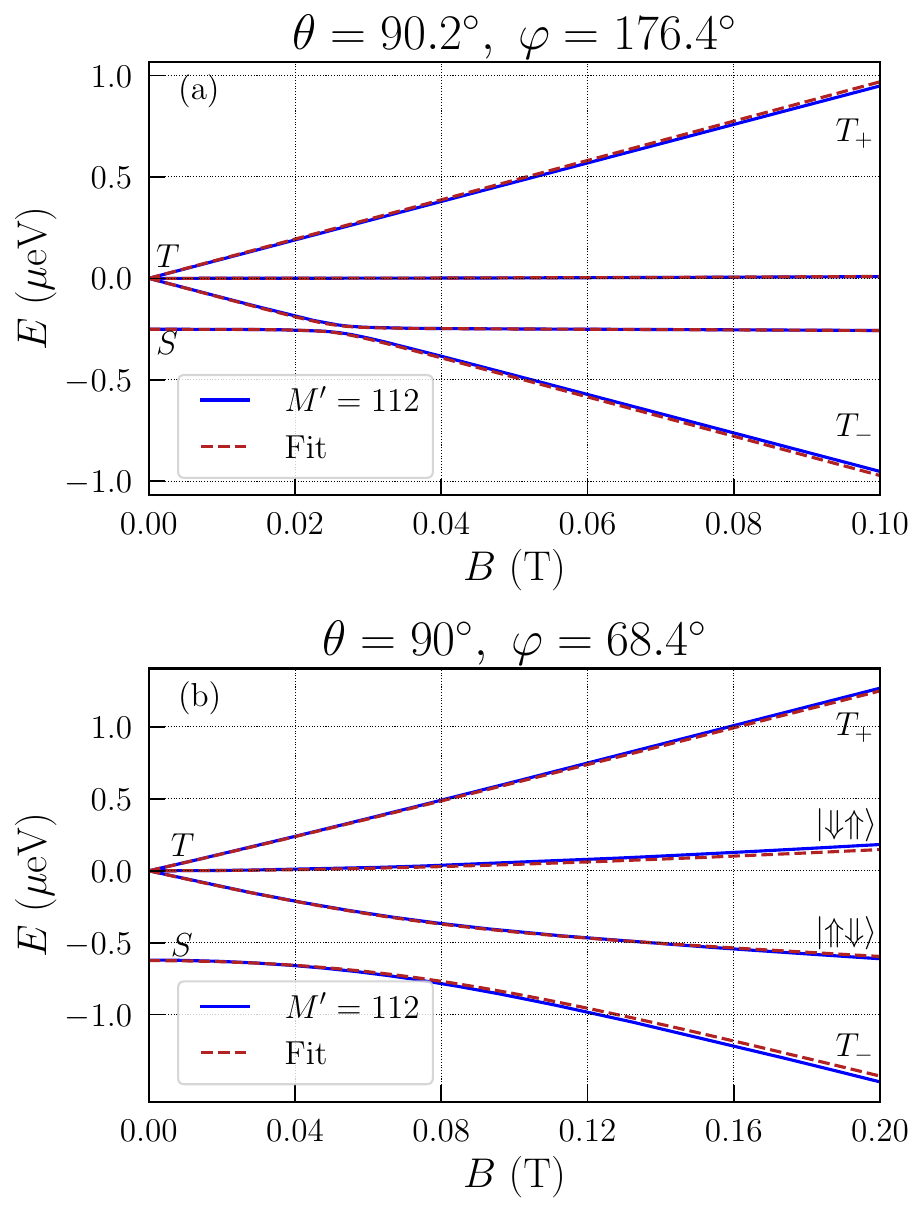}
\caption{Singlet and triplet energies as a function of the magnetic field amplitude in (a) circular (B gates grounded) and (b) squeezed dots ($V_{\mathrm{B}_1}=V_{\mathrm{B}_5}=50$\,mV and $V_{\mathrm{B}_2}=V_{\mathrm{B}_6}=-15$\,mV). The bias point is $\delta V_\mathrm{d}=6$\,mV, $V_\mathrm{J}=-15$ mV for the circular dot and $\delta V_\mathrm{d}=6.1$\,mV, $V_\mathrm{J}=-19.8$ mV for the squeezed dot. The orientation of the magnetic field, characterized by the angles $\theta$ and $\varphi$ defined in Fig.~\ref{fig:device}, is chosen to maximize the gap $\Delta_{ST_-}$ at the $S/T_-$ anti-crossing. The blue line is the numerical data computed in the dressed $M^\prime=112$ basis set, while the dashed red line is the fit with the Hamiltonian~\eqref{eq:H}-\eqref{eq:T}.}
\label{fig:spectrumB}
\end{figure}

For a given $\vec{b}$, we can also choose the $\Uparrow$ and $\Downarrow$ pseudo-spins that diagonalize the Zeeman Hamiltonian in each dot \cite{Geyer24}. Then, 
\begin{subequations}
\begin{align}
H_L&=\frac{1}{2}(+\varepsilon I_2+\mu_B g_L^*B\sigma_z) \\
H_R&=\frac{1}{2}(-\varepsilon I_2+\mu_B g_R^*B\sigma_z)
\end{align}
\end{subequations}
and we define the spin conserving $t_\mathrm{sc}$ and spin flipping $t_\mathrm{sf}$ from the tunneling block in this ``qubit frame'':
\begin{equation}
T=\begin{bmatrix}
t_\mathrm{sc} & t_\mathrm{sf} \\
-t_\mathrm{sf}^* & t_\mathrm{sc}^*
\end{bmatrix}\,.
\end{equation}
We emphasize that $t_\mathrm{sc}$ and $t_\mathrm{sf}$ depend on the orientation $\vec{b}$ of the magnetic field (although $t=\sqrt{|t_\mathrm{sc}|^2+|t_\mathrm{sf}|^2}$ does not), and that $t_\mathrm{sc}$ can always be made real by an appropriate choice of phases for the $L$ and $R$ orbitals. In fact, $|t_\mathrm{sc}|$ and $|t_\mathrm{sf}|$ have very simple expressions as a function of the $g$-matrices $\tilde{g}_L$ and $\tilde{g}_R$ in a spin-orbit frame (see also note \cite{Note8}):
\begin{subequations}
\label{eq:tsf}
\begin{align}
|t_\mathrm{sc}|&=|t\cos(\Theta_{LR}/2)| \\
|t_\mathrm{sf}|&=|t\sin(\Theta_{LR}/2)|\,,
\end{align}
\end{subequations}
where $\Theta_{LR}$ is the angle between the Larmor vectors $\tilde{\vec{\omega}}_L=\mu_B\tilde{g}_L\vec{B}/\hbar$ and $\tilde{\vec{\omega}}_R=\mu_B\tilde{g}_R\vec{B}/\hbar$. In general, $t_\mathrm{sf}$ picks contribution from the mismatch of the principal $g$-factors and magnetic axes in the two dots (due to different confinement and strains) and from Rashba/Dresselhaus SOC (see later discussion) \cite{Stepanenko12,froning_strong_SO_2021,jirovec_Freqs_ST-Ge,mutter_all-electrical_ST-Ge,yu2022strong,Jirovec23}. We can next write the two-particle Hamiltonian in the $\{S(2,0),S(0,2),S(1,1),T_0(1,1),T_+(1,1),T_-(1,1)\}$ basis set built upon the $\{\ket{L\Uparrow},\ket{L\Downarrow},\ket{R\Uparrow}\ket{R\Downarrow}\}$ orbitals of the qubit frame
\begin{widetext}
\begin{equation}
\hat{H}_2=\begin{bmatrix}    
U+\varepsilon & 0 & \sqrt{2}t_\mathrm{sc} & 0 & -t_\mathrm{sf}^* & -t_\mathrm{sf} \\
0 & U-\varepsilon & \sqrt{2}t_\mathrm{sc} & 0 & -t_\mathrm{sf}^* & -t_\mathrm{sf} \\
\sqrt{2}t_\mathrm{sc} & \sqrt{2}t_\mathrm{sc} & 0 & \mu_BB\delta g^* & 0 & 0 \\
0 & 0 & \mu_BB\delta g^* & 0 & 0 & 0 \\
-t_\mathrm{sf} & -t_\mathrm{sf} & 0 & 0 & \mu_B B\bar{g}^* & 0 \\
-t_\mathrm{sf}^* & -t_\mathrm{sf}^* & 0 & 0 & 0 & - \mu_B B\bar{g}^* 
\end{bmatrix}\,,
\label{eq:6bandH}
\end{equation}
\end{widetext}
where $\delta g^*=(g_\mathrm{L}^*-g_\mathrm{R}^*)/2$ is the half difference and $\bar{g}^*=(g_\mathrm{L}^*+g_\mathrm{R}^*)/2$ the average of the $g$-factors of the dots. When $B\to0$, this Hamiltonian is the same (up to a unitary transform) as Eq.~\eqref{eq:Hexchange} with $t=\sqrt{t_\mathrm{sc}^2+|t_\mathrm{sf}|^2}$ and $J_c$ neglected. The magnetic field splits the $T_+(1,1)$ and $T_-(1,1)$ states by $\Delta E=2\mu_B B\bar{g}^*$, and mixes the $S(1,1)$ and $T_0(1,1)$ states with strength $\Delta E_Z=2\mu_B B\delta g^*$. This gives rise to a competition between the Zeeman and exchange interactions. At very low magnetic fields, the ground-state is the mixed $S(1,1)/S(0,2)/S(2,0)$ singlet split from the $T_0(1,1)$ triplet by the exchange energy $J$. The magnetic field tends to break this singlet into anti-symmetric $\ket{\Uparrow\Downarrow}$ and $\ket{\Downarrow\Uparrow}$ states, which become the approximate eigenstates at large $\Delta E_Z\gg J$. In the regime $0\ll\varepsilon\ll U$ [where $S(1,1)$ weakly mixes with $S(0,2)$] and at high magnetic field $\Delta E_Z\gg J$, the eigenenergies are actually
\begin{subequations}
\begin{align}
E_{T^-}&\approx -\mu_B B\bar{g}^* \\
E_{\Downarrow\Uparrow}&\approx -\mu_B B\delta g^*-J_\parallel/2 \\
E_{\Uparrow\Downarrow}&\approx +\mu_B B\delta g^*-J_\parallel/2 \\
E_{T^+}&\approx +\mu_B B\bar{g}^* \,,
\end{align}
\end{subequations}
where
\begin{align}
J_\parallel&\approx\frac{1}{2}\left(-\delta\varepsilon+\sqrt{\delta\varepsilon^2+8t_\mathrm{sc}^2}\right)-2\frac{|t_\mathrm{sf}|^2}{\delta\varepsilon} \nonumber \\ 
&\approx\frac{2t^2-4|t_\mathrm{sf}|^2}{\delta\varepsilon}
\label{eq:jsoc}
\end{align}
is the net exchange interaction resulting from the $S(1,1)/S(0,2)$ mixing by $t_\mathrm{sc}$ and the $T_\pm(1,1)/S(0,2)$ mixing by $t_\mathrm{sf}$ (with $\delta\varepsilon=U-\varepsilon$). Note that the exchange splitting $J_\parallel$ at high magnetic field is generally different from $J$ (unless $t_\mathrm{sf}=0$) and dependent (like $t_\mathrm{sc}$ and $t_\mathrm{sf}$) on the orientation of the magnetic field \cite{Hetenyi20,Geyer24}. The crossover between the low and high field regimes is well visible in Fig.~\ref{fig:spectrumB}b, which displays the singlet and triplet energies as a function of the magnetic field amplitude in a squeezed double dot with different $g$-factors (see details below). The exchange shift $-J_\parallel/2$ of the $\ket{\Uparrow\Downarrow}$ and $\ket{\Downarrow\Uparrow}$ states can be exploited to perform, e.g., conditional rotations \cite{Hendrickx21}. 

Another effect of SOC is the appearance of a $ST_-$ anti-crossing at finite magnetic field. This anti-crossing is relevant for the manipulation of singlet-triplet qubits \cite{jirovec_Freqs_ST-Ge,rooney_strains_ST-Ge_ST_-}, and for readout schemes based on Pauli spin blockade \cite{Kelly2025,danon_pauli_2009} (see section \ref{sec:tdsimus}). The $T_-(1,1)$ state is indeed coupled by $t_\mathrm{sf}$ to $S(0,2)$ and $S(2,0)$, then by $t_\mathrm{sc}$ to $S(1,1)$, which is concurrently mixed with $T_0(1,1)$ by $\delta g^*$. In the regime $0\ll\varepsilon\ll U$ and $|t_\mathrm{sf}|\ll t_\mathrm{sc}$, the magnetic field $B_{ST_-}$ and gap $\Delta_{ST_-}$ at the anti-crossing read
\begin{subequations}
\begin{align}
\mu_B\bar{g}^*B_{ST_-}&\approx J_\parallel\frac{(\bar{g}^*)^2}{\left(\delta g^*\sin\frac{\Omega}{2}\right)^2+g^*_Lg^*_R}
\nonumber \\
\Delta_{ST_-}&\approx 2|t_\mathrm{sf}||\sin\frac{\Omega}{2}||\cos\frac{\Omega^\prime}{2}|\,,
\end{align}
\end{subequations}
with:
\begin{subequations}
\begin{align}
\Omega&=\arctan_2(-2\sqrt{2}t_\mathrm{sc}, \delta\varepsilon) \\
\Omega^\prime&=\arctan_2(\Delta E_Z, J_\parallel)\,.
\end{align}
\end{subequations}
The angle $\Omega$ quantifies the admixture of $S(0,2)$ in $S(1,1)$ by tunneling (which opens $\Delta_{ST_-}$), and $\Omega^\prime$ the admixture of $T_0(1,1)$ in $S(1,1)$ by the imbalance of $g$-factors (which closes $\Delta_{ST_-}$).

\begin{figure}[t]
\centering
\includegraphics[width=.9\linewidth]{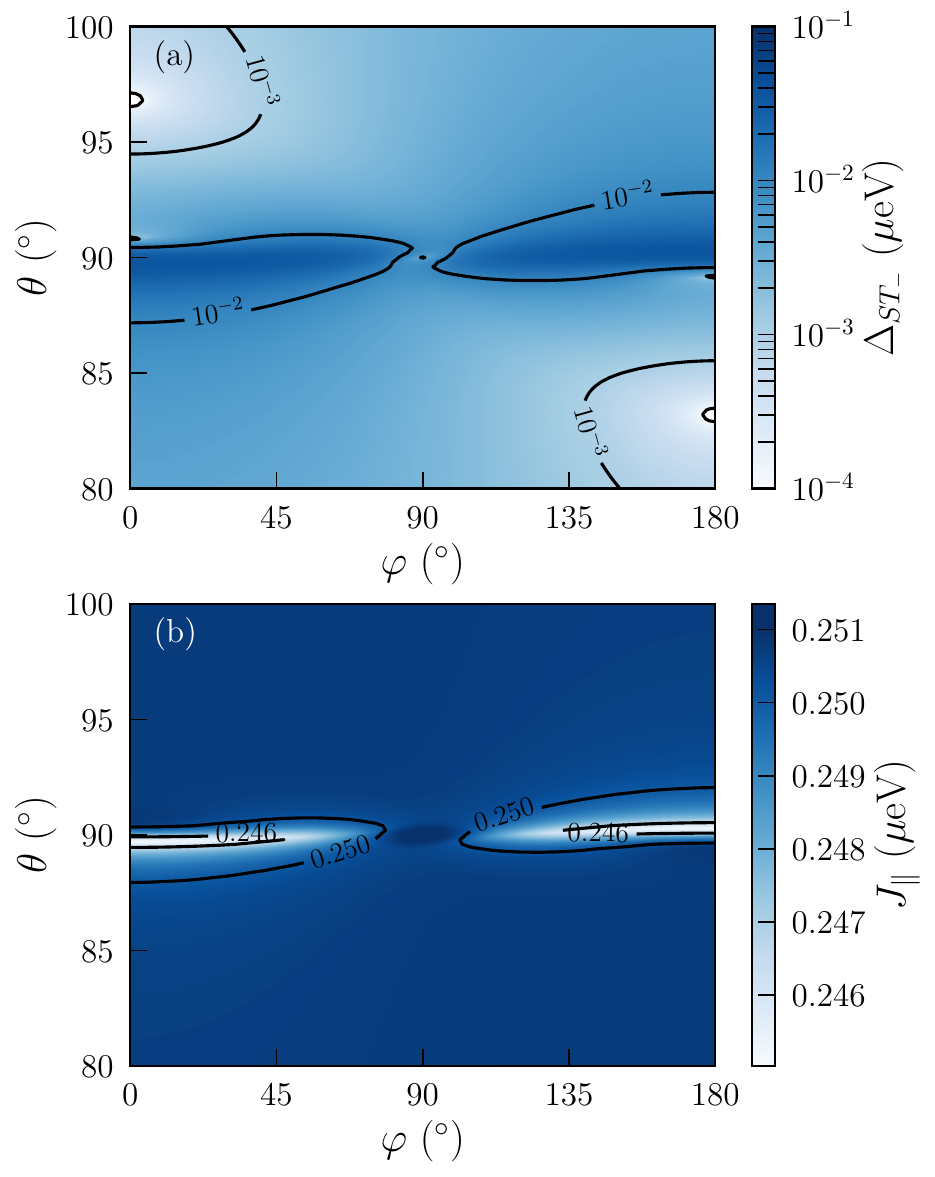}
\caption{(a) $\Delta_\mathrm{ST_-}$ and (b) $J_\parallel$ as a function of the orientation $(\theta,\varphi)$ of the magnetic field, computed with the dressed $M^\prime=112$ basis set in circular dots (all B gates grounded). The bias point is $\delta V_\mathrm{d}=6$\,mV and $V_\mathrm{J}=-15$ mV.}
\label{fig:mapsym}
\end{figure}

\begin{figure}[t]
\centering
\includegraphics[width=.9\linewidth]{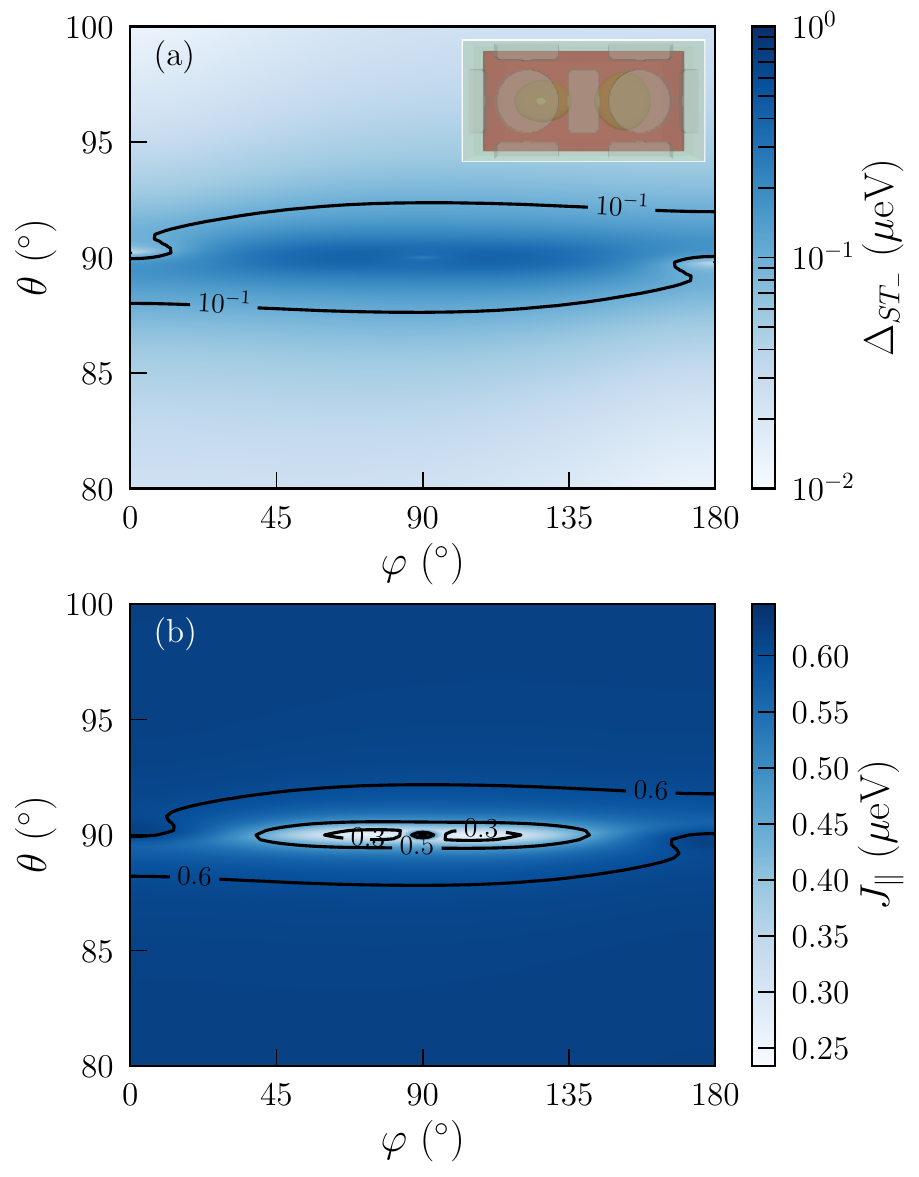}
\caption{(a) $\Delta_\mathrm{ST_-}$ and (b) $J_\parallel$ as a function of the orientation $(\theta,\varphi)$ of the magnetic field, computed with the dressed $M^\prime=112$ basis set in squeezed dots ($V_{\mathrm{B}_1}=V_{\mathrm{B}_5}=50$\,mV and $V_{\mathrm{B}_2}=V_{\mathrm{B}_6}=-15$\,mV). The bias point is $\delta V_\mathrm{d}=6.1$\,mV and $V_\mathrm{J}=-19.8$ mV. The inset in (a) shows the isodensity surface that encloses 90\% of the interacting holes charge at zero magnetic field and zero detuning.}
\label{fig:mapasym}
\end{figure}

As an illustration, we plot in Figs.~\ref{fig:mapsym} and \ref{fig:mapasym} the maps of $J_\parallel$ and $\Delta_{ST_-}$ as a function of the orientation of the magnetic field in two relevant cases: the same quasi-circular dots as in the previous figures (B gates grounded), and slightly squeezed dots ($V_{\mathrm{B}_1}=V_{\mathrm{B}_5}=50$\,mV and $V_{\mathrm{B}_2}=V_{\mathrm{B}_6}=-15$\,mV) \footnote{For the squeezed dots, $\delta V_\mathrm{d}=\delta V_\mathrm{L}-\delta V_\mathrm{R}$ with $\delta V_\mathrm{R}=-1.035\delta V_\mathrm{L}$.}. \hl{We include here the inhomogeneous strains imposed by the differential thermal contraction of metals and insulators, in order to unveil their effects on the exchange interaction and $ST_-$ splitting.} The maps are computed from the dependence of the ground-state singlet and triplet energies on the magnetic field amplitude, illustrated in Fig.~\ref{fig:spectrumB} at specific magnetic field orientations. The field $B_{ST_-}$ and gap $\Delta_{ST_-}$ of the $ST_-$ anti-crossing are found by dichotomy, then the anisotropic exchange energy is estimated as $J_\parallel=E_{T^+}+E_{T^-}-E_{\Uparrow\Downarrow}-E_{\Downarrow\Uparrow}$ at magnetic field $B=4B_{ST_-}$. The use of the dressed basis set enables efficient extractions of these quantities and the production of very detailed maps. 

\begin{table*}[t]
    \caption{Fitted values for the tunnel couplings ($\mu$eV) and $g$-matrices of the circular and squeezed dots.}
    \label{tb:Fits}
    \centering
    \begin{tabular}{*{11}{|c}|}
        \cline{2-11}       
        \multicolumn{1}{c|}{}& \multirow{2}{*}{$t_1$} & \multirow{2}{*}{$t_2$} & \multicolumn{2}{c|}{$g_{xx}$} & \multicolumn{2}{c|}{$g_{yy}$} & \multicolumn{2}{c|}{$g_{zz}$} & \multicolumn{2}{c|}{$g_{zx}$}\\        
        \cline{4-11}
        \multicolumn{1}{c|}{}& & & L & R & L & R & L & R & L & R\\
        \hline
        Circular dots & $-8.27$ & $-0.055$ & $0.173$ & $0.147$ & $-0.104$ & $-0.115$ & $14.23$ & $14.19$ & $-0.0143$ & $0.0123$\\
        \hline
        Squeezed dots & $-11.3$ & $-0.084$ & $0.229$ & $0.120$ & $-0.020$ & $-0.141$ & $13.32$ & $12.82$ & $-0.0141$ & $0.0177$\\
        \hline
        \end{tabular}
\end{table*}

In both circular and squeezed dots, the maps show significant anisotropy, $J-J_\parallel$ and $\Delta_{ST_-}$ being maximum for in-plane magnetic fields. This is consistent with Eq.~\eqref{eq:6bandH} and a spin-flip tunneling matrix element $t_\mathrm{sf}$ that peaks in-plane. In order to rationalize these trends, we have fitted the $g_\mathrm{L}$, $g_\mathrm{R}$ matrices and tunneling block $T$ of the model Hamiltonian [Eqs.~\eqref{eq:H}-\eqref{eq:T}] to the numerical data. The solution is not unique, however, as discussed above. Yet if we would start from pure, real $\nu=\pm 3/2$ envelopes as L/R states, and deal with HH/LH mixing as a perturbation \cite{martinez2022hole,Abadillo2023}, the $g_{L/R}$ matrices would take the following form given the symmetry of the device (a $\sigma_{xz}$ mirror plane):
\begin{equation}
\tilde{g}_{L/R}=\begin{bmatrix}
g_{xx} & 0 & \approx 0 \\
0 & g_{yy} & 0 \\
g_{zx} & 0 & g_{zz} 
\end{bmatrix}\,.
\end{equation}
In this pseudo-spin basis set, symmetry also imposes that $t_1$ and $t_2$ be real. With the constraints $g_{zz}>0$ and $t_1<0$, the fit is unique and can be easily interpreted with respect to perturbation theory. In particular, $t_2$ shall describe the effects of $\propto p_x\sigma_y$ Rashba interactions along the way between the dots.

The fitted values of the tunnel couplings $t_1$, $t_2$ and $g$-matrix elements are reported in Table \ref{tb:Fits}. The symmetry between the circular L and R dots is actually broken by the finite detuning $V_\mathrm{d}=6$\,meV, so that the $g$-matrices are slightly different in the two dots. The numerical ($M^\prime=112$) and model electronic structures are compared at specific magnetic field orientations in Fig.~\ref{fig:spectrumB}. The agreement between the numerics and model is satisfactory, but not perfect for all magnetic field orientations (and the parameters dependent on the bias point) owing to the couplings with higher excited singlets and triplets (see section \ref{sec:physics}). Nevertheless, the model Hamiltonian provides useful insights into the physics of the DQD.

Let us start with the effective $g$-factors. They are best discussed from the factorization of the $g$-matrices \cite{Crippa18,Venitucci18}
\begin{equation}
g=Ug_dV^\dagger\,,
\end{equation}
where $g_d$ is the diagonal matrix of principal $g$-factors $g_1$, $g_2$, $g_3$, and $V$ is the matrix with the corresponding principal magnetic axes $\vec{v}_1$, $\vec{v}_2$, $\vec{v}_3$ as columns. Both are independent on the pseudo-spin basis set and fully characterize the effective $g$-factors $g^*=|g_dV^\dagger\vec{b}|$ of the dots. $U$ describes a change of pseudo-spin basis set that puts the Zeeman Hamiltonian in the form $H_Z=\mu_B(g_1B_1\sigma_x+g_2B_2\sigma_y+g_3B_3\sigma_z)/2$ with $B_i$ the components of the magnetic field in the principal axes set \cite{Venitucci18}. In the present cases, the principal $g$-factors $g_1\approx g_{xx}$, $g_2=g_{yy}$ and $g_3\approx g_{zz}$ coincide (at least up to third decimal place) with the diagonal elements of the $g$-matrices $g_L$ and $g_R$. The out-of-plane $g_{zz}$ is always much larger than $g_{xx}$ and $g_{yy}$ (which have opposite signs). This anisotropy is characteristic of heavy-holes in germanium. The principal axes $\vec{v}_1\approx\vec{x}$, $\vec{v}_2=\vec{y}$, $\vec{v}_3\approx\vec{z}$ are the cartesian axes slightly rotated around $\vec{y}$ by $\delta\theta\approx g_{zx}/g_{zz}$ (up to $0.08^\circ$ in the right squeezed dot). The rotation angles have thus opposite sign in the left and right dot. These rotations result from the coupling between the in- and out-of-plane motions (non-separability of the confinement potential) and from the effects of the shear strains $\varepsilon_{xz}$ on the Zeeman Hamiltonian of the holes \cite{martinez2022hole,Abadillo2023}. Although small, their effects are amplified by the large ratio between $g_{zz}$ and $g_{xx}$. They slightly skew the maps of Fig.~\ref{fig:mapasym} and (more visibly) Fig.~\ref{fig:mapsym}. 

The tunnel coupling $t_2$ is more than two orders of magnitude smaller than $t_1$. This suggests that the Rashba SOC is weak in the present devices. Its strength can alternatively be characterized by the spin-orbit length $\ell_\mathrm{so}$ such that the pseudo-spin of the hole rotates by an angle $\alpha=2d/\ell_\mathrm{so}$ when tunneling over a distance $d$ \cite{Aleiner01,Levitov03,yu2022strong}. This spin-orbit length can be estimated as $\ell_\mathrm{so}\approx a_\mathrm{2D}|t_1/t_2|\approx 25\,\mu$m, which is much longer than the inter-dot separation $a_\mathrm{2D}=180$\,nm. As a matter of fact, Rashba interactions arise from the breaking of the inversion symmetry by the structure and applied electric field and by the inhomogeneous shear strains \cite{Winkler03,Marcellina17,Abadillo2023}. However, the cubic Rashba interaction is small at low vertical electric fields in such devices \cite{Rodriguez2023}. Moreover, the linear Rashba interaction arising from the gradient of shear strains is almost zero on average on the way from one dot to the other \hl{(the gradient of $\varepsilon_{xz}$ changes sign between the dots as the strains are similar in the two dots)} \cite{Abadillo2023}. 

Therefore, the main features of Figs.~\ref{fig:mapsym} and \ref{fig:mapasym} are essentially preserved if $t_2$ is neglected, in which case Eq.~\eqref{eq:tsf} holds with the $g$-matrices of Table~\ref{tb:Fits}. $t_\mathrm{sf}$ thus primarily results from the mismatch between the principal $g$-factors and (to a lesser extent) magnetic axes of the two dots. It is almost zero when the magnetic field is along the principal $\approx\vec{x}$, $\vec{y}$ or $\approx\vec{z}$ axes (because the pseudo-spin precession axes $\vec{\omega}_L$ and $\vec{\omega}_R$ remain aligned despite the different principal $g$-factors). Moreover, $\vec{\omega}_L$ and $\vec{\omega}_R$ get locked onto the $\vec{z}$ axis once the magnetic field goes out-of-plane owing to the large $g_{zz}\gg g_{xx},\,|g_{yy}|$. Consequently, $t_\mathrm{sf}$, $\Delta_{ST_-}$ and the deviations $J-J_\parallel$ are significant only near the equatorial plane. Leaving out $g_{zx}$, $|t_\mathrm{sf}|$ is actually maximal when $\theta\approx 90^\circ$ and \footnote{These expressions apply for positive $g_{xx}$'s and negative $g_{yy}$'s.}
\begin{equation}
\varphi=\arccos\left(\pm\sqrt{\frac{g_{yy}^Lg_{yy}^R}{g_{xx}^Lg_{xx}^R+g_{yy}^Lg_{yy}^R}}\right)
\end{equation}
where it reaches:
\begin{equation}
|t_\mathrm{sf}|_\mathrm{max}=|t|\sqrt{\frac{1}{2}+\frac{\sqrt{g_{xx}^Lg_{yy}^Lg_{xx}^Rg_{yy}^R}}{g_{xx}^Lg_{yy}^R+g_{xx}^Rg_{yy}^L}}\,.
\end{equation}
Using these expressions, we find $|t_\mathrm{sf}|_\mathrm{max}=0.53\,\mu$eV at $\varphi=\pm(90\pm 34.49)^\circ$ in the quasi-circular dots and $|t_\mathrm{sf}|_\mathrm{max}= 5.63\,\mu$eV at $\varphi=\pm(90\pm 17.62)^\circ$ in the squeezed dots, in fair agreement with the estimates drawn from Eq.~\eqref{eq:jsoc} on Figs.~\ref{fig:mapsym} and \ref{fig:mapasym}. The effects of SOC are thus stronger in the squeezed dots where the mismatch between the principal $g$-factors is larger. \hl{Although they can have a strong impact on the Rabi frequencies of single dots \cite{Abadillo2023}, inhomogeneous cool-down strains play a secondary role on the anisotropic exchange interactions and $ST_-$ splitting especially when the mismatch between the principal $g$-factors is significant.}

\section{Time-dependent simulations}
\label{sec:tdsimus}

To conclude, we show that the dressed basis set is suitable for time-dependent many-body simulations.

\begin{figure}[t]
\centering
\includegraphics[width=.9\linewidth]{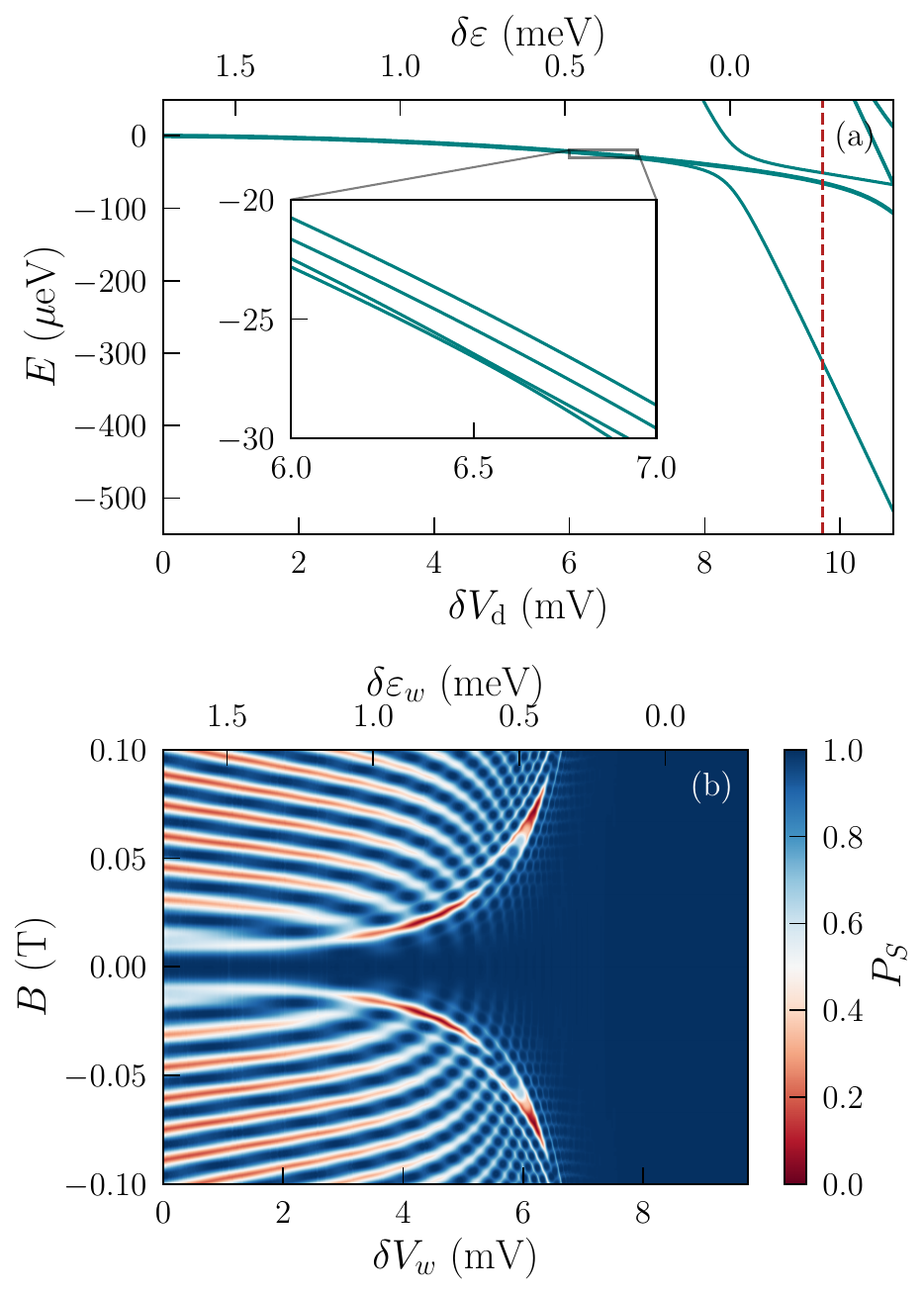}
\caption{(a) Electronic structure of the same squeezed DQD as in Fig.~\ref{fig:mapasym}, as a function of the detuning voltage $\delta V_\mathrm{d}$ ($V_\mathrm{J}=-19.8$\,meV, $B=100$\,mT). The dashed vertical line indicates the position of the measurement point in the time-dependent simulations. The inset is a close-up on the $ST_-$ anti-crossing. (b) Funnel pattern computed in the squeezed DQD\hl{, as a function of the detuning voltage $\delta V_w$ at the wait point and magnetic field amplitude $B$}. The magnetic field is oriented along $\vec{x}$ in both panels \hl{and the Hamiltonians are computed in the dressed $M^\prime=112$ basis set}. The top scales are the detuning energy with respect to the $S(1,1)/S(0,2)$ anti-crossing.}
\label{fig:funnel}
\end{figure}

As an illustration, we compute the spin funnel of the singlet-triplet qubit based on the squeezed DQD of the previous section \cite{jirovec_Freqs_ST-Ge,Jirovec23}. The electronic structure of this qubit at magnetic field $B_x=100$\,mT is plotted as a function of the detuning voltage in Fig.~\ref{fig:funnel}a. The $S(1,1)$ anti-crosses the $S(0,2)$ state at $\delta V_\mathrm{d}=\delta V_\mathrm{d}^*=8.37$\,mV. We start from the $S(0,2)$ state at $\delta V_\mathrm{d}>\delta V_\mathrm{d}^*$, then pulse diabatically (rise time $\tau_r=2$\,ns) to some \hl{smaller detuning $\delta V_\mathrm{d}=\delta V_w<\delta V_\mathrm{d}^*$} in order to initialize the qubit in a singlet state. We next wait for $\tau_w=65$\,ns and pulse diabatically back to the measurement point $\delta V_\mathrm{d}=9.74$\,mV where the triplets are blocked. We finally measure the probability $P_S$ to return to the $S(0,2)$ state after this sequence. The time-dependent Schr\"odinger equation is solved on a grid of times $t=t_0,\,...,\,t_N$; for that purpose, the evolution operator $U(t_{n+1},\,t_n)=\exp[i\hat{H}(t_{n+1/2})(t_n-t_{n+1})/\hbar]$ is computed in each time interval $[t_n,\,t_{n+1}]$ from the exact diagonalization of the Hamiltonian $\hat{H}(t_{n+1/2})$ [with $t_{n+1/2}=(t_n+t_{n+1})/2$]. \hl{This allows for large time steps at the wait point where the Hamiltonian is constant. $\hat{H}(t)$ can either be computed in the dressed ($M^\prime=112$) or full CI ($M=4560$) basis sets.} We do not account for decoherence and relaxation in this calculation. 

The singlet return probability $P_S$ is plotted as a function of \hl{$\delta V_w$ and $B$} in Fig.~\ref{fig:funnel}b. The magnetic field is oriented along $\vec{x}$. This figure, \hl{computed in the dressed basis set,} exhibits the characteristic funnel structure concurring with the $ST_-$ anti-crossing. At these points of the \hl{$(\delta V_w, B)$} plane, the $ST_-$ mixing is strong and the initial singlet state undergoes rapid $ST_-$ oscillations leading to a drop in $P_S$. The width of the funnel lines is, therefore, proportional to the singlet-triplet gap $\Delta_{ST_-}$. Along these lines $P_S$ is also modulated by the total, $\propto\tau_w$ phase accumulated during the $ST_-$ oscillations. 

There is a second set of lines clearly visible outside the funnel. They correspond to $ST_0$ oscillations \hl{at large enough magnetic field and detuning energy $\delta\varepsilon_w$ at the wait point}. Indeed, the $S$ and $T_0$ states get mixed by the imbalance of Zeeman splittings $\Delta E_Z\propto (g_L^*-g_R^*)B$ that competes with the exchange energy. This gives rise to a modulation of the return probability essentially periodic with the magnetic field deep in the $(1,1)$ charge state. 

\begin{figure}[t]
\centering
\includegraphics[width=.9\linewidth]{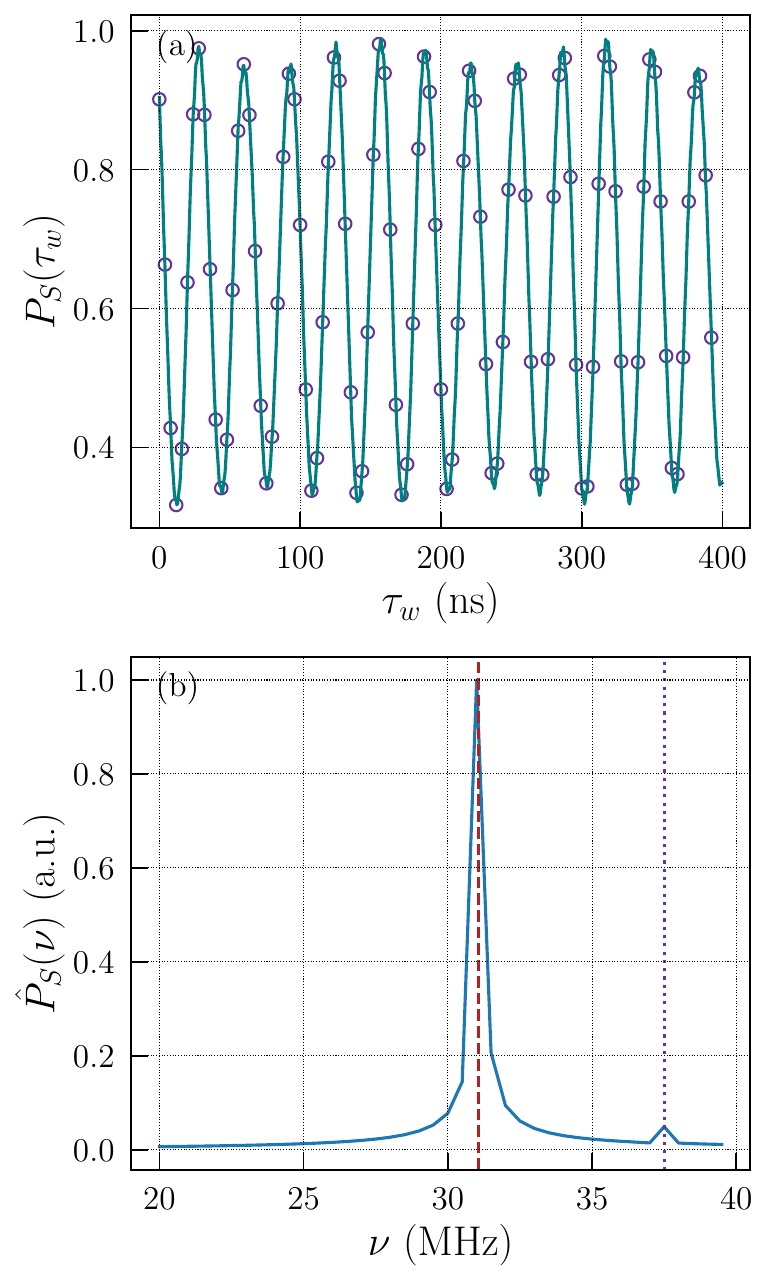}
\caption{(a) Singlet return probability $P_S(\tau_w)$ as a function of the wait time $\tau_w$ at $\delta V_w=0$ ($B=25$\,mT). \hl{The line is computed in the dressed basis set, and the dots in the full CI basis set.} (b) Fourier transform ${\hat P}_S(\nu)$ of $P_S(\tau_w)$ \hl{in the dressed basis set}. The vertical dashed line is the expected frequency of the $ST_0$ oscillations, $\nu=31.08$\,MHz, and the vertical dotted line, the expected frequency of the $\ket{\Downarrow\Uparrow}/T_-$ oscillations, $\nu=37.5$\,MHz.}
\label{fig:ST0osc}
\end{figure}

Additionally, we plot in Fig.~\ref{fig:ST0osc} the return probability $P_S(\tau_w)$ as a function of the wait time $\tau_w$ at \hl{$\delta V_w=0$}, and its Fourier transform ${\hat P}_S(\nu)$. The magnetic field is $B_x=25$\,mT. At this operation point, the qubit shall primarily undergo $ST_0$ oscillations. Fig.~\ref{fig:ST0osc} highlights different features. First, the $ST_0$ oscillations are incomplete ($P_S$ does not reach zero) due to the finite exchange at zero detuning. Indeed, $S$ and $T_0$ are not fully split into $\ket{\Uparrow\Downarrow}$ and $\ket{\Downarrow\Uparrow}$ states, so that an initial singlet state does not rotate around an axis of the equatorial plane of the $ST_0$ Bloch sphere that would bring it through the $T_0$ state \cite{jirovec_Freqs_ST-Ge}. Second, there is a clear beating between a main oscillation at $\nu=31.08$\,MHz and a secondary oscillation at $\nu=37.5$\,MHz. The former is the expected $ST_0$ oscillation, while the frequency of the latter matches the $\ket{\Downarrow\Uparrow}/T_-$ splitting. This results from a small residual $ST_-$ mixing, and from non fully diabatic passages through the nearby $ST_-$ anti-crossing, which give rise to state preparation and measurement errors [$P_S(0)<1$]. 

\hl{As illustrated in Fig.~\ref{fig:ST0osc}a, the time-dependent calculations in the dressed basis set are in very good agreement with those in the full CI basis set, but are 55\,000 times faster} \footnote{We emphasize, though, that the cost of the method we have chosen to solve the time-dependent Schrodinger equation scales as $M^3$ (with $M$ the dimension of the basis set). Other methods (Trotter, ...) may show a better ($\propto M^2$) scaling. Anyhow, the gain in the dressed basis set remains very significant.}. The dressed basis achieves, therefore, an excellent balance between accuracy and efficiency. Such maps and time traces are experimentally used to characterize the electronic structure of singlet-triplet qubits and demonstrate the existence of detuning-dependent $ST_0$ oscillations needed for their manipulation \cite{jirovec_Freqs_ST-Ge}. The optimization of the operation point of this singlet-triplet qubit, as well as decoherence go beyond the scope of this paper, and will be addressed in future works.

\section{Conclusions}

We have investigated the exchange interactions in double quantum dots with an efficient computational method. For that purpose, we have built a dressed basis set of two-particle wave functions that captures the main correlations that build up in full CI method. This dressed basis set can reproduce the low-energy singlet and triplet states over the whole operational gate voltages range with as few as a hundred basis functions, about fifty times less than in the CI. This speeds up intensive many-body calculations such as time-dependent simulations. We have applied this methodology to a double hole quantum dot in a germanium heterostructure. We have discussed the physics of the exchange interactions in this double dot, and we have highlighted the prominent role of Coulomb correlations. We have also analyzed the interplay between confinement, strains and Rashba interactions at finite magnetic field. We find that the spin-flip tunneling terms responsible for anisotropic exchange and singlet-triplet mixings are essentially driven by the imbalance of $g$-factors in this system, the inter-dot Rashba interactions being small even in the presence of inhomogeneous strains. \hl{The methodology of this work can be extended to at least three dots (but at an increasing cost for the CI calculations needed to set-up the dressed basis set), and shall also capture long-range spin-dependent dipole interactions in the presence of strong spin-orbit coupling \cite{Flindt06,Trif07,Trifunovic12}.} This opens the way to the modeling of complex operations in Loss-DiVincenzo, singlet-triplet and exchange-only qubits with realistic layouts, disorder and noise.

\section*{Acknowledgements}

This work was supported by the ``France 2030'' program (PEPR PRESQUILE-ANR-22-PETQ-0002), by the French National Research Agency (project InGeQT), and by the Horizon Europe Framework Program (grant agreement 101174557 QLSI2).

\appendix

\section{\hl{Luttinger-Kohn Hamiltonian}}
\label{app:LKH}
\hl{The HH and LH envelope functions, $\varphi_n^\nu$ fulfill a set of differential equations defined by the Luttinger-Kohn Hamiltonian:}
\begin{equation}
H_\mathrm{LK}(\{V_\mathrm{G}^0\},\vec{B})=H_\mathrm{K}+H_\varepsilon+H_\mathrm{Z}+V_t(\vec{r})\mathbb{1}_4\,,
\label{eq:H4KP}
\end{equation}
\hl{where $H_\mathrm{K}$ is the kinetic energy, $H_\varepsilon$ describes the effects of strains, $H_\mathrm{Z}$ is the Zeeman Hamiltonian and $\mathbb{1}_4$ is the $4\times4$ identity matrix. $H_\mathrm{K}$ and $H_\varepsilon$ share the same generic form in the $J_z=\{+\tfrac{3}{2},+\tfrac{1}{2},-\tfrac{1}{2},-\tfrac{3}{2}\}$ basis set:}
\begin{equation}
H_\mathrm{K/\varepsilon}=\begin{pmatrix}
P+Q & -S & R & 0 \\
-S^\dagger & P-Q & 0 & R \\
R^\dagger & 0 & P-Q & S \\
0 & R^\dagger & S^\dagger & P+Q
\end{pmatrix}
\label{eq:LK}\,,
\end{equation}
\hl{where, for $H_\mathrm{K}$,}
\begin{subequations}
\begin{align}
P_\mathrm{K}&=\frac{1}{2m_0}\gamma_1(p_x^2+p_y^2+p_z^2) \\
Q_\mathrm{K}&=\frac{1}{2m_0}\gamma_2(p_x^2+p_y^2-2p_z^2) \\
R_\mathrm{K}&=\frac{1}{2m_0}\sqrt{3}\left[-\gamma_2(p_x^2-p_y^2)+2i\gamma_3\{p_x,\,p_y\}\right] \\
S_\mathrm{K}&=\frac{1}{2m_0}2\sqrt{3}\gamma_3\{p_x-ip_y,\,p_z\}\,,
\end{align}
\end{subequations}
\hl{with $\{A,\,B\}=\tfrac{1}{2}(AB+BA)$, and, for $H_\varepsilon$,}
\begin{subequations}
\begin{align}
P_\varepsilon&=-a_v(\varepsilon_{xx}+\varepsilon_{yy}+\varepsilon_{zz}) \\
Q_\varepsilon&=-\frac{1}{2}b_v(\varepsilon_{xx}+\varepsilon_{yy}-2\varepsilon_{zz}) \\
R_\varepsilon&=\frac{\sqrt{3}}{2}b_v(\varepsilon_{xx}-\varepsilon_{yy})-id_v\varepsilon_{xy} \\
S_\varepsilon&=-d_v(\varepsilon_{xz}-i\varepsilon_{yz})\,.
\end{align}
\end{subequations}
\hl{Here $\mathbf{p}$ is the momentum, $m_0$ is the free electron mass, and $\gamma_1$, $\gamma_2$, $\gamma_3$ are the Luttinger parameters that characterize the hole masses. The $\varepsilon_{\alpha\beta}$ are the strains; $a_v$ is the hydrostatic, $b_v$ the uniaxial and $d_v$ the shear deformation potential of the valence band. The form of Eq.}~(\ref{eq:LK})\hl{, which couples different $J_z$'s through the $R$ and $S$ terms, embodies the action of SOC in the valence band. The Zeeman Hamiltonian $H_\mathrm{Z}=2\mu_B(\kappa\mathbf{B}\cdot\mathbf{J}+q\mathbf{B}\cdot\mathbf{J}^3)$ describes the action of the magnetic field on the Bloch functions, with $\mathbf{J}$ the spin $\tfrac{3}{2}$ operator, $\mathbf{J}^3\equiv(J_x^3,J_y^3,J_z^3)$, $\mu_B$ the Bohr magneton, and $\kappa$, $q$ the isotropic and cubic Zeeman parameters. The action of $\mathbf{B}$ on the envelopes of the hole is accounted for by the substitution $\mathbf{p}\to-i\hbar\boldsymbol{\nabla}+e\mathbf{A}$ in $H_\mathrm{K}$, with $\mathbf{A}=\frac{1}{2}\mathbf{B}\times\mathbf{r}$ the magnetic vector potential.}

\section{Implementation details}
\label{app:details}

\subsection{Convergence and time-reversal symmetry}

The calculated exchange interactions (e.g., Fig.~\ref{fig:JVJ}) may become noisy when the barrier is closed ($V_\mathrm{J}\approx 0$\,V) and the exchange energy gets in the neV range. Actually, all quantities (single-particle energies, $U_{ijkl}$'s, etc.) must be converged with a better-than-neV accuracy to address this range, which is very demanding (and close to the computational limits, the spectrum of the FD hamiltonian spanning tens of eVs). It is also important not to break time-reversal symmetry (${\cal T}$) in the CI calculations, to avoid spurious mixings between singlets and triplets resulting in meaningless exchange splittings. To prevent (small) symmetry breakings by numerical inaccuracies, we always add $\ket{\psi_1}$ and ${\cal T}\ket{\psi_1}$ instead of both members $\ket{\psi_1}$ and $\ket{\psi_2}$ of a given Kramers pair to the single-particle basis set, and re-enforce {\it a posteriori} the time-reversal symmetry relations on the $W_{ijkl}$'s (namely, average the elements expected equal or conjugate by permutation of the indices). 

We also emphasize that the three members of a triplet must always be included in the dressed CI basis set in order not to break time-reversal symmetry.

\subsection{Gauge invariance}

Strict gauge invariance can hardly be enforced in a finite basis set as the latter can not accommodate arbitrary phase variations if the vector potential $\vec{A}$ gets large. Practically, we did not, however, experience significant differences with usual gauges (symmetric $\vec{A}=\vec{B}\times\vec{r}/2$, Landau, etc...) centered on the DQD. Moreover, the model is gauge-invariant in the linear response regime where the matrices $N_\mathrm{\alpha\beta}$ can be discarded, if the elements of the $M_\mathrm\alpha$'s coupling different Kramers pairs of $H_0$ can also be neglected \cite{Abadillo2023}. The linear response (the $g$-matrix formalism) holds in the present cases \cite{Venitucci18}. We did, therefore, neglect the $N_\mathrm{\alpha\beta}$'s in the all calculations.

\bibliography{Refs.bib}

\begin{thebibliography}{143}%
\makeatletter
\providecommand \@ifxundefined [1]{%
 \@ifx{#1\undefined}
}%
\providecommand \@ifnum [1]{%
 \ifnum #1\expandafter \@firstoftwo
 \else \expandafter \@secondoftwo
 \fi
}%
\providecommand \@ifx [1]{%
 \ifx #1\expandafter \@firstoftwo
 \else \expandafter \@secondoftwo
 \fi
}%
\providecommand \natexlab [1]{#1}%
\providecommand \enquote  [1]{``#1''}%
\providecommand \bibnamefont  [1]{#1}%
\providecommand \bibfnamefont [1]{#1}%
\providecommand \citenamefont [1]{#1}%
\providecommand \href@noop [0]{\@secondoftwo}%
\providecommand \href [0]{\begingroup \@sanitize@url \@href}%
\providecommand \@href[1]{\@@startlink{#1}\@@href}%
\providecommand \@@href[1]{\endgroup#1\@@endlink}%
\providecommand \@sanitize@url [0]{\catcode `\\12\catcode `\$12\catcode `\&12\catcode `\#12\catcode `\^12\catcode `\_12\catcode `\%12\relax}%
\providecommand \@@startlink[1]{}%
\providecommand \@@endlink[0]{}%
\providecommand \url  [0]{\begingroup\@sanitize@url \@url }%
\providecommand \@url [1]{\endgroup\@href {#1}{\urlprefix }}%
\providecommand \urlprefix  [0]{URL }%
\providecommand \Eprint [0]{\href }%
\providecommand \doibase [0]{https://doi.org/}%
\providecommand \selectlanguage [0]{\@gobble}%
\providecommand \bibinfo  [0]{\@secondoftwo}%
\providecommand \bibfield  [0]{\@secondoftwo}%
\providecommand \translation [1]{[#1]}%
\providecommand \BibitemOpen [0]{}%
\providecommand \bibitemStop [0]{}%
\providecommand \bibitemNoStop [0]{.\EOS\space}%
\providecommand \EOS [0]{\spacefactor3000\relax}%
\providecommand \BibitemShut  [1]{\csname bibitem#1\endcsname}%
\let\auto@bib@innerbib\@empty
%</preamble>
\bibitem [{\citenamefont {Loss}\ and\ \citenamefont {DiVincenzo}(1998)}]{Loss98}%
  \BibitemOpen
  \bibfield  {author} {\bibinfo {author} {\bibfnamefont {D.}~\bibnamefont {Loss}}\ and\ \bibinfo {author} {\bibfnamefont {D.~P.}\ \bibnamefont {DiVincenzo}},\ }\bibfield  {title} {\bibinfo {title} {{Quantum computation with quantum dots}},\ }\href {https://doi.org/10.1103/PhysRevA.57.120} {\bibfield  {journal} {\bibinfo  {journal} {Physical Review A}\ }\textbf {\bibinfo {volume} {57}},\ \bibinfo {pages} {120} (\bibinfo {year} {1998})}\BibitemShut {NoStop}%
\bibitem [{\citenamefont {Stano}\ and\ \citenamefont {Loss}(2022)}]{Stano22}%
  \BibitemOpen
  \bibfield  {author} {\bibinfo {author} {\bibfnamefont {P.}~\bibnamefont {Stano}}\ and\ \bibinfo {author} {\bibfnamefont {D.}~\bibnamefont {Loss}},\ }\bibfield  {title} {\bibinfo {title} {Review of performance metrics of spin qubits in gated semiconducting nanostructures},\ }\href {https://doi.org/10.1038/s42254-022-00484-w} {\bibfield  {journal} {\bibinfo  {journal} {Nature Reviews Physics}\ }\textbf {\bibinfo {volume} {4}},\ \bibinfo {pages} {672} (\bibinfo {year} {2022})}\BibitemShut {NoStop}%
\bibitem [{\citenamefont {Burkard}\ \emph {et~al.}(2023)\citenamefont {Burkard}, \citenamefont {Ladd}, \citenamefont {Pan}, \citenamefont {Nichol},\ and\ \citenamefont {Petta}}]{Burkard2023Review}%
  \BibitemOpen
  \bibfield  {author} {\bibinfo {author} {\bibfnamefont {G.}~\bibnamefont {Burkard}}, \bibinfo {author} {\bibfnamefont {T.~D.}\ \bibnamefont {Ladd}}, \bibinfo {author} {\bibfnamefont {A.}~\bibnamefont {Pan}}, \bibinfo {author} {\bibfnamefont {J.~M.}\ \bibnamefont {Nichol}},\ and\ \bibinfo {author} {\bibfnamefont {J.~R.}\ \bibnamefont {Petta}},\ }\bibfield  {title} {\bibinfo {title} {Semiconductor spin qubits},\ }\href {https://doi.org/10.1103/RevModPhys.95.025003} {\bibfield  {journal} {\bibinfo  {journal} {Review of Modern Physics}\ }\textbf {\bibinfo {volume} {95}},\ \bibinfo {pages} {025003} (\bibinfo {year} {2023})}\BibitemShut {NoStop}%
\bibitem [{\citenamefont {Fang}\ \emph {et~al.}(2023)\citenamefont {Fang}, \citenamefont {Philippopoulos}, \citenamefont {Culcer}, \citenamefont {Coish},\ and\ \citenamefont {Chesi}}]{Fang2023Review}%
  \BibitemOpen
  \bibfield  {author} {\bibinfo {author} {\bibfnamefont {Y.}~\bibnamefont {Fang}}, \bibinfo {author} {\bibfnamefont {P.}~\bibnamefont {Philippopoulos}}, \bibinfo {author} {\bibfnamefont {D.}~\bibnamefont {Culcer}}, \bibinfo {author} {\bibfnamefont {W.~A.}\ \bibnamefont {Coish}},\ and\ \bibinfo {author} {\bibfnamefont {S.}~\bibnamefont {Chesi}},\ }\bibfield  {title} {\bibinfo {title} {Recent advances in hole-spin qubits},\ }\href {https://doi.org/10.1088/2633-4356/acb87e} {\bibfield  {journal} {\bibinfo  {journal} {Materials for Quantum Technology}\ }\textbf {\bibinfo {volume} {3}},\ \bibinfo {pages} {012003} (\bibinfo {year} {2023})}\BibitemShut {NoStop}%
\bibitem [{\citenamefont {Itoh}\ and\ \citenamefont {Watanabe}(2014)}]{Itoh14}%
  \BibitemOpen
  \bibfield  {author} {\bibinfo {author} {\bibfnamefont {K.~M.}\ \bibnamefont {Itoh}}\ and\ \bibinfo {author} {\bibfnamefont {H.}~\bibnamefont {Watanabe}},\ }\bibfield  {title} {\bibinfo {title} {Isotope engineering of silicon and diamond for quantum computing and sensing applications},\ }\href {https://doi.org/10.1557/mrc.2014.32} {\bibfield  {journal} {\bibinfo  {journal} {MRS Communications}\ }\textbf {\bibinfo {volume} {4}},\ \bibinfo {pages} {143} (\bibinfo {year} {2014})}\BibitemShut {NoStop}%
\bibitem [{\citenamefont {Cvitkovich}\ \emph {et~al.}(2024)\citenamefont {Cvitkovich}, \citenamefont {Stano}, \citenamefont {Wilhelmer}, \citenamefont {Waldh\"or}, \citenamefont {Loss}, \citenamefont {Niquet},\ and\ \citenamefont {Grasser}}]{Cvitkovich24}%
  \BibitemOpen
  \bibfield  {author} {\bibinfo {author} {\bibfnamefont {L.}~\bibnamefont {Cvitkovich}}, \bibinfo {author} {\bibfnamefont {P.}~\bibnamefont {Stano}}, \bibinfo {author} {\bibfnamefont {C.}~\bibnamefont {Wilhelmer}}, \bibinfo {author} {\bibfnamefont {D.}~\bibnamefont {Waldh\"or}}, \bibinfo {author} {\bibfnamefont {D.}~\bibnamefont {Loss}}, \bibinfo {author} {\bibfnamefont {Y.-M.}\ \bibnamefont {Niquet}},\ and\ \bibinfo {author} {\bibfnamefont {T.}~\bibnamefont {Grasser}},\ }\bibfield  {title} {\bibinfo {title} {Coherence limit due to hyperfine interaction with nuclei in the barrier material of $\mathrm{Si}$ spin qubits},\ }\href {https://doi.org/10.1103/PhysRevApplied.22.064089} {\bibfield  {journal} {\bibinfo  {journal} {Physical Review Applied}\ }\textbf {\bibinfo {volume} {22}},\ \bibinfo {pages} {064089} (\bibinfo {year} {2024})}\BibitemShut {NoStop}%
\bibitem [{\citenamefont {Yoneda}\ \emph {et~al.}(2018)\citenamefont {Yoneda}, \citenamefont {Takeda}, \citenamefont {Otsuka}, \citenamefont {Nakajima}, \citenamefont {Delbecq}, \citenamefont {Allison}, \citenamefont {Honda}, \citenamefont {Kodera}, \citenamefont {Oda}, \citenamefont {Hoshi}, \citenamefont {Usami}, \citenamefont {Itoh},\ and\ \citenamefont {Tarucha}}]{Yoneda18}%
  \BibitemOpen
  \bibfield  {author} {\bibinfo {author} {\bibfnamefont {J.}~\bibnamefont {Yoneda}}, \bibinfo {author} {\bibfnamefont {K.}~\bibnamefont {Takeda}}, \bibinfo {author} {\bibfnamefont {T.}~\bibnamefont {Otsuka}}, \bibinfo {author} {\bibfnamefont {T.}~\bibnamefont {Nakajima}}, \bibinfo {author} {\bibfnamefont {M.~R.}\ \bibnamefont {Delbecq}}, \bibinfo {author} {\bibfnamefont {G.}~\bibnamefont {Allison}}, \bibinfo {author} {\bibfnamefont {T.}~\bibnamefont {Honda}}, \bibinfo {author} {\bibfnamefont {T.}~\bibnamefont {Kodera}}, \bibinfo {author} {\bibfnamefont {S.}~\bibnamefont {Oda}}, \bibinfo {author} {\bibfnamefont {Y.}~\bibnamefont {Hoshi}}, \bibinfo {author} {\bibfnamefont {N.}~\bibnamefont {Usami}}, \bibinfo {author} {\bibfnamefont {K.~M.}\ \bibnamefont {Itoh}},\ and\ \bibinfo {author} {\bibfnamefont {S.}~\bibnamefont {Tarucha}},\ }\bibfield  {title} {\bibinfo {title} {A quantum-dot spin qubit with coherence limited by charge noise and fidelity higher than 99.9\%},\ }\href
  {https://doi.org/https://doi.org/10.1038/s41565-017-0014-x} {\bibfield  {journal} {\bibinfo  {journal} {Nature Nanotechnology}\ }\textbf {\bibinfo {volume} {13}},\ \bibinfo {pages} {102} (\bibinfo {year} {2018})}\BibitemShut {NoStop}%
\bibitem [{\citenamefont {Philips}\ \emph {et~al.}(2022)\citenamefont {Philips}, \citenamefont {Madzik}, \citenamefont {Amitonov}, \citenamefont {de~Snoo}, \citenamefont {Russ}, \citenamefont {Kalhor}, \citenamefont {Volk}, \citenamefont {Lawrie}, \citenamefont {Brousse}, \citenamefont {Tryputen}, \citenamefont {Wuetz}, \citenamefont {Sammak}, \citenamefont {Veldhorst}, \citenamefont {Scappucci},\ and\ \citenamefont {Vandersypen}}]{Philips22}%
  \BibitemOpen
  \bibfield  {author} {\bibinfo {author} {\bibfnamefont {S.~G.~J.}\ \bibnamefont {Philips}}, \bibinfo {author} {\bibfnamefont {M.~T.}\ \bibnamefont {Madzik}}, \bibinfo {author} {\bibfnamefont {S.~V.}\ \bibnamefont {Amitonov}}, \bibinfo {author} {\bibfnamefont {S.~L.}\ \bibnamefont {de~Snoo}}, \bibinfo {author} {\bibfnamefont {M.}~\bibnamefont {Russ}}, \bibinfo {author} {\bibfnamefont {N.}~\bibnamefont {Kalhor}}, \bibinfo {author} {\bibfnamefont {C.}~\bibnamefont {Volk}}, \bibinfo {author} {\bibfnamefont {W.~I.~L.}\ \bibnamefont {Lawrie}}, \bibinfo {author} {\bibfnamefont {D.}~\bibnamefont {Brousse}}, \bibinfo {author} {\bibfnamefont {L.}~\bibnamefont {Tryputen}}, \bibinfo {author} {\bibfnamefont {B.~P.}\ \bibnamefont {Wuetz}}, \bibinfo {author} {\bibfnamefont {A.}~\bibnamefont {Sammak}}, \bibinfo {author} {\bibfnamefont {M.}~\bibnamefont {Veldhorst}}, \bibinfo {author} {\bibfnamefont {G.}~\bibnamefont {Scappucci}},\ and\ \bibinfo {author} {\bibfnamefont {L.~M.~K.}\ \bibnamefont {Vandersypen}},\ }\bibfield
  {title} {\bibinfo {title} {Universal control of a six-qubit quantum processor in silicon},\ }\href {https://doi.org/10.1038/s41586-022-05117-x} {\bibfield  {journal} {\bibinfo  {journal} {Nature}\ }\textbf {\bibinfo {volume} {609}},\ \bibinfo {pages} {919} (\bibinfo {year} {2022})}\BibitemShut {NoStop}%
\bibitem [{\citenamefont {Mills}\ \emph {et~al.}(2022)\citenamefont {Mills}, \citenamefont {Guinn}, \citenamefont {Gullans}, \citenamefont {Sigillito}, \citenamefont {Feldman}, \citenamefont {Nielsen},\ and\ \citenamefont {Petta}}]{Mills22}%
  \BibitemOpen
  \bibfield  {author} {\bibinfo {author} {\bibfnamefont {A.~R.}\ \bibnamefont {Mills}}, \bibinfo {author} {\bibfnamefont {C.~R.}\ \bibnamefont {Guinn}}, \bibinfo {author} {\bibfnamefont {M.~J.}\ \bibnamefont {Gullans}}, \bibinfo {author} {\bibfnamefont {A.~J.}\ \bibnamefont {Sigillito}}, \bibinfo {author} {\bibfnamefont {M.~M.}\ \bibnamefont {Feldman}}, \bibinfo {author} {\bibfnamefont {E.}~\bibnamefont {Nielsen}},\ and\ \bibinfo {author} {\bibfnamefont {J.~R.}\ \bibnamefont {Petta}},\ }\bibfield  {title} {\bibinfo {title} {Two-qubit silicon quantum processor with operation fidelity exceeding 99\%},\ }\href {https://doi.org/10.1126/sciadv.abn5130} {\bibfield  {journal} {\bibinfo  {journal} {Science Advances}\ }\textbf {\bibinfo {volume} {8}},\ \bibinfo {pages} {eabn5130} (\bibinfo {year} {2022})}\BibitemShut {NoStop}%
\bibitem [{\citenamefont {Xue}\ \emph {et~al.}(2022)\citenamefont {Xue}, \citenamefont {Russ}, \citenamefont {Samkharadze}, \citenamefont {Undseth}, \citenamefont {Sammak}, \citenamefont {Scappucci},\ and\ \citenamefont {Vandersypen}}]{Xue22}%
  \BibitemOpen
  \bibfield  {author} {\bibinfo {author} {\bibfnamefont {X.}~\bibnamefont {Xue}}, \bibinfo {author} {\bibfnamefont {M.}~\bibnamefont {Russ}}, \bibinfo {author} {\bibfnamefont {N.}~\bibnamefont {Samkharadze}}, \bibinfo {author} {\bibfnamefont {B.}~\bibnamefont {Undseth}}, \bibinfo {author} {\bibfnamefont {A.}~\bibnamefont {Sammak}}, \bibinfo {author} {\bibfnamefont {G.}~\bibnamefont {Scappucci}},\ and\ \bibinfo {author} {\bibfnamefont {L.~M.~K.}\ \bibnamefont {Vandersypen}},\ }\bibfield  {title} {\bibinfo {title} {Quantum logic with spin qubits crossing the surface code threshold},\ }\href {https://doi.org/10.1038/s41586-021-04273-w} {\bibfield  {journal} {\bibinfo  {journal} {Nature}\ }\textbf {\bibinfo {volume} {601}},\ \bibinfo {pages} {343} (\bibinfo {year} {2022})}\BibitemShut {NoStop}%
\bibitem [{\citenamefont {Noiri}\ \emph {et~al.}(2022)\citenamefont {Noiri}, \citenamefont {Takeda}, \citenamefont {Nakajima}, \citenamefont {Kobayashi}, \citenamefont {Sammak}, \citenamefont {Scappucci},\ and\ \citenamefont {Tarucha}}]{Noiri22}%
  \BibitemOpen
  \bibfield  {author} {\bibinfo {author} {\bibfnamefont {A.}~\bibnamefont {Noiri}}, \bibinfo {author} {\bibfnamefont {K.}~\bibnamefont {Takeda}}, \bibinfo {author} {\bibfnamefont {T.}~\bibnamefont {Nakajima}}, \bibinfo {author} {\bibfnamefont {T.}~\bibnamefont {Kobayashi}}, \bibinfo {author} {\bibfnamefont {A.}~\bibnamefont {Sammak}}, \bibinfo {author} {\bibfnamefont {G.}~\bibnamefont {Scappucci}},\ and\ \bibinfo {author} {\bibfnamefont {S.}~\bibnamefont {Tarucha}},\ }\bibfield  {title} {\bibinfo {title} {Fast universal quantum gate above the fault-tolerance threshold in silicon},\ }\href {https://doi.org/10.1038/s41586-021-04182-y} {\bibfield  {journal} {\bibinfo  {journal} {Nature}\ }\textbf {\bibinfo {volume} {601}},\ \bibinfo {pages} {338} (\bibinfo {year} {2022})}\BibitemShut {NoStop}%
\bibitem [{\citenamefont {Huang}\ \emph {et~al.}(2024)\citenamefont {Huang}, \citenamefont {Su}, \citenamefont {Lim}, \citenamefont {Feng}, \citenamefont {Van~Straaten}, \citenamefont {Severin}, \citenamefont {Gilbert}, \citenamefont {Dumoulin~Stuyck}, \citenamefont {Tanttu}, \citenamefont {Serrano}, \citenamefont {Cifuentes}, \citenamefont {Hansen}, \citenamefont {Seedhouse}, \citenamefont {Vahapoglu}, \citenamefont {Leon}, \citenamefont {Abrosimov}, \citenamefont {Pohl}, \citenamefont {Thewalt}, \citenamefont {Hudson}, \citenamefont {Escott}, \citenamefont {Ares}, \citenamefont {Bartlett}, \citenamefont {Morello}, \citenamefont {Saraiva}, \citenamefont {Laucht}, \citenamefont {Dzurak},\ and\ \citenamefont {Yang}}]{Huang24}%
  \BibitemOpen
  \bibfield  {author} {\bibinfo {author} {\bibfnamefont {J.~Y.}\ \bibnamefont {Huang}}, \bibinfo {author} {\bibfnamefont {R.~Y.}\ \bibnamefont {Su}}, \bibinfo {author} {\bibfnamefont {W.~H.}\ \bibnamefont {Lim}}, \bibinfo {author} {\bibfnamefont {M.}~\bibnamefont {Feng}}, \bibinfo {author} {\bibfnamefont {B.}~\bibnamefont {Van~Straaten}}, \bibinfo {author} {\bibfnamefont {B.}~\bibnamefont {Severin}}, \bibinfo {author} {\bibfnamefont {W.}~\bibnamefont {Gilbert}}, \bibinfo {author} {\bibfnamefont {N.}~\bibnamefont {Dumoulin~Stuyck}}, \bibinfo {author} {\bibfnamefont {T.}~\bibnamefont {Tanttu}}, \bibinfo {author} {\bibfnamefont {S.}~\bibnamefont {Serrano}}, \bibinfo {author} {\bibfnamefont {J.~D.}\ \bibnamefont {Cifuentes}}, \bibinfo {author} {\bibfnamefont {I.}~\bibnamefont {Hansen}}, \bibinfo {author} {\bibfnamefont {A.~E.}\ \bibnamefont {Seedhouse}}, \bibinfo {author} {\bibfnamefont {E.}~\bibnamefont {Vahapoglu}}, \bibinfo {author} {\bibfnamefont {R.~C.~C.}\ \bibnamefont {Leon}}, \bibinfo {author}
  {\bibfnamefont {N.~V.}\ \bibnamefont {Abrosimov}}, \bibinfo {author} {\bibfnamefont {H.-J.}\ \bibnamefont {Pohl}}, \bibinfo {author} {\bibfnamefont {M.~L.~W.}\ \bibnamefont {Thewalt}}, \bibinfo {author} {\bibfnamefont {F.~E.}\ \bibnamefont {Hudson}}, \bibinfo {author} {\bibfnamefont {C.~C.}\ \bibnamefont {Escott}}, \bibinfo {author} {\bibfnamefont {N.}~\bibnamefont {Ares}}, \bibinfo {author} {\bibfnamefont {S.~D.}\ \bibnamefont {Bartlett}}, \bibinfo {author} {\bibfnamefont {A.}~\bibnamefont {Morello}}, \bibinfo {author} {\bibfnamefont {A.}~\bibnamefont {Saraiva}}, \bibinfo {author} {\bibfnamefont {A.}~\bibnamefont {Laucht}}, \bibinfo {author} {\bibfnamefont {A.~S.}\ \bibnamefont {Dzurak}},\ and\ \bibinfo {author} {\bibfnamefont {C.~H.}\ \bibnamefont {Yang}},\ }\bibfield  {title} {\bibinfo {title} {High-fidelity spin qubit operation and algorithmic initialization above 1 {K}},\ }\href {https://doi.org/10.1038/s41586-024-07160-2} {\bibfield  {journal} {\bibinfo  {journal} {Nature}\ }\textbf {\bibinfo {volume}
  {627}},\ \bibinfo {pages} {772} (\bibinfo {year} {2024})}\BibitemShut {NoStop}%
\bibitem [{\citenamefont {Steinacker}\ \emph {et~al.}(2024)\citenamefont {Steinacker}, \citenamefont {Stuyck}, \citenamefont {Lim}, \citenamefont {Tanttu}, \citenamefont {Feng}, \citenamefont {Nickl}, \citenamefont {Serrano}, \citenamefont {Candido}, \citenamefont {Cifuentes}, \citenamefont {Hudson}, \citenamefont {Chan}, \citenamefont {Kubicek}, \citenamefont {Jussot}, \citenamefont {Canvel}, \citenamefont {Beyne}, \citenamefont {Shimura}, \citenamefont {Loo}, \citenamefont {Godfrin}, \citenamefont {Raes}, \citenamefont {Baudot}, \citenamefont {Wan}, \citenamefont {Laucht}, \citenamefont {Yang}, \citenamefont {Saraiva}, \citenamefont {Escott}, \citenamefont {Greve},\ and\ \citenamefont {Dzurak}}]{Steinacker24}%
  \BibitemOpen
  \bibfield  {author} {\bibinfo {author} {\bibfnamefont {P.}~\bibnamefont {Steinacker}}, \bibinfo {author} {\bibfnamefont {N.~D.}\ \bibnamefont {Stuyck}}, \bibinfo {author} {\bibfnamefont {W.~H.}\ \bibnamefont {Lim}}, \bibinfo {author} {\bibfnamefont {T.}~\bibnamefont {Tanttu}}, \bibinfo {author} {\bibfnamefont {M.}~\bibnamefont {Feng}}, \bibinfo {author} {\bibfnamefont {A.}~\bibnamefont {Nickl}}, \bibinfo {author} {\bibfnamefont {S.}~\bibnamefont {Serrano}}, \bibinfo {author} {\bibfnamefont {M.}~\bibnamefont {Candido}}, \bibinfo {author} {\bibfnamefont {J.~D.}\ \bibnamefont {Cifuentes}}, \bibinfo {author} {\bibfnamefont {F.~E.}\ \bibnamefont {Hudson}}, \bibinfo {author} {\bibfnamefont {K.~W.}\ \bibnamefont {Chan}}, \bibinfo {author} {\bibfnamefont {S.}~\bibnamefont {Kubicek}}, \bibinfo {author} {\bibfnamefont {J.}~\bibnamefont {Jussot}}, \bibinfo {author} {\bibfnamefont {Y.}~\bibnamefont {Canvel}}, \bibinfo {author} {\bibfnamefont {S.}~\bibnamefont {Beyne}}, \bibinfo {author} {\bibfnamefont {Y.}~\bibnamefont
  {Shimura}}, \bibinfo {author} {\bibfnamefont {R.}~\bibnamefont {Loo}}, \bibinfo {author} {\bibfnamefont {C.}~\bibnamefont {Godfrin}}, \bibinfo {author} {\bibfnamefont {B.}~\bibnamefont {Raes}}, \bibinfo {author} {\bibfnamefont {S.}~\bibnamefont {Baudot}}, \bibinfo {author} {\bibfnamefont {D.}~\bibnamefont {Wan}}, \bibinfo {author} {\bibfnamefont {A.}~\bibnamefont {Laucht}}, \bibinfo {author} {\bibfnamefont {C.~H.}\ \bibnamefont {Yang}}, \bibinfo {author} {\bibfnamefont {A.}~\bibnamefont {Saraiva}}, \bibinfo {author} {\bibfnamefont {C.~C.}\ \bibnamefont {Escott}}, \bibinfo {author} {\bibfnamefont {K.~D.}\ \bibnamefont {Greve}},\ and\ \bibinfo {author} {\bibfnamefont {A.~S.}\ \bibnamefont {Dzurak}},\ }\bibfield  {title} {\bibinfo {title} {A 300 mm foundry silicon spin qubit unit cell exceeding 99\% fidelity in all operations},\ }\href {https://arxiv.org/abs/2410.15590} {\bibfield  {journal} {\bibinfo  {journal} {arXiv:2410.15590}\ } (\bibinfo {year} {2024})}\BibitemShut {NoStop}%
\bibitem [{\citenamefont {Maurand}\ \emph {et~al.}(2016)\citenamefont {Maurand}, \citenamefont {Jehl}, \citenamefont {Kotekar-Patil}, \citenamefont {Corna}, \citenamefont {Bohuslavskyi}, \citenamefont {Lavi\'{e}ville}, \citenamefont {Hutin}, \citenamefont {Barraud}, \citenamefont {Vinet}, \citenamefont {Sanquer},\ and\ \citenamefont {de~Franceschi}}]{Maurand16}%
  \BibitemOpen
  \bibfield  {author} {\bibinfo {author} {\bibfnamefont {R.}~\bibnamefont {Maurand}}, \bibinfo {author} {\bibfnamefont {X.}~\bibnamefont {Jehl}}, \bibinfo {author} {\bibfnamefont {D.}~\bibnamefont {Kotekar-Patil}}, \bibinfo {author} {\bibfnamefont {A.}~\bibnamefont {Corna}}, \bibinfo {author} {\bibfnamefont {H.}~\bibnamefont {Bohuslavskyi}}, \bibinfo {author} {\bibfnamefont {R.}~\bibnamefont {Lavi\'{e}ville}}, \bibinfo {author} {\bibfnamefont {L.}~\bibnamefont {Hutin}}, \bibinfo {author} {\bibfnamefont {S.}~\bibnamefont {Barraud}}, \bibinfo {author} {\bibfnamefont {M.}~\bibnamefont {Vinet}}, \bibinfo {author} {\bibfnamefont {M.}~\bibnamefont {Sanquer}},\ and\ \bibinfo {author} {\bibfnamefont {S.}~\bibnamefont {de~Franceschi}},\ }\bibfield  {title} {\bibinfo {title} {A {CMOS} silicon spin qubit},\ }\href {https://doi.org/10.1038/ncomms13575} {\bibfield  {journal} {\bibinfo  {journal} {Nature Communications}\ }\textbf {\bibinfo {volume} {7}},\ \bibinfo {pages} {13575} (\bibinfo {year} {2016})}\BibitemShut
  {NoStop}%
\bibitem [{\citenamefont {Watzinger}\ \emph {et~al.}(2018)\citenamefont {Watzinger}, \citenamefont {Kuku\v{c}ka}, \citenamefont {Vuku\v{s}i\'c}, \citenamefont {Gao}, \citenamefont {Wang}, \citenamefont {Sch\"affler}, \citenamefont {Zhang},\ and\ \citenamefont {Katsaros}}]{Watzinger18}%
  \BibitemOpen
  \bibfield  {author} {\bibinfo {author} {\bibfnamefont {H.}~\bibnamefont {Watzinger}}, \bibinfo {author} {\bibfnamefont {J.}~\bibnamefont {Kuku\v{c}ka}}, \bibinfo {author} {\bibfnamefont {L.}~\bibnamefont {Vuku\v{s}i\'c}}, \bibinfo {author} {\bibfnamefont {F.}~\bibnamefont {Gao}}, \bibinfo {author} {\bibfnamefont {T.}~\bibnamefont {Wang}}, \bibinfo {author} {\bibfnamefont {F.}~\bibnamefont {Sch\"affler}}, \bibinfo {author} {\bibfnamefont {J.-J.}\ \bibnamefont {Zhang}},\ and\ \bibinfo {author} {\bibfnamefont {G.}~\bibnamefont {Katsaros}},\ }\bibfield  {title} {\bibinfo {title} {A germanium hole spin qubit},\ }\href {https://doi.org/10.1038/s41467-018-06418-4} {\bibfield  {journal} {\bibinfo  {journal} {Nature Communications}\ }\textbf {\bibinfo {volume} {9}},\ \bibinfo {pages} {3902} (\bibinfo {year} {2018})}\BibitemShut {NoStop}%
\bibitem [{\citenamefont {Hendrickx}\ \emph {et~al.}(2020{\natexlab{a}})\citenamefont {Hendrickx}, \citenamefont {Lawrie}, \citenamefont {Petit}, \citenamefont {Sammak}, \citenamefont {Scappucci},\ and\ \citenamefont {Veldhorst}}]{Hendrickx20b}%
  \BibitemOpen
  \bibfield  {author} {\bibinfo {author} {\bibfnamefont {N.~W.}\ \bibnamefont {Hendrickx}}, \bibinfo {author} {\bibfnamefont {W.~I.~L.}\ \bibnamefont {Lawrie}}, \bibinfo {author} {\bibfnamefont {L.}~\bibnamefont {Petit}}, \bibinfo {author} {\bibfnamefont {A.}~\bibnamefont {Sammak}}, \bibinfo {author} {\bibfnamefont {G.}~\bibnamefont {Scappucci}},\ and\ \bibinfo {author} {\bibfnamefont {M.}~\bibnamefont {Veldhorst}},\ }\bibfield  {title} {\bibinfo {title} {A single-hole spin qubit},\ }\href {https://doi.org/10.1038/s41467-020-17211-7} {\bibfield  {journal} {\bibinfo  {journal} {Nature Communications}\ }\textbf {\bibinfo {volume} {11}},\ \bibinfo {pages} {3478} (\bibinfo {year} {2020}{\natexlab{a}})}\BibitemShut {NoStop}%
\bibitem [{\citenamefont {Hendrickx}\ \emph {et~al.}(2020{\natexlab{b}})\citenamefont {Hendrickx}, \citenamefont {Franke}, \citenamefont {Sammak}, \citenamefont {Scappucci},\ and\ \citenamefont {Veldhorst}}]{Hendrickx20}%
  \BibitemOpen
  \bibfield  {author} {\bibinfo {author} {\bibfnamefont {N.~W.}\ \bibnamefont {Hendrickx}}, \bibinfo {author} {\bibfnamefont {D.~P.}\ \bibnamefont {Franke}}, \bibinfo {author} {\bibfnamefont {A.}~\bibnamefont {Sammak}}, \bibinfo {author} {\bibfnamefont {G.}~\bibnamefont {Scappucci}},\ and\ \bibinfo {author} {\bibfnamefont {M.}~\bibnamefont {Veldhorst}},\ }\bibfield  {title} {\bibinfo {title} {{Fast two-qubit logic with holes in germanium}},\ }\href {https://doi.org/10.1038/s41586-019-1919-3} {\bibfield  {journal} {\bibinfo  {journal} {Nature}\ }\textbf {\bibinfo {volume} {577}},\ \bibinfo {pages} {487} (\bibinfo {year} {2020}{\natexlab{b}})}\BibitemShut {NoStop}%
\bibitem [{\citenamefont {Froning}\ \emph {et~al.}(2021{\natexlab{a}})\citenamefont {Froning}, \citenamefont {Camenzind}, \citenamefont {van~der Molen}, \citenamefont {Li}, \citenamefont {Bakkers}, \citenamefont {Zumbühl},\ and\ \citenamefont {Braakman}}]{Froning21}%
  \BibitemOpen
  \bibfield  {author} {\bibinfo {author} {\bibfnamefont {F.~N.~M.}\ \bibnamefont {Froning}}, \bibinfo {author} {\bibfnamefont {L.~C.}\ \bibnamefont {Camenzind}}, \bibinfo {author} {\bibfnamefont {O.~A.~H.}\ \bibnamefont {van~der Molen}}, \bibinfo {author} {\bibfnamefont {A.}~\bibnamefont {Li}}, \bibinfo {author} {\bibfnamefont {E.~P. A.~M.}\ \bibnamefont {Bakkers}}, \bibinfo {author} {\bibfnamefont {D.~M.}\ \bibnamefont {Zumbühl}},\ and\ \bibinfo {author} {\bibfnamefont {F.~R.}\ \bibnamefont {Braakman}},\ }\bibfield  {title} {\bibinfo {title} {Ultrafast hole spin qubit with gate-tunable spin–orbit switch functionality},\ }\href {https://doi.org/10.1038/s41565-020-00828-6} {\bibfield  {journal} {\bibinfo  {journal} {Nature Nanotechnology}\ }\textbf {\bibinfo {volume} {16}},\ \bibinfo {pages} {308} (\bibinfo {year} {2021}{\natexlab{a}})}\BibitemShut {NoStop}%
\bibitem [{\citenamefont {Hendrickx}\ \emph {et~al.}(2021)\citenamefont {Hendrickx}, \citenamefont {Lawrie~William}, \citenamefont {Russ}, \citenamefont {van Riggelen}, \citenamefont {de~Snoo}, \citenamefont {Schouten}, \citenamefont {Sammak}, \citenamefont {Scappucci},\ and\ \citenamefont {Veldhorst}}]{Hendrickx21}%
  \BibitemOpen
  \bibfield  {author} {\bibinfo {author} {\bibfnamefont {N.~W.}\ \bibnamefont {Hendrickx}}, \bibinfo {author} {\bibfnamefont {I.~L.}\ \bibnamefont {Lawrie~William}}, \bibinfo {author} {\bibfnamefont {M.}~\bibnamefont {Russ}}, \bibinfo {author} {\bibfnamefont {F.}~\bibnamefont {van Riggelen}}, \bibinfo {author} {\bibfnamefont {S.~L.}\ \bibnamefont {de~Snoo}}, \bibinfo {author} {\bibfnamefont {R.~N.}\ \bibnamefont {Schouten}}, \bibinfo {author} {\bibfnamefont {A.}~\bibnamefont {Sammak}}, \bibinfo {author} {\bibfnamefont {G.}~\bibnamefont {Scappucci}},\ and\ \bibinfo {author} {\bibfnamefont {M.}~\bibnamefont {Veldhorst}},\ }\bibfield  {title} {\bibinfo {title} {A four-qubit germanium quantum processor},\ }\href {https://doi.org/10.1038/s41586-021-03332-6} {\bibfield  {journal} {\bibinfo  {journal} {Nature}\ }\textbf {\bibinfo {volume} {591}},\ \bibinfo {pages} {580} (\bibinfo {year} {2021})}\BibitemShut {NoStop}%
\bibitem [{\citenamefont {Camenzind}\ \emph {et~al.}(2022)\citenamefont {Camenzind}, \citenamefont {Geyer}, \citenamefont {Fuhrer}, \citenamefont {Warburton}, \citenamefont {Zumbühl},\ and\ \citenamefont {Kuhlmann}}]{Camenzind22}%
  \BibitemOpen
  \bibfield  {author} {\bibinfo {author} {\bibfnamefont {L.~C.}\ \bibnamefont {Camenzind}}, \bibinfo {author} {\bibfnamefont {S.}~\bibnamefont {Geyer}}, \bibinfo {author} {\bibfnamefont {A.}~\bibnamefont {Fuhrer}}, \bibinfo {author} {\bibfnamefont {R.~J.}\ \bibnamefont {Warburton}}, \bibinfo {author} {\bibfnamefont {D.~M.}\ \bibnamefont {Zumbühl}},\ and\ \bibinfo {author} {\bibfnamefont {A.~V.}\ \bibnamefont {Kuhlmann}},\ }\bibfield  {title} {\bibinfo {title} {A hole spin qubit in a fin field-effect transistor above 4 kelvin},\ }\href {https://doi.org/10.1038/s41928-022-00722-0} {\bibfield  {journal} {\bibinfo  {journal} {Nature Electronics}\ }\textbf {\bibinfo {volume} {5}},\ \bibinfo {pages} {178} (\bibinfo {year} {2022})}\BibitemShut {NoStop}%
\bibitem [{\citenamefont {Wang}\ \emph {et~al.}(2022)\citenamefont {Wang}, \citenamefont {Xu}, \citenamefont {Gao}, \citenamefont {Liu}, \citenamefont {Ma}, \citenamefont {Zhang}, \citenamefont {Wang}, \citenamefont {Cao}, \citenamefont {Wang}, \citenamefont {Zhang}, \citenamefont {Culcer}, \citenamefont {Hu}, \citenamefont {Jiang}, \citenamefont {Li}, \citenamefont {Guo},\ and\ \citenamefont {Guo}}]{wang2022ultrafast}%
  \BibitemOpen
  \bibfield  {author} {\bibinfo {author} {\bibfnamefont {K.}~\bibnamefont {Wang}}, \bibinfo {author} {\bibfnamefont {G.}~\bibnamefont {Xu}}, \bibinfo {author} {\bibfnamefont {F.}~\bibnamefont {Gao}}, \bibinfo {author} {\bibfnamefont {H.}~\bibnamefont {Liu}}, \bibinfo {author} {\bibfnamefont {R.-L.}\ \bibnamefont {Ma}}, \bibinfo {author} {\bibfnamefont {X.}~\bibnamefont {Zhang}}, \bibinfo {author} {\bibfnamefont {Z.}~\bibnamefont {Wang}}, \bibinfo {author} {\bibfnamefont {G.}~\bibnamefont {Cao}}, \bibinfo {author} {\bibfnamefont {T.}~\bibnamefont {Wang}}, \bibinfo {author} {\bibfnamefont {J.-J.}\ \bibnamefont {Zhang}}, \bibinfo {author} {\bibfnamefont {D.}~\bibnamefont {Culcer}}, \bibinfo {author} {\bibfnamefont {X.}~\bibnamefont {Hu}}, \bibinfo {author} {\bibfnamefont {H.-W.}\ \bibnamefont {Jiang}}, \bibinfo {author} {\bibfnamefont {H.-O.}\ \bibnamefont {Li}}, \bibinfo {author} {\bibfnamefont {G.-C.}\ \bibnamefont {Guo}},\ and\ \bibinfo {author} {\bibfnamefont {G.-P.}\ \bibnamefont {Guo}},\ }\bibfield
  {title} {\bibinfo {title} {Ultrafast coherent control of a hole spin qubit in a germanium quantum dot},\ }\href {https://doi.org/10.1038/s41467-021-27880-7} {\bibfield  {journal} {\bibinfo  {journal} {Nature Communications}\ }\textbf {\bibinfo {volume} {13}},\ \bibinfo {pages} {206} (\bibinfo {year} {2022})}\BibitemShut {NoStop}%
\bibitem [{\citenamefont {Lawrie}\ \emph {et~al.}(2023)\citenamefont {Lawrie}, \citenamefont {Rimbach-Russ}, \citenamefont {Riggelen}, \citenamefont {Hendrickx}, \citenamefont {Snoo}, \citenamefont {Sammak}, \citenamefont {Scappucci}, \citenamefont {Helsen},\ and\ \citenamefont {Veldhorst}}]{Lawrie23}%
  \BibitemOpen
  \bibfield  {author} {\bibinfo {author} {\bibfnamefont {W.~I.~L.}\ \bibnamefont {Lawrie}}, \bibinfo {author} {\bibfnamefont {M.}~\bibnamefont {Rimbach-Russ}}, \bibinfo {author} {\bibfnamefont {F.~v.}\ \bibnamefont {Riggelen}}, \bibinfo {author} {\bibfnamefont {N.~W.}\ \bibnamefont {Hendrickx}}, \bibinfo {author} {\bibfnamefont {S.~L.~d.}\ \bibnamefont {Snoo}}, \bibinfo {author} {\bibfnamefont {A.}~\bibnamefont {Sammak}}, \bibinfo {author} {\bibfnamefont {G.}~\bibnamefont {Scappucci}}, \bibinfo {author} {\bibfnamefont {J.}~\bibnamefont {Helsen}},\ and\ \bibinfo {author} {\bibfnamefont {M.}~\bibnamefont {Veldhorst}},\ }\bibfield  {title} {\bibinfo {title} {Simultaneous single-qubit driving of semiconductor spin qubits at the fault-tolerant threshold},\ }\href {https://doi.org/10.1038/s41467-023-39334-3} {\bibfield  {journal} {\bibinfo  {journal} {Nature Communications}\ }\textbf {\bibinfo {volume} {14}},\ \bibinfo {pages} {3617} (\bibinfo {year} {2023})}\BibitemShut {NoStop}%
\bibitem [{\citenamefont {John}\ \emph {et~al.}(2025)\citenamefont {John}, \citenamefont {Yu}, \citenamefont {van Straaten}, \citenamefont {Rodríguez-Mena}, \citenamefont {Rodríguez}, \citenamefont {Oosterhout}, \citenamefont {Stehouwer}, \citenamefont {Scappucci}, \citenamefont {Bosco}, \citenamefont {Rimbach-Russ}, \citenamefont {Niquet}, \citenamefont {Borsoi},\ and\ \citenamefont {Veldhorst}}]{Valentin25}%
  \BibitemOpen
  \bibfield  {author} {\bibinfo {author} {\bibfnamefont {V.}~\bibnamefont {John}}, \bibinfo {author} {\bibfnamefont {C.~X.}\ \bibnamefont {Yu}}, \bibinfo {author} {\bibfnamefont {B.}~\bibnamefont {van Straaten}}, \bibinfo {author} {\bibfnamefont {E.~A.}\ \bibnamefont {Rodríguez-Mena}}, \bibinfo {author} {\bibfnamefont {M.}~\bibnamefont {Rodríguez}}, \bibinfo {author} {\bibfnamefont {S.}~\bibnamefont {Oosterhout}}, \bibinfo {author} {\bibfnamefont {L.~E.~A.}\ \bibnamefont {Stehouwer}}, \bibinfo {author} {\bibfnamefont {G.}~\bibnamefont {Scappucci}}, \bibinfo {author} {\bibfnamefont {S.}~\bibnamefont {Bosco}}, \bibinfo {author} {\bibfnamefont {M.}~\bibnamefont {Rimbach-Russ}}, \bibinfo {author} {\bibfnamefont {Y.-M.}\ \bibnamefont {Niquet}}, \bibinfo {author} {\bibfnamefont {F.}~\bibnamefont {Borsoi}},\ and\ \bibinfo {author} {\bibfnamefont {M.}~\bibnamefont {Veldhorst}},\ }\bibfield  {title} {\bibinfo {title} {A two-dimensional 10-qubit array in germanium with robust and localised qubit control},\ }\href
  {https://arxiv.org/abs/2412.16044} {\bibfield  {journal} {\bibinfo  {journal} {arXiv:2412.16044}\ } (\bibinfo {year} {2025})}\BibitemShut {NoStop}%
\bibitem [{\citenamefont {Winkler}(2003)}]{Winkler03}%
  \BibitemOpen
  \bibfield  {author} {\bibinfo {author} {\bibfnamefont {R.}~\bibnamefont {Winkler}},\ }\href {https://doi.org/10.1007/b13586} {\emph {\bibinfo {title} {Spin-orbit coupling in two-dimensional electron and hole systems}}}\ (\bibinfo  {publisher} {Springer},\ \bibinfo {address} {Berlin},\ \bibinfo {year} {2003})\BibitemShut {NoStop}%
\bibitem [{\citenamefont {Wang}\ \emph {et~al.}(2021)\citenamefont {Wang}, \citenamefont {Marcellina}, \citenamefont {Hamilton}, \citenamefont {Cullen}, \citenamefont {Rogge}, \citenamefont {Salfi},\ and\ \citenamefont {Culcer}}]{Wang21}%
  \BibitemOpen
  \bibfield  {author} {\bibinfo {author} {\bibfnamefont {Z.}~\bibnamefont {Wang}}, \bibinfo {author} {\bibfnamefont {E.}~\bibnamefont {Marcellina}}, \bibinfo {author} {\bibfnamefont {A.~R.}\ \bibnamefont {Hamilton}}, \bibinfo {author} {\bibfnamefont {J.~H.}\ \bibnamefont {Cullen}}, \bibinfo {author} {\bibfnamefont {S.}~\bibnamefont {Rogge}}, \bibinfo {author} {\bibfnamefont {J.}~\bibnamefont {Salfi}},\ and\ \bibinfo {author} {\bibfnamefont {D.}~\bibnamefont {Culcer}},\ }\bibfield  {title} {\bibinfo {title} {Optimal operation points for ultrafast, highly coherent {Ge} hole spin-orbit qubits},\ }\href {https://doi.org/10.1038/s41534-021-00386-2} {\bibfield  {journal} {\bibinfo  {journal} {npj Quantum Information}\ }\textbf {\bibinfo {volume} {7}},\ \bibinfo {pages} {54} (\bibinfo {year} {2021})}\BibitemShut {NoStop}%
\bibitem [{\citenamefont {Piot}\ \emph {et~al.}(2022)\citenamefont {Piot}, \citenamefont {Brun}, \citenamefont {Schmitt}, \citenamefont {Zihlmann}, \citenamefont {Michal}, \citenamefont {Apra}, \citenamefont {Abadillo-Uriel}, \citenamefont {Jehl}, \citenamefont {Bertrand}, \citenamefont {Niebojewski}, \citenamefont {Hutin}, \citenamefont {Vinet}, \citenamefont {Urdampilleta}, \citenamefont {Meunier}, \citenamefont {Niquet}, \citenamefont {Maurand},\ and\ \citenamefont {De~Franceschi}}]{Piot22}%
  \BibitemOpen
  \bibfield  {author} {\bibinfo {author} {\bibfnamefont {N.}~\bibnamefont {Piot}}, \bibinfo {author} {\bibfnamefont {B.}~\bibnamefont {Brun}}, \bibinfo {author} {\bibfnamefont {V.}~\bibnamefont {Schmitt}}, \bibinfo {author} {\bibfnamefont {S.}~\bibnamefont {Zihlmann}}, \bibinfo {author} {\bibfnamefont {V.~P.}\ \bibnamefont {Michal}}, \bibinfo {author} {\bibfnamefont {A.}~\bibnamefont {Apra}}, \bibinfo {author} {\bibfnamefont {J.~C.}\ \bibnamefont {Abadillo-Uriel}}, \bibinfo {author} {\bibfnamefont {X.}~\bibnamefont {Jehl}}, \bibinfo {author} {\bibfnamefont {B.}~\bibnamefont {Bertrand}}, \bibinfo {author} {\bibfnamefont {H.}~\bibnamefont {Niebojewski}}, \bibinfo {author} {\bibfnamefont {L.}~\bibnamefont {Hutin}}, \bibinfo {author} {\bibfnamefont {M.}~\bibnamefont {Vinet}}, \bibinfo {author} {\bibfnamefont {M.}~\bibnamefont {Urdampilleta}}, \bibinfo {author} {\bibfnamefont {T.}~\bibnamefont {Meunier}}, \bibinfo {author} {\bibfnamefont {Y.-M.}\ \bibnamefont {Niquet}}, \bibinfo {author} {\bibfnamefont
  {R.}~\bibnamefont {Maurand}},\ and\ \bibinfo {author} {\bibfnamefont {S.}~\bibnamefont {De~Franceschi}},\ }\bibfield  {title} {\bibinfo {title} {A single hole spin with enhanced coherence in natural silicon},\ }\href {https://doi.org/10.1038/s41565-022-01196-z} {\bibfield  {journal} {\bibinfo  {journal} {Nature Nanotechnology}\ }\textbf {\bibinfo {volume} {17}},\ \bibinfo {pages} {1072} (\bibinfo {year} {2022})}\BibitemShut {NoStop}%
\bibitem [{\citenamefont {Hendrickx}\ \emph {et~al.}(2024)\citenamefont {Hendrickx}, \citenamefont {Massai}, \citenamefont {Mergenthaler}, \citenamefont {Schupp}, \citenamefont {Paredes}, \citenamefont {Bedell}, \citenamefont {Salis},\ and\ \citenamefont {Fuhrer}}]{Hendrickx2024}%
  \BibitemOpen
  \bibfield  {author} {\bibinfo {author} {\bibfnamefont {N.~W.}\ \bibnamefont {Hendrickx}}, \bibinfo {author} {\bibfnamefont {L.}~\bibnamefont {Massai}}, \bibinfo {author} {\bibfnamefont {M.}~\bibnamefont {Mergenthaler}}, \bibinfo {author} {\bibfnamefont {F.}~\bibnamefont {Schupp}}, \bibinfo {author} {\bibfnamefont {S.}~\bibnamefont {Paredes}}, \bibinfo {author} {\bibfnamefont {S.~W.}\ \bibnamefont {Bedell}}, \bibinfo {author} {\bibfnamefont {G.}~\bibnamefont {Salis}},\ and\ \bibinfo {author} {\bibfnamefont {A.}~\bibnamefont {Fuhrer}},\ }\bibfield  {title} {\bibinfo {title} {Sweet-spot operation of a germanium hole spin qubit with highly anisotropic noise sensitivity},\ }\href {https://doi.org/10.1038/s41563-024-01857-5} {\bibfield  {journal} {\bibinfo  {journal} {Nature Materials}\ }\textbf {\bibinfo {volume} {23}},\ \bibinfo {pages} {920} (\bibinfo {year} {2024})}\BibitemShut {NoStop}%
\bibitem [{\citenamefont {Mauro}\ \emph {et~al.}(2024)\citenamefont {Mauro}, \citenamefont {Rodr\'{\i}guez-Mena}, \citenamefont {Bassi}, \citenamefont {Schmitt},\ and\ \citenamefont {Niquet}}]{Mauro24}%
  \BibitemOpen
  \bibfield  {author} {\bibinfo {author} {\bibfnamefont {L.}~\bibnamefont {Mauro}}, \bibinfo {author} {\bibfnamefont {E.~A.}\ \bibnamefont {Rodr\'{\i}guez-Mena}}, \bibinfo {author} {\bibfnamefont {M.}~\bibnamefont {Bassi}}, \bibinfo {author} {\bibfnamefont {V.}~\bibnamefont {Schmitt}},\ and\ \bibinfo {author} {\bibfnamefont {Y.-M.}\ \bibnamefont {Niquet}},\ }\bibfield  {title} {\bibinfo {title} {Geometry of the dephasing sweet spots of spin-orbit qubits},\ }\href {https://doi.org/10.1103/PhysRevB.109.155406} {\bibfield  {journal} {\bibinfo  {journal} {Physical Review B}\ }\textbf {\bibinfo {volume} {109}},\ \bibinfo {pages} {155406} (\bibinfo {year} {2024})}\BibitemShut {NoStop}%
\bibitem [{\citenamefont {Bassi}\ \emph {et~al.}(2024)\citenamefont {Bassi}, \citenamefont {Rodrıguez-Mena}, \citenamefont {Brun}, \citenamefont {Zihlmann}, \citenamefont {Nguyen}, \citenamefont {Champain}, \citenamefont {Abadillo-Uriel}, \citenamefont {Bertrand}, \citenamefont {Niebojewski}, \citenamefont {Maurand}, \citenamefont {Niquet}, \citenamefont {Jehl}, \citenamefont {Franceschi},\ and\ \citenamefont {Schmitt}}]{Bassi24}%
  \BibitemOpen
  \bibfield  {author} {\bibinfo {author} {\bibfnamefont {M.}~\bibnamefont {Bassi}}, \bibinfo {author} {\bibfnamefont {E.-A.}\ \bibnamefont {Rodrıguez-Mena}}, \bibinfo {author} {\bibfnamefont {B.}~\bibnamefont {Brun}}, \bibinfo {author} {\bibfnamefont {S.}~\bibnamefont {Zihlmann}}, \bibinfo {author} {\bibfnamefont {T.}~\bibnamefont {Nguyen}}, \bibinfo {author} {\bibfnamefont {V.}~\bibnamefont {Champain}}, \bibinfo {author} {\bibfnamefont {J.~C.}\ \bibnamefont {Abadillo-Uriel}}, \bibinfo {author} {\bibfnamefont {B.}~\bibnamefont {Bertrand}}, \bibinfo {author} {\bibfnamefont {H.}~\bibnamefont {Niebojewski}}, \bibinfo {author} {\bibfnamefont {R.}~\bibnamefont {Maurand}}, \bibinfo {author} {\bibfnamefont {Y.-M.}\ \bibnamefont {Niquet}}, \bibinfo {author} {\bibfnamefont {X.}~\bibnamefont {Jehl}}, \bibinfo {author} {\bibfnamefont {S.~D.}\ \bibnamefont {Franceschi}},\ and\ \bibinfo {author} {\bibfnamefont {V.}~\bibnamefont {Schmitt}},\ }\bibfield  {title} {\bibinfo {title} {Optimal operation of hole spin qubits},\
  }\href {https://arxiv.org/abs/2412.13069} {\bibfield  {journal} {\bibinfo  {journal} {arXiv:2310.05902}\ } (\bibinfo {year} {2024})}\BibitemShut {NoStop}%
\bibitem [{\citenamefont {Sammak}\ \emph {et~al.}(2019)\citenamefont {Sammak}, \citenamefont {Sabbagh}, \citenamefont {Hendrickx}, \citenamefont {Lodari}, \citenamefont {Paquelet~Wuetz}, \citenamefont {Tosato}, \citenamefont {Yeoh}, \citenamefont {Bollani}, \citenamefont {Virgilio}, \citenamefont {Schubert}, \citenamefont {Zaumseil}, \citenamefont {Capellini}, \citenamefont {Veldhorst},\ and\ \citenamefont {Scappucci}}]{Sammak19}%
  \BibitemOpen
  \bibfield  {author} {\bibinfo {author} {\bibfnamefont {A.}~\bibnamefont {Sammak}}, \bibinfo {author} {\bibfnamefont {D.}~\bibnamefont {Sabbagh}}, \bibinfo {author} {\bibfnamefont {N.~W.}\ \bibnamefont {Hendrickx}}, \bibinfo {author} {\bibfnamefont {M.}~\bibnamefont {Lodari}}, \bibinfo {author} {\bibfnamefont {B.}~\bibnamefont {Paquelet~Wuetz}}, \bibinfo {author} {\bibfnamefont {A.}~\bibnamefont {Tosato}}, \bibinfo {author} {\bibfnamefont {L.}~\bibnamefont {Yeoh}}, \bibinfo {author} {\bibfnamefont {M.}~\bibnamefont {Bollani}}, \bibinfo {author} {\bibfnamefont {M.}~\bibnamefont {Virgilio}}, \bibinfo {author} {\bibfnamefont {M.~A.}\ \bibnamefont {Schubert}}, \bibinfo {author} {\bibfnamefont {P.}~\bibnamefont {Zaumseil}}, \bibinfo {author} {\bibfnamefont {G.}~\bibnamefont {Capellini}}, \bibinfo {author} {\bibfnamefont {M.}~\bibnamefont {Veldhorst}},\ and\ \bibinfo {author} {\bibfnamefont {G.}~\bibnamefont {Scappucci}},\ }\bibfield  {title} {\bibinfo {title} {Shallow and undoped germanium quantum wells: A
  playground for spin and hybrid quantum technology},\ }\href {https://doi.org/10.1002/adfm.201807613} {\bibfield  {journal} {\bibinfo  {journal} {Advanced Functional Materials}\ }\textbf {\bibinfo {volume} {29}},\ \bibinfo {pages} {1807613} (\bibinfo {year} {2019})}\BibitemShut {NoStop}%
\bibitem [{\citenamefont {Scappucci}\ \emph {et~al.}(2021)\citenamefont {Scappucci}, \citenamefont {Kloeffel}, \citenamefont {Zwanenburg}, \citenamefont {Loss}, \citenamefont {Myronov}, \citenamefont {Zhang}, \citenamefont {De~Franceschi}, \citenamefont {Katsaros},\ and\ \citenamefont {Veldhorst}}]{Scappucci20}%
  \BibitemOpen
  \bibfield  {author} {\bibinfo {author} {\bibfnamefont {G.}~\bibnamefont {Scappucci}}, \bibinfo {author} {\bibfnamefont {C.}~\bibnamefont {Kloeffel}}, \bibinfo {author} {\bibfnamefont {F.~A.}\ \bibnamefont {Zwanenburg}}, \bibinfo {author} {\bibfnamefont {D.}~\bibnamefont {Loss}}, \bibinfo {author} {\bibfnamefont {M.}~\bibnamefont {Myronov}}, \bibinfo {author} {\bibfnamefont {J.-J.}\ \bibnamefont {Zhang}}, \bibinfo {author} {\bibfnamefont {S.}~\bibnamefont {De~Franceschi}}, \bibinfo {author} {\bibfnamefont {G.}~\bibnamefont {Katsaros}},\ and\ \bibinfo {author} {\bibfnamefont {M.}~\bibnamefont {Veldhorst}},\ }\bibfield  {title} {\bibinfo {title} {The germanium quantum information route},\ }\href {https://doi.org/10.1038/s41578-020-00262-z} {\bibfield  {journal} {\bibinfo  {journal} {Nature Reviews Materials}\ }\textbf {\bibinfo {volume} {6}},\ \bibinfo {pages} {926} (\bibinfo {year} {2021})}\BibitemShut {NoStop}%
\bibitem [{\citenamefont {Wang}\ \emph {et~al.}(2023)\citenamefont {Wang}, \citenamefont {Déprez}, \citenamefont {Tidjani}, \citenamefont {Lawrie}, \citenamefont {Hendrickx}, \citenamefont {Sammak}, \citenamefont {Scappucci},\ and\ \citenamefont {Veldhorst}}]{Wang2023}%
  \BibitemOpen
  \bibfield  {author} {\bibinfo {author} {\bibfnamefont {C.-A.}\ \bibnamefont {Wang}}, \bibinfo {author} {\bibfnamefont {C.}~\bibnamefont {Déprez}}, \bibinfo {author} {\bibfnamefont {H.}~\bibnamefont {Tidjani}}, \bibinfo {author} {\bibfnamefont {W.~I.~L.}\ \bibnamefont {Lawrie}}, \bibinfo {author} {\bibfnamefont {N.~W.}\ \bibnamefont {Hendrickx}}, \bibinfo {author} {\bibfnamefont {A.}~\bibnamefont {Sammak}}, \bibinfo {author} {\bibfnamefont {G.}~\bibnamefont {Scappucci}},\ and\ \bibinfo {author} {\bibfnamefont {M.}~\bibnamefont {Veldhorst}},\ }\bibfield  {title} {\bibinfo {title} {Probing resonating valence bonds on a programmable germanium quantum simulator},\ }\href {https://doi.org/10.1038/s41534-023-00727-3} {\bibfield  {journal} {\bibinfo  {journal} {npj Quantum Information}\ }\textbf {\bibinfo {volume} {9}},\ \bibinfo {pages} {58} (\bibinfo {year} {2023})}\BibitemShut {NoStop}%
\bibitem [{\citenamefont {Borsoi}\ \emph {et~al.}(2024)\citenamefont {Borsoi}, \citenamefont {Hendrickx}, \citenamefont {John}, \citenamefont {Meyer}, \citenamefont {Motz}, \citenamefont {van Riggelen}, \citenamefont {Sammak}, \citenamefont {de~Snoo}, \citenamefont {Scappucci},\ and\ \citenamefont {Veldhorst}}]{Borsoi24}%
  \BibitemOpen
  \bibfield  {author} {\bibinfo {author} {\bibfnamefont {F.}~\bibnamefont {Borsoi}}, \bibinfo {author} {\bibfnamefont {N.~W.}\ \bibnamefont {Hendrickx}}, \bibinfo {author} {\bibfnamefont {V.}~\bibnamefont {John}}, \bibinfo {author} {\bibfnamefont {M.}~\bibnamefont {Meyer}}, \bibinfo {author} {\bibfnamefont {S.}~\bibnamefont {Motz}}, \bibinfo {author} {\bibfnamefont {F.}~\bibnamefont {van Riggelen}}, \bibinfo {author} {\bibfnamefont {A.}~\bibnamefont {Sammak}}, \bibinfo {author} {\bibfnamefont {S.~L.}\ \bibnamefont {de~Snoo}}, \bibinfo {author} {\bibfnamefont {G.}~\bibnamefont {Scappucci}},\ and\ \bibinfo {author} {\bibfnamefont {M.}~\bibnamefont {Veldhorst}},\ }\bibfield  {title} {\bibinfo {title} {Shared control of a 16 semiconductor quantum dot crossbar array},\ }\href {https://doi.org/10.1038/s41565-023-01491-3} {\bibfield  {journal} {\bibinfo  {journal} {Nature Nanotechnology}\ }\textbf {\bibinfo {volume} {19}},\ \bibinfo {pages} {21} (\bibinfo {year} {2024})}\BibitemShut {NoStop}%
\bibitem [{\citenamefont {van Riggelen-Doelman}\ \emph {et~al.}(2024)\citenamefont {van Riggelen-Doelman}, \citenamefont {Wang}, \citenamefont {de~Snoo}, \citenamefont {Lawrie}, \citenamefont {Hendrickx}, \citenamefont {Rimbach-Russ}, \citenamefont {Sammak}, \citenamefont {Scappucci}, \citenamefont {Déprez},\ and\ \citenamefont {Veldhorst}}]{van_riggelen_shuttling_2024}%
  \BibitemOpen
  \bibfield  {author} {\bibinfo {author} {\bibfnamefont {F.}~\bibnamefont {van Riggelen-Doelman}}, \bibinfo {author} {\bibfnamefont {C.-A.}\ \bibnamefont {Wang}}, \bibinfo {author} {\bibfnamefont {S.~L.}\ \bibnamefont {de~Snoo}}, \bibinfo {author} {\bibfnamefont {W.~I.~L.}\ \bibnamefont {Lawrie}}, \bibinfo {author} {\bibfnamefont {N.~W.}\ \bibnamefont {Hendrickx}}, \bibinfo {author} {\bibfnamefont {M.}~\bibnamefont {Rimbach-Russ}}, \bibinfo {author} {\bibfnamefont {A.}~\bibnamefont {Sammak}}, \bibinfo {author} {\bibfnamefont {G.}~\bibnamefont {Scappucci}}, \bibinfo {author} {\bibfnamefont {C.}~\bibnamefont {Déprez}},\ and\ \bibinfo {author} {\bibfnamefont {M.}~\bibnamefont {Veldhorst}},\ }\bibfield  {title} {\bibinfo {title} {Coherent spin qubit shuttling through germanium quantum dots},\ }\href {https://doi.org/10.1038/s41467-024-49358-y} {\bibfield  {journal} {\bibinfo  {journal} {Nature Communications}\ }\textbf {\bibinfo {volume} {15}},\ \bibinfo {pages} {5716} (\bibinfo {year} {2024})}\BibitemShut
  {NoStop}%
\bibitem [{\citenamefont {Wang}\ \emph {et~al.}(2024{\natexlab{a}})\citenamefont {Wang}, \citenamefont {John}, \citenamefont {Tidjani}, \citenamefont {Yu}, \citenamefont {Ivlev}, \citenamefont {Déprez}, \citenamefont {van Riggelen-Doelman}, \citenamefont {Woods}, \citenamefont {Hendrickx}, \citenamefont {Lawrie}, \citenamefont {Stehouwer}, \citenamefont {Oosterhout}, \citenamefont {Sammak}, \citenamefont {Friesen}, \citenamefont {Scappucci}, \citenamefont {de~Snoo}, \citenamefont {Rimbach-Russ}, \citenamefont {Borsoi},\ and\ \citenamefont {Veldhorst}}]{Wang2024}%
  \BibitemOpen
  \bibfield  {author} {\bibinfo {author} {\bibfnamefont {C.-A.}\ \bibnamefont {Wang}}, \bibinfo {author} {\bibfnamefont {V.}~\bibnamefont {John}}, \bibinfo {author} {\bibfnamefont {H.}~\bibnamefont {Tidjani}}, \bibinfo {author} {\bibfnamefont {C.~X.}\ \bibnamefont {Yu}}, \bibinfo {author} {\bibfnamefont {A.~S.}\ \bibnamefont {Ivlev}}, \bibinfo {author} {\bibfnamefont {C.}~\bibnamefont {Déprez}}, \bibinfo {author} {\bibfnamefont {F.}~\bibnamefont {van Riggelen-Doelman}}, \bibinfo {author} {\bibfnamefont {B.~D.}\ \bibnamefont {Woods}}, \bibinfo {author} {\bibfnamefont {N.~W.}\ \bibnamefont {Hendrickx}}, \bibinfo {author} {\bibfnamefont {W.~I.~L.}\ \bibnamefont {Lawrie}}, \bibinfo {author} {\bibfnamefont {L.~E.~A.}\ \bibnamefont {Stehouwer}}, \bibinfo {author} {\bibfnamefont {S.~D.}\ \bibnamefont {Oosterhout}}, \bibinfo {author} {\bibfnamefont {A.}~\bibnamefont {Sammak}}, \bibinfo {author} {\bibfnamefont {M.}~\bibnamefont {Friesen}}, \bibinfo {author} {\bibfnamefont {G.}~\bibnamefont {Scappucci}}, \bibinfo
  {author} {\bibfnamefont {S.~L.}\ \bibnamefont {de~Snoo}}, \bibinfo {author} {\bibfnamefont {M.}~\bibnamefont {Rimbach-Russ}}, \bibinfo {author} {\bibfnamefont {F.}~\bibnamefont {Borsoi}},\ and\ \bibinfo {author} {\bibfnamefont {M.}~\bibnamefont {Veldhorst}},\ }\bibfield  {title} {\bibinfo {title} {Operating semiconductor quantum processors with hopping spins},\ }\href {https://doi.org/10.1126/science.ado5915} {\bibfield  {journal} {\bibinfo  {journal} {Science}\ }\textbf {\bibinfo {volume} {385}},\ \bibinfo {pages} {447} (\bibinfo {year} {2024}{\natexlab{a}})}\BibitemShut {NoStop}%
\bibitem [{\citenamefont {Rashba}\ and\ \citenamefont {Efros}(2003)}]{Rashba03}%
  \BibitemOpen
  \bibfield  {author} {\bibinfo {author} {\bibfnamefont {E.~I.}\ \bibnamefont {Rashba}}\ and\ \bibinfo {author} {\bibfnamefont {A.~L.}\ \bibnamefont {Efros}},\ }\bibfield  {title} {\bibinfo {title} {Orbital mechanisms of electron-spin manipulation by an electric field},\ }\href {https://doi.org/10.1103/PhysRevLett.91.126405} {\bibfield  {journal} {\bibinfo  {journal} {Physical Review Letters}\ }\textbf {\bibinfo {volume} {91}},\ \bibinfo {pages} {126405} (\bibinfo {year} {2003})}\BibitemShut {NoStop}%
\bibitem [{\citenamefont {Golovach}\ \emph {et~al.}(2006)\citenamefont {Golovach}, \citenamefont {Borhani},\ and\ \citenamefont {Loss}}]{Golovach06}%
  \BibitemOpen
  \bibfield  {author} {\bibinfo {author} {\bibfnamefont {V.~N.}\ \bibnamefont {Golovach}}, \bibinfo {author} {\bibfnamefont {M.}~\bibnamefont {Borhani}},\ and\ \bibinfo {author} {\bibfnamefont {D.}~\bibnamefont {Loss}},\ }\bibfield  {title} {\bibinfo {title} {Electric-dipole-induced spin resonance in quantum dots},\ }\href {https://doi.org/10.1103/PhysRevB.74.165319} {\bibfield  {journal} {\bibinfo  {journal} {Physical Review B}\ }\textbf {\bibinfo {volume} {74}},\ \bibinfo {pages} {165319} (\bibinfo {year} {2006})}\BibitemShut {NoStop}%
\bibitem [{\citenamefont {Kato}\ \emph {et~al.}(2003)\citenamefont {Kato}, \citenamefont {Myers}, \citenamefont {Driscoll}, \citenamefont {Gossard}, \citenamefont {Levy},\ and\ \citenamefont {Awschalom}}]{Kato03}%
  \BibitemOpen
  \bibfield  {author} {\bibinfo {author} {\bibfnamefont {Y.}~\bibnamefont {Kato}}, \bibinfo {author} {\bibfnamefont {R.~C.}\ \bibnamefont {Myers}}, \bibinfo {author} {\bibfnamefont {D.~C.}\ \bibnamefont {Driscoll}}, \bibinfo {author} {\bibfnamefont {A.~C.}\ \bibnamefont {Gossard}}, \bibinfo {author} {\bibfnamefont {J.}~\bibnamefont {Levy}},\ and\ \bibinfo {author} {\bibfnamefont {D.~D.}\ \bibnamefont {Awschalom}},\ }\bibfield  {title} {\bibinfo {title} {Gigahertz electron spin manipulation using voltage-controlled $g$-tensor modulation},\ }\href {https://doi.org/10.1126/science.1080880} {\bibfield  {journal} {\bibinfo  {journal} {Science}\ }\textbf {\bibinfo {volume} {299}},\ \bibinfo {pages} {1201} (\bibinfo {year} {2003})}\BibitemShut {NoStop}%
\bibitem [{\citenamefont {Crippa}\ \emph {et~al.}(2018)\citenamefont {Crippa}, \citenamefont {Maurand}, \citenamefont {Bourdet}, \citenamefont {Kotekar-Patil}, \citenamefont {Amisse}, \citenamefont {Jehl}, \citenamefont {Sanquer}, \citenamefont {Laviéville}, \citenamefont {Bohuslavskyi}, \citenamefont {Hutin}, \citenamefont {Barraud}, \citenamefont {Vinet}, \citenamefont {Niquet},\ and\ \citenamefont {{De Franceschi}}}]{Crippa18}%
  \BibitemOpen
  \bibfield  {author} {\bibinfo {author} {\bibfnamefont {A.}~\bibnamefont {Crippa}}, \bibinfo {author} {\bibfnamefont {R.}~\bibnamefont {Maurand}}, \bibinfo {author} {\bibfnamefont {L.}~\bibnamefont {Bourdet}}, \bibinfo {author} {\bibfnamefont {D.}~\bibnamefont {Kotekar-Patil}}, \bibinfo {author} {\bibfnamefont {A.}~\bibnamefont {Amisse}}, \bibinfo {author} {\bibfnamefont {X.}~\bibnamefont {Jehl}}, \bibinfo {author} {\bibfnamefont {M.}~\bibnamefont {Sanquer}}, \bibinfo {author} {\bibfnamefont {R.}~\bibnamefont {Laviéville}}, \bibinfo {author} {\bibfnamefont {H.}~\bibnamefont {Bohuslavskyi}}, \bibinfo {author} {\bibfnamefont {L.}~\bibnamefont {Hutin}}, \bibinfo {author} {\bibfnamefont {S.}~\bibnamefont {Barraud}}, \bibinfo {author} {\bibfnamefont {M.}~\bibnamefont {Vinet}}, \bibinfo {author} {\bibfnamefont {Y.-M.}\ \bibnamefont {Niquet}},\ and\ \bibinfo {author} {\bibfnamefont {S.}~\bibnamefont {{De Franceschi}}},\ }\bibfield  {title} {\bibinfo {title} {Electrical spin driving by $g$-matrix modulation in
  spin-orbit qubits},\ }\href {https://doi.org/10.1103/PhysRevLett.120.137702} {\bibfield  {journal} {\bibinfo  {journal} {Physical Review Letters}\ }\textbf {\bibinfo {volume} {120}},\ \bibinfo {pages} {137702} (\bibinfo {year} {2018})}\BibitemShut {NoStop}%
\bibitem [{\citenamefont {Kloeffel}\ \emph {et~al.}(2011)\citenamefont {Kloeffel}, \citenamefont {Trif},\ and\ \citenamefont {Loss}}]{Kloeffel11}%
  \BibitemOpen
  \bibfield  {author} {\bibinfo {author} {\bibfnamefont {C.}~\bibnamefont {Kloeffel}}, \bibinfo {author} {\bibfnamefont {M.}~\bibnamefont {Trif}},\ and\ \bibinfo {author} {\bibfnamefont {D.}~\bibnamefont {Loss}},\ }\bibfield  {title} {\bibinfo {title} {Strong spin-orbit interaction and helical hole states in {Ge/Si} nanowires},\ }\href {https://doi.org/10.1103/PhysRevB.84.195314} {\bibfield  {journal} {\bibinfo  {journal} {Physical Review B}\ }\textbf {\bibinfo {volume} {84}},\ \bibinfo {pages} {195314} (\bibinfo {year} {2011})}\BibitemShut {NoStop}%
\bibitem [{\citenamefont {Kloeffel}\ \emph {et~al.}(2013)\citenamefont {Kloeffel}, \citenamefont {Trif}, \citenamefont {Stano},\ and\ \citenamefont {Loss}}]{Kloeffel13}%
  \BibitemOpen
  \bibfield  {author} {\bibinfo {author} {\bibfnamefont {C.}~\bibnamefont {Kloeffel}}, \bibinfo {author} {\bibfnamefont {M.}~\bibnamefont {Trif}}, \bibinfo {author} {\bibfnamefont {P.}~\bibnamefont {Stano}},\ and\ \bibinfo {author} {\bibfnamefont {D.}~\bibnamefont {Loss}},\ }\bibfield  {title} {\bibinfo {title} {{Circuit QED with hole-spin qubits in Ge/Si nanowire quantum dots}},\ }\href {https://doi.org/10.1103/PhysRevB.88.241405} {\bibfield  {journal} {\bibinfo  {journal} {Physical Review B}\ }\textbf {\bibinfo {volume} {88}},\ \bibinfo {pages} {241405} (\bibinfo {year} {2013})}\BibitemShut {NoStop}%
\bibitem [{\citenamefont {Marcellina}\ \emph {et~al.}(2017)\citenamefont {Marcellina}, \citenamefont {Hamilton}, \citenamefont {Winkler},\ and\ \citenamefont {Culcer}}]{Marcellina17}%
  \BibitemOpen
  \bibfield  {author} {\bibinfo {author} {\bibfnamefont {E.}~\bibnamefont {Marcellina}}, \bibinfo {author} {\bibfnamefont {A.~R.}\ \bibnamefont {Hamilton}}, \bibinfo {author} {\bibfnamefont {R.}~\bibnamefont {Winkler}},\ and\ \bibinfo {author} {\bibfnamefont {D.}~\bibnamefont {Culcer}},\ }\bibfield  {title} {\bibinfo {title} {Spin-orbit interactions in inversion-asymmetric two-dimensional hole systems: A variational analysis},\ }\href {https://doi.org/10.1103/PhysRevB.95.075305} {\bibfield  {journal} {\bibinfo  {journal} {Physical Review B}\ }\textbf {\bibinfo {volume} {95}},\ \bibinfo {pages} {075305} (\bibinfo {year} {2017})}\BibitemShut {NoStop}%
\bibitem [{\citenamefont {Venitucci}\ \emph {et~al.}(2018)\citenamefont {Venitucci}, \citenamefont {Bourdet}, \citenamefont {Pouzada},\ and\ \citenamefont {Niquet}}]{Venitucci18}%
  \BibitemOpen
  \bibfield  {author} {\bibinfo {author} {\bibfnamefont {B.}~\bibnamefont {Venitucci}}, \bibinfo {author} {\bibfnamefont {L.}~\bibnamefont {Bourdet}}, \bibinfo {author} {\bibfnamefont {D.}~\bibnamefont {Pouzada}},\ and\ \bibinfo {author} {\bibfnamefont {Y.-M.}\ \bibnamefont {Niquet}},\ }\bibfield  {title} {\bibinfo {title} {Electrical manipulation of semiconductor spin qubits within the $g$-matrix formalism},\ }\href {https://doi.org/10.1103/PhysRevB.98.155319} {\bibfield  {journal} {\bibinfo  {journal} {Physical Review B}\ }\textbf {\bibinfo {volume} {98}},\ \bibinfo {pages} {155319} (\bibinfo {year} {2018})}\BibitemShut {NoStop}%
\bibitem [{\citenamefont {Li}\ \emph {et~al.}(2020)\citenamefont {Li}, \citenamefont {Venitucci},\ and\ \citenamefont {Niquet}}]{Li20}%
  \BibitemOpen
  \bibfield  {author} {\bibinfo {author} {\bibfnamefont {J.}~\bibnamefont {Li}}, \bibinfo {author} {\bibfnamefont {B.}~\bibnamefont {Venitucci}},\ and\ \bibinfo {author} {\bibfnamefont {Y.-M.}\ \bibnamefont {Niquet}},\ }\bibfield  {title} {\bibinfo {title} {Hole-phonon interactions in quantum dots: Effects of phonon confinement and encapsulation materials on spin-orbit qubits},\ }\href {https://link.aps.org/doi/10.1103/PhysRevB.102.075415} {\bibfield  {journal} {\bibinfo  {journal} {Physical Review B}\ }\textbf {\bibinfo {volume} {102}},\ \bibinfo {pages} {075415} (\bibinfo {year} {2020})}\BibitemShut {NoStop}%
\bibitem [{\citenamefont {Terrazos}\ \emph {et~al.}(2021)\citenamefont {Terrazos}, \citenamefont {Marcellina}, \citenamefont {Wang}, \citenamefont {Coppersmith}, \citenamefont {Friesen}, \citenamefont {Hamilton}, \citenamefont {Hu}, \citenamefont {Koiller}, \citenamefont {Saraiva}, \citenamefont {Culcer},\ and\ \citenamefont {Capaz}}]{Terrazos21}%
  \BibitemOpen
  \bibfield  {author} {\bibinfo {author} {\bibfnamefont {L.~A.}\ \bibnamefont {Terrazos}}, \bibinfo {author} {\bibfnamefont {E.}~\bibnamefont {Marcellina}}, \bibinfo {author} {\bibfnamefont {Z.}~\bibnamefont {Wang}}, \bibinfo {author} {\bibfnamefont {S.~N.}\ \bibnamefont {Coppersmith}}, \bibinfo {author} {\bibfnamefont {M.}~\bibnamefont {Friesen}}, \bibinfo {author} {\bibfnamefont {A.~R.}\ \bibnamefont {Hamilton}}, \bibinfo {author} {\bibfnamefont {X.}~\bibnamefont {Hu}}, \bibinfo {author} {\bibfnamefont {B.}~\bibnamefont {Koiller}}, \bibinfo {author} {\bibfnamefont {A.~L.}\ \bibnamefont {Saraiva}}, \bibinfo {author} {\bibfnamefont {D.}~\bibnamefont {Culcer}},\ and\ \bibinfo {author} {\bibfnamefont {R.~B.}\ \bibnamefont {Capaz}},\ }\bibfield  {title} {\bibinfo {title} {Theory of hole-spin qubits in strained germanium quantum dots},\ }\href {https://doi.org/10.1103/PhysRevB.103.125201} {\bibfield  {journal} {\bibinfo  {journal} {Physical Review B}\ }\textbf {\bibinfo {volume} {103}},\ \bibinfo {pages} {125201}
  (\bibinfo {year} {2021})}\BibitemShut {NoStop}%
\bibitem [{\citenamefont {Michal}\ \emph {et~al.}(2021)\citenamefont {Michal}, \citenamefont {Venitucci},\ and\ \citenamefont {Niquet}}]{Michal21}%
  \BibitemOpen
  \bibfield  {author} {\bibinfo {author} {\bibfnamefont {V.~P.}\ \bibnamefont {Michal}}, \bibinfo {author} {\bibfnamefont {B.}~\bibnamefont {Venitucci}},\ and\ \bibinfo {author} {\bibfnamefont {Y.-M.}\ \bibnamefont {Niquet}},\ }\bibfield  {title} {\bibinfo {title} {Longitudinal and transverse electric field manipulation of hole spin-orbit qubits in one-dimensional channels},\ }\href {https://doi.org/10.1103/PhysRevB.103.045305} {\bibfield  {journal} {\bibinfo  {journal} {Physical Review B}\ }\textbf {\bibinfo {volume} {103}},\ \bibinfo {pages} {045305} (\bibinfo {year} {2021})}\BibitemShut {NoStop}%
\bibitem [{\citenamefont {Bosco}\ \emph {et~al.}(2021)\citenamefont {Bosco}, \citenamefont {Benito}, \citenamefont {Adelsberger},\ and\ \citenamefont {Loss}}]{Bosco21b}%
  \BibitemOpen
  \bibfield  {author} {\bibinfo {author} {\bibfnamefont {S.}~\bibnamefont {Bosco}}, \bibinfo {author} {\bibfnamefont {M.}~\bibnamefont {Benito}}, \bibinfo {author} {\bibfnamefont {C.}~\bibnamefont {Adelsberger}},\ and\ \bibinfo {author} {\bibfnamefont {D.}~\bibnamefont {Loss}},\ }\bibfield  {title} {\bibinfo {title} {Squeezed hole spin qubits in {Ge} quantum dots with ultrafast gates at low power},\ }\href {https://doi.org/10.1103/PhysRevB.104.115425} {\bibfield  {journal} {\bibinfo  {journal} {Physical Review B}\ }\textbf {\bibinfo {volume} {104}},\ \bibinfo {pages} {115425} (\bibinfo {year} {2021})}\BibitemShut {NoStop}%
\bibitem [{\citenamefont {Martinez}\ \emph {et~al.}(2022)\citenamefont {Martinez}, \citenamefont {Abadillo-Uriel}, \citenamefont {Rodr\'{\i}guez-Mena},\ and\ \citenamefont {Niquet}}]{martinez2022hole}%
  \BibitemOpen
  \bibfield  {author} {\bibinfo {author} {\bibfnamefont {B.}~\bibnamefont {Martinez}}, \bibinfo {author} {\bibfnamefont {J.~C.}\ \bibnamefont {Abadillo-Uriel}}, \bibinfo {author} {\bibfnamefont {E.~A.}\ \bibnamefont {Rodr\'{\i}guez-Mena}},\ and\ \bibinfo {author} {\bibfnamefont {Y.-M.}\ \bibnamefont {Niquet}},\ }\bibfield  {title} {\bibinfo {title} {Hole spin manipulation in inhomogeneous and nonseparable electric fields},\ }\href {https://doi.org/10.1103/PhysRevB.106.235426} {\bibfield  {journal} {\bibinfo  {journal} {Physical Review B}\ }\textbf {\bibinfo {volume} {106}},\ \bibinfo {pages} {235426} (\bibinfo {year} {2022})}\BibitemShut {NoStop}%
\bibitem [{\citenamefont {Sarkar}\ \emph {et~al.}(2023)\citenamefont {Sarkar}, \citenamefont {Wang}, \citenamefont {Rendell}, \citenamefont {Hendrickx}, \citenamefont {Veldhorst}, \citenamefont {Scappucci}, \citenamefont {Khalifa}, \citenamefont {Salfi}, \citenamefont {Saraiva}, \citenamefont {Dzurak}, \citenamefont {Hamilton},\ and\ \citenamefont {Culcer}}]{Sarkar23}%
  \BibitemOpen
  \bibfield  {author} {\bibinfo {author} {\bibfnamefont {A.}~\bibnamefont {Sarkar}}, \bibinfo {author} {\bibfnamefont {Z.}~\bibnamefont {Wang}}, \bibinfo {author} {\bibfnamefont {M.}~\bibnamefont {Rendell}}, \bibinfo {author} {\bibfnamefont {N.~W.}\ \bibnamefont {Hendrickx}}, \bibinfo {author} {\bibfnamefont {M.}~\bibnamefont {Veldhorst}}, \bibinfo {author} {\bibfnamefont {G.}~\bibnamefont {Scappucci}}, \bibinfo {author} {\bibfnamefont {M.}~\bibnamefont {Khalifa}}, \bibinfo {author} {\bibfnamefont {J.}~\bibnamefont {Salfi}}, \bibinfo {author} {\bibfnamefont {A.}~\bibnamefont {Saraiva}}, \bibinfo {author} {\bibfnamefont {A.~S.}\ \bibnamefont {Dzurak}}, \bibinfo {author} {\bibfnamefont {A.~R.}\ \bibnamefont {Hamilton}},\ and\ \bibinfo {author} {\bibfnamefont {D.}~\bibnamefont {Culcer}},\ }\bibfield  {title} {\bibinfo {title} {Electrical operation of planar {Ge} hole spin qubits in an in-plane magnetic field},\ }\href {https://doi.org/10.1103/PhysRevB.108.245301} {\bibfield  {journal} {\bibinfo  {journal}
  {Physical Review B}\ }\textbf {\bibinfo {volume} {108}},\ \bibinfo {pages} {245301} (\bibinfo {year} {2023})}\BibitemShut {NoStop}%
\bibitem [{\citenamefont {Abadillo-Uriel}\ \emph {et~al.}(2023)\citenamefont {Abadillo-Uriel}, \citenamefont {Rodr\'{\i}guez-Mena}, \citenamefont {Martinez},\ and\ \citenamefont {Niquet}}]{Abadillo2023}%
  \BibitemOpen
  \bibfield  {author} {\bibinfo {author} {\bibfnamefont {J.~C.}\ \bibnamefont {Abadillo-Uriel}}, \bibinfo {author} {\bibfnamefont {E.~A.}\ \bibnamefont {Rodr\'{\i}guez-Mena}}, \bibinfo {author} {\bibfnamefont {B.}~\bibnamefont {Martinez}},\ and\ \bibinfo {author} {\bibfnamefont {Y.-M.}\ \bibnamefont {Niquet}},\ }\bibfield  {title} {\bibinfo {title} {Hole-spin driving by strain-induced spin-orbit interactions},\ }\href {https://doi.org/10.1103/PhysRevLett.131.097002} {\bibfield  {journal} {\bibinfo  {journal} {Physical Review Letters}\ }\textbf {\bibinfo {volume} {131}},\ \bibinfo {pages} {097002} (\bibinfo {year} {2023})}\BibitemShut {NoStop}%
\bibitem [{\citenamefont {Rodr\'{\i}guez-Mena}\ \emph {et~al.}(2023)\citenamefont {Rodr\'{\i}guez-Mena}, \citenamefont {Abadillo-Uriel}, \citenamefont {Veste}, \citenamefont {Martinez}, \citenamefont {Li}, \citenamefont {Skl\'enard},\ and\ \citenamefont {Niquet}}]{Rodriguez2023}%
  \BibitemOpen
  \bibfield  {author} {\bibinfo {author} {\bibfnamefont {E.~A.}\ \bibnamefont {Rodr\'{\i}guez-Mena}}, \bibinfo {author} {\bibfnamefont {J.~C.}\ \bibnamefont {Abadillo-Uriel}}, \bibinfo {author} {\bibfnamefont {G.}~\bibnamefont {Veste}}, \bibinfo {author} {\bibfnamefont {B.}~\bibnamefont {Martinez}}, \bibinfo {author} {\bibfnamefont {J.}~\bibnamefont {Li}}, \bibinfo {author} {\bibfnamefont {B.}~\bibnamefont {Skl\'enard}},\ and\ \bibinfo {author} {\bibfnamefont {Y.-M.}\ \bibnamefont {Niquet}},\ }\bibfield  {title} {\bibinfo {title} {Linear-in-momentum spin orbit interactions in planar {Ge/GeSi} heterostructures and spin qubits},\ }\href {https://doi.org/10.1103/PhysRevB.108.205416} {\bibfield  {journal} {\bibinfo  {journal} {Physical Review B}\ }\textbf {\bibinfo {volume} {108}},\ \bibinfo {pages} {205416} (\bibinfo {year} {2023})}\BibitemShut {NoStop}%
\bibitem [{\citenamefont {Wang}\ \emph {et~al.}(2024{\natexlab{b}})\citenamefont {Wang}, \citenamefont {Ercan}, \citenamefont {Gyure}, \citenamefont {Scappucci}, \citenamefont {Veldhorst},\ and\ \citenamefont {Rimbach-Russ}}]{Wang24b}%
  \BibitemOpen
  \bibfield  {author} {\bibinfo {author} {\bibfnamefont {C.-A.}\ \bibnamefont {Wang}}, \bibinfo {author} {\bibfnamefont {H.~E.}\ \bibnamefont {Ercan}}, \bibinfo {author} {\bibfnamefont {M.~F.}\ \bibnamefont {Gyure}}, \bibinfo {author} {\bibfnamefont {G.}~\bibnamefont {Scappucci}}, \bibinfo {author} {\bibfnamefont {M.}~\bibnamefont {Veldhorst}},\ and\ \bibinfo {author} {\bibfnamefont {M.}~\bibnamefont {Rimbach-Russ}},\ }\bibfield  {title} {\bibinfo {title} {Modeling of planar germanium hole qubits in electric and magnetic fields},\ }\href {https://doi.org/10.1038/s41534-024-00897-8} {\bibfield  {journal} {\bibinfo  {journal} {npj Quantum Information}\ }\textbf {\bibinfo {volume} {10}},\ \bibinfo {pages} {102} (\bibinfo {year} {2024}{\natexlab{b}})}\BibitemShut {NoStop}%
\bibitem [{\citenamefont {Stano}\ and\ \citenamefont {Loss}(2025)}]{Stano25}%
  \BibitemOpen
  \bibfield  {author} {\bibinfo {author} {\bibfnamefont {P.}~\bibnamefont {Stano}}\ and\ \bibinfo {author} {\bibfnamefont {D.}~\bibnamefont {Loss}},\ }\bibfield  {title} {\bibinfo {title} {Quantification of the heavy-hole--light-hole mixing in two-dimensional hole gases},\ }\href {https://doi.org/10.1103/PhysRevB.111.115301} {\bibfield  {journal} {\bibinfo  {journal} {Physical Review B}\ }\textbf {\bibinfo {volume} {111}},\ \bibinfo {pages} {115301} (\bibinfo {year} {2025})}\BibitemShut {NoStop}%
\bibitem [{\citenamefont {Mauro}\ \emph {et~al.}(2025)\citenamefont {Mauro}, \citenamefont {Rodr\'{\i}guez-Mena}, \citenamefont {Martinez},\ and\ \citenamefont {Niquet}}]{Mauro25}%
  \BibitemOpen
  \bibfield  {author} {\bibinfo {author} {\bibfnamefont {L.}~\bibnamefont {Mauro}}, \bibinfo {author} {\bibfnamefont {E.~A.}\ \bibnamefont {Rodr\'{\i}guez-Mena}}, \bibinfo {author} {\bibfnamefont {B.}~\bibnamefont {Martinez}},\ and\ \bibinfo {author} {\bibfnamefont {Y.-M.}\ \bibnamefont {Niquet}},\ }\bibfield  {title} {\bibinfo {title} {Strain engineering in $\mathrm{Ge}$/$\mathrm{Ge}\text{\ensuremath{-}}\mathrm{Si}$ spin-qubit heterostructures},\ }\href {https://doi.org/10.1103/PhysRevApplied.23.024057} {\bibfield  {journal} {\bibinfo  {journal} {Physical Review Applied}\ }\textbf {\bibinfo {volume} {23}},\ \bibinfo {pages} {024057} (\bibinfo {year} {2025})}\BibitemShut {NoStop}%
\bibitem [{\citenamefont {Burkard}\ \emph {et~al.}(1999)\citenamefont {Burkard}, \citenamefont {Loss},\ and\ \citenamefont {DiVincenzo}}]{Burkard99}%
  \BibitemOpen
  \bibfield  {author} {\bibinfo {author} {\bibfnamefont {G.}~\bibnamefont {Burkard}}, \bibinfo {author} {\bibfnamefont {D.}~\bibnamefont {Loss}},\ and\ \bibinfo {author} {\bibfnamefont {D.~P.}\ \bibnamefont {DiVincenzo}},\ }\bibfield  {title} {\bibinfo {title} {Coupled quantum dots as quantum gates},\ }\href {https://doi.org/10.1103/PhysRevB.59.2070} {\bibfield  {journal} {\bibinfo  {journal} {Physical Review B}\ }\textbf {\bibinfo {volume} {59}},\ \bibinfo {pages} {2070} (\bibinfo {year} {1999})}\BibitemShut {NoStop}%
\bibitem [{\citenamefont {Petta}\ \emph {et~al.}(2005)\citenamefont {Petta}, \citenamefont {Johnson}, \citenamefont {Taylor}, \citenamefont {Laird}, \citenamefont {Yacoby}, \citenamefont {Lukin}, \citenamefont {Marcus}, \citenamefont {Hanson},\ and\ \citenamefont {Gossard}}]{Petta05}%
  \BibitemOpen
  \bibfield  {author} {\bibinfo {author} {\bibfnamefont {J.~R.}\ \bibnamefont {Petta}}, \bibinfo {author} {\bibfnamefont {A.~C.}\ \bibnamefont {Johnson}}, \bibinfo {author} {\bibfnamefont {J.~M.}\ \bibnamefont {Taylor}}, \bibinfo {author} {\bibfnamefont {E.~A.}\ \bibnamefont {Laird}}, \bibinfo {author} {\bibfnamefont {A.}~\bibnamefont {Yacoby}}, \bibinfo {author} {\bibfnamefont {M.~D.}\ \bibnamefont {Lukin}}, \bibinfo {author} {\bibfnamefont {C.~M.}\ \bibnamefont {Marcus}}, \bibinfo {author} {\bibfnamefont {M.~P.}\ \bibnamefont {Hanson}},\ and\ \bibinfo {author} {\bibfnamefont {A.~C.}\ \bibnamefont {Gossard}},\ }\bibfield  {title} {\bibinfo {title} {{Coherent Manipulation of Coupled Electron Spins in Semiconductor Quantum Dots}},\ }\href {https://doi.org/10.1126/science.1116955} {\bibfield  {journal} {\bibinfo  {journal} {Science}\ }\textbf {\bibinfo {volume} {309}},\ \bibinfo {pages} {2180} (\bibinfo {year} {2005})}\BibitemShut {NoStop}%
\bibitem [{\citenamefont {Jock}\ \emph {et~al.}(2018)\citenamefont {Jock}, \citenamefont {Jacobson}, \citenamefont {Harvey-Collard}, \citenamefont {Mounce}, \citenamefont {Srinivasa}, \citenamefont {Ward}, \citenamefont {Anderson}, \citenamefont {Manginell}, \citenamefont {Wendt}, \citenamefont {Rudolph}, \citenamefont {Pluym}, \citenamefont {Gamble}, \citenamefont {Baczewski}, \citenamefont {Witzel},\ and\ \citenamefont {Carroll}}]{Jock18}%
  \BibitemOpen
  \bibfield  {author} {\bibinfo {author} {\bibfnamefont {R.~M.}\ \bibnamefont {Jock}}, \bibinfo {author} {\bibfnamefont {N.~T.}\ \bibnamefont {Jacobson}}, \bibinfo {author} {\bibfnamefont {P.}~\bibnamefont {Harvey-Collard}}, \bibinfo {author} {\bibfnamefont {A.~M.}\ \bibnamefont {Mounce}}, \bibinfo {author} {\bibfnamefont {V.}~\bibnamefont {Srinivasa}}, \bibinfo {author} {\bibfnamefont {D.~R.}\ \bibnamefont {Ward}}, \bibinfo {author} {\bibfnamefont {J.}~\bibnamefont {Anderson}}, \bibinfo {author} {\bibfnamefont {R.}~\bibnamefont {Manginell}}, \bibinfo {author} {\bibfnamefont {J.~R.}\ \bibnamefont {Wendt}}, \bibinfo {author} {\bibfnamefont {M.}~\bibnamefont {Rudolph}}, \bibinfo {author} {\bibfnamefont {T.}~\bibnamefont {Pluym}}, \bibinfo {author} {\bibfnamefont {J.~K.}\ \bibnamefont {Gamble}}, \bibinfo {author} {\bibfnamefont {A.~D.}\ \bibnamefont {Baczewski}}, \bibinfo {author} {\bibfnamefont {W.~M.}\ \bibnamefont {Witzel}},\ and\ \bibinfo {author} {\bibfnamefont {M.~S.}\ \bibnamefont {Carroll}},\ }\bibfield
  {title} {\bibinfo {title} {A silicon metal-oxide-semiconductor electron spin-orbit qubit},\ }\href {https://doi.org/10.1038/s41467-018-04200-0} {\bibfield  {journal} {\bibinfo  {journal} {Nature Communications}\ }\textbf {\bibinfo {volume} {9}},\ \bibinfo {pages} {1768} (\bibinfo {year} {2018})}\BibitemShut {NoStop}%
\bibitem [{\citenamefont {Takeda}\ \emph {et~al.}(2020)\citenamefont {Takeda}, \citenamefont {Noiri}, \citenamefont {Yoneda}, \citenamefont {Nakajima},\ and\ \citenamefont {Tarucha}}]{Takeda20}%
  \BibitemOpen
  \bibfield  {author} {\bibinfo {author} {\bibfnamefont {K.}~\bibnamefont {Takeda}}, \bibinfo {author} {\bibfnamefont {A.}~\bibnamefont {Noiri}}, \bibinfo {author} {\bibfnamefont {J.}~\bibnamefont {Yoneda}}, \bibinfo {author} {\bibfnamefont {T.}~\bibnamefont {Nakajima}},\ and\ \bibinfo {author} {\bibfnamefont {S.}~\bibnamefont {Tarucha}},\ }\bibfield  {title} {\bibinfo {title} {Resonantly {Driven} {Singlet}-{Triplet} {Spin} {Qubit} in {Silicon}},\ }\href {https://doi.org/10.1103/PhysRevLett.124.117701} {\bibfield  {journal} {\bibinfo  {journal} {Physical Review Letters}\ }\textbf {\bibinfo {volume} {124}},\ \bibinfo {pages} {117701} (\bibinfo {year} {2020})}\BibitemShut {NoStop}%
\bibitem [{\citenamefont {Jirovec}\ \emph {et~al.}(2021)\citenamefont {Jirovec}, \citenamefont {Hofmann}, \citenamefont {Ballabio}, \citenamefont {Mutter}, \citenamefont {Tavani}, \citenamefont {Botifoll}, \citenamefont {Crippa}, \citenamefont {Kukucka}, \citenamefont {Sagi}, \citenamefont {Martins}, \citenamefont {Saez-Mollejo}, \citenamefont {Prieto}, \citenamefont {Borovkov}, \citenamefont {Arbiol}, \citenamefont {Chrastina}, \citenamefont {Isella},\ and\ \citenamefont {Katsaros}}]{jirovec_Freqs_ST-Ge}%
  \BibitemOpen
  \bibfield  {author} {\bibinfo {author} {\bibfnamefont {D.}~\bibnamefont {Jirovec}}, \bibinfo {author} {\bibfnamefont {A.}~\bibnamefont {Hofmann}}, \bibinfo {author} {\bibfnamefont {A.}~\bibnamefont {Ballabio}}, \bibinfo {author} {\bibfnamefont {P.~M.}\ \bibnamefont {Mutter}}, \bibinfo {author} {\bibfnamefont {G.}~\bibnamefont {Tavani}}, \bibinfo {author} {\bibfnamefont {M.}~\bibnamefont {Botifoll}}, \bibinfo {author} {\bibfnamefont {A.}~\bibnamefont {Crippa}}, \bibinfo {author} {\bibfnamefont {J.}~\bibnamefont {Kukucka}}, \bibinfo {author} {\bibfnamefont {O.}~\bibnamefont {Sagi}}, \bibinfo {author} {\bibfnamefont {F.}~\bibnamefont {Martins}}, \bibinfo {author} {\bibfnamefont {J.}~\bibnamefont {Saez-Mollejo}}, \bibinfo {author} {\bibfnamefont {I.}~\bibnamefont {Prieto}}, \bibinfo {author} {\bibfnamefont {M.}~\bibnamefont {Borovkov}}, \bibinfo {author} {\bibfnamefont {J.}~\bibnamefont {Arbiol}}, \bibinfo {author} {\bibfnamefont {D.}~\bibnamefont {Chrastina}}, \bibinfo {author} {\bibfnamefont {G.}~\bibnamefont
  {Isella}},\ and\ \bibinfo {author} {\bibfnamefont {G.}~\bibnamefont {Katsaros}},\ }\bibfield  {title} {\bibinfo {title} {A singlet-triplet hole spin qubit in planar {Ge}},\ }\href {https://doi.org/10.1038/s41563-021-01022-2} {\bibfield  {journal} {\bibinfo  {journal} {Nature Materials}\ }\textbf {\bibinfo {volume} {20}},\ \bibinfo {pages} {1106} (\bibinfo {year} {2021})}\BibitemShut {NoStop}%
\bibitem [{\citenamefont {Jirovec}\ \emph {et~al.}(2022)\citenamefont {Jirovec}, \citenamefont {Mutter}, \citenamefont {Hofmann}, \citenamefont {Crippa}, \citenamefont {Rychetsky}, \citenamefont {Craig}, \citenamefont {Kukucka}, \citenamefont {Martins}, \citenamefont {Ballabio}, \citenamefont {Ares}, \citenamefont {Chrastina}, \citenamefont {Isella}, \citenamefont {Burkard},\ and\ \citenamefont {Katsaros}}]{Jirovec23}%
  \BibitemOpen
  \bibfield  {author} {\bibinfo {author} {\bibfnamefont {D.}~\bibnamefont {Jirovec}}, \bibinfo {author} {\bibfnamefont {P.~M.}\ \bibnamefont {Mutter}}, \bibinfo {author} {\bibfnamefont {A.}~\bibnamefont {Hofmann}}, \bibinfo {author} {\bibfnamefont {A.}~\bibnamefont {Crippa}}, \bibinfo {author} {\bibfnamefont {M.}~\bibnamefont {Rychetsky}}, \bibinfo {author} {\bibfnamefont {D.~L.}\ \bibnamefont {Craig}}, \bibinfo {author} {\bibfnamefont {J.}~\bibnamefont {Kukucka}}, \bibinfo {author} {\bibfnamefont {F.}~\bibnamefont {Martins}}, \bibinfo {author} {\bibfnamefont {A.}~\bibnamefont {Ballabio}}, \bibinfo {author} {\bibfnamefont {N.}~\bibnamefont {Ares}}, \bibinfo {author} {\bibfnamefont {D.}~\bibnamefont {Chrastina}}, \bibinfo {author} {\bibfnamefont {G.}~\bibnamefont {Isella}}, \bibinfo {author} {\bibfnamefont {G.}~\bibnamefont {Burkard}},\ and\ \bibinfo {author} {\bibfnamefont {G.}~\bibnamefont {Katsaros}},\ }\bibfield  {title} {\bibinfo {title} {Dynamics of hole singlet-triplet qubits with large $g$-factor
  differences},\ }\href {https://doi.org/10.1103/PhysRevLett.128.126803} {\bibfield  {journal} {\bibinfo  {journal} {Physical Review Letters}\ }\textbf {\bibinfo {volume} {128}},\ \bibinfo {pages} {126803} (\bibinfo {year} {2022})}\BibitemShut {NoStop}%
\bibitem [{\citenamefont {Saez-Mollejo}\ \emph {et~al.}(2024)\citenamefont {Saez-Mollejo}, \citenamefont {Jirovec}, \citenamefont {Schell}, \citenamefont {Kukucka}, \citenamefont {Calcaterra}, \citenamefont {Chrastina}, \citenamefont {Isella}, \citenamefont {Rimbach-Russ}, \citenamefont {Bosco},\ and\ \citenamefont {Katsaros}}]{Saezmollejo24}%
  \BibitemOpen
  \bibfield  {author} {\bibinfo {author} {\bibfnamefont {J.}~\bibnamefont {Saez-Mollejo}}, \bibinfo {author} {\bibfnamefont {D.}~\bibnamefont {Jirovec}}, \bibinfo {author} {\bibfnamefont {Y.}~\bibnamefont {Schell}}, \bibinfo {author} {\bibfnamefont {J.}~\bibnamefont {Kukucka}}, \bibinfo {author} {\bibfnamefont {S.}~\bibnamefont {Calcaterra}}, \bibinfo {author} {\bibfnamefont {D.}~\bibnamefont {Chrastina}}, \bibinfo {author} {\bibfnamefont {G.}~\bibnamefont {Isella}}, \bibinfo {author} {\bibfnamefont {M.}~\bibnamefont {Rimbach-Russ}}, \bibinfo {author} {\bibfnamefont {S.}~\bibnamefont {Bosco}},\ and\ \bibinfo {author} {\bibfnamefont {G.}~\bibnamefont {Katsaros}},\ }\bibfield  {title} {\bibinfo {title} {Exchange anisotropies in microwave-driven singlet-triplet qubits},\ }\href {https://arxiv.org/abs/2408.03224} {\bibfield  {journal} {\bibinfo  {journal} {arXiv:2408.03224}\ } (\bibinfo {year} {2024})}\BibitemShut {NoStop}%
\bibitem [{\citenamefont {Rooney}\ \emph {et~al.}(2025)\citenamefont {Rooney}, \citenamefont {Luo}, \citenamefont {Stehouwer}, \citenamefont {Scappucci}, \citenamefont {Veldhorst},\ and\ \citenamefont {Jiang}}]{rooney_strains_ST-Ge_ST_-}%
  \BibitemOpen
  \bibfield  {author} {\bibinfo {author} {\bibfnamefont {J.}~\bibnamefont {Rooney}}, \bibinfo {author} {\bibfnamefont {Z.}~\bibnamefont {Luo}}, \bibinfo {author} {\bibfnamefont {L.~E.~A.}\ \bibnamefont {Stehouwer}}, \bibinfo {author} {\bibfnamefont {G.}~\bibnamefont {Scappucci}}, \bibinfo {author} {\bibfnamefont {M.}~\bibnamefont {Veldhorst}},\ and\ \bibinfo {author} {\bibfnamefont {H.-W.}\ \bibnamefont {Jiang}},\ }\bibfield  {title} {\bibinfo {title} {Gate modulation of the hole singlet-triplet qubit frequency in germanium},\ }\href {https://doi.org/10.1038/s41534-024-00953-3} {\bibfield  {journal} {\bibinfo  {journal} {npj Quantum Information}\ }\textbf {\bibinfo {volume} {11}},\ \bibinfo {pages} {15} (\bibinfo {year} {2025})}\BibitemShut {NoStop}%
\bibitem [{\citenamefont {Liles}\ \emph {et~al.}(2024)\citenamefont {Liles}, \citenamefont {Halverson}, \citenamefont {Wang}, \citenamefont {Shamim}, \citenamefont {Eggli}, \citenamefont {Jin}, \citenamefont {Hillier}, \citenamefont {Kumar}, \citenamefont {Vorreiter}, \citenamefont {Rendell}, \citenamefont {Huang}, \citenamefont {Escott}, \citenamefont {Hudson}, \citenamefont {Lim}, \citenamefont {Culcer}, \citenamefont {Dzurak},\ and\ \citenamefont {Hamilton}}]{Liles24}%
  \BibitemOpen
  \bibfield  {author} {\bibinfo {author} {\bibfnamefont {S.~D.}\ \bibnamefont {Liles}}, \bibinfo {author} {\bibfnamefont {D.~J.}\ \bibnamefont {Halverson}}, \bibinfo {author} {\bibfnamefont {Z.}~\bibnamefont {Wang}}, \bibinfo {author} {\bibfnamefont {A.}~\bibnamefont {Shamim}}, \bibinfo {author} {\bibfnamefont {R.~S.}\ \bibnamefont {Eggli}}, \bibinfo {author} {\bibfnamefont {I.~K.}\ \bibnamefont {Jin}}, \bibinfo {author} {\bibfnamefont {J.}~\bibnamefont {Hillier}}, \bibinfo {author} {\bibfnamefont {K.}~\bibnamefont {Kumar}}, \bibinfo {author} {\bibfnamefont {I.}~\bibnamefont {Vorreiter}}, \bibinfo {author} {\bibfnamefont {M.~J.}\ \bibnamefont {Rendell}}, \bibinfo {author} {\bibfnamefont {J.~Y.}\ \bibnamefont {Huang}}, \bibinfo {author} {\bibfnamefont {C.~C.}\ \bibnamefont {Escott}}, \bibinfo {author} {\bibfnamefont {F.~E.}\ \bibnamefont {Hudson}}, \bibinfo {author} {\bibfnamefont {W.~H.}\ \bibnamefont {Lim}}, \bibinfo {author} {\bibfnamefont {D.}~\bibnamefont {Culcer}}, \bibinfo {author} {\bibfnamefont
  {A.~S.}\ \bibnamefont {Dzurak}},\ and\ \bibinfo {author} {\bibfnamefont {A.~R.}\ \bibnamefont {Hamilton}},\ }\bibfield  {title} {\bibinfo {title} {A singlet-triplet hole-spin qubit in mos silicon},\ }\href {https://doi.org/10.1038/s41467-024-51902-9} {\bibfield  {journal} {\bibinfo  {journal} {Nature Communications}\ }\textbf {\bibinfo {volume} {15}},\ \bibinfo {pages} {7690} (\bibinfo {year} {2024})}\BibitemShut {NoStop}%
\bibitem [{\citenamefont {Zhang}\ \emph {et~al.}(2025)\citenamefont {Zhang}, \citenamefont {Morozova}, \citenamefont {Rimbach-Russ}, \citenamefont {Jirovec}, \citenamefont {Hsiao}, \citenamefont {Fariña}, \citenamefont {Wang}, \citenamefont {Oosterhout}, \citenamefont {Sammak}, \citenamefont {Scappucci}, \citenamefont {Veldhorst},\ and\ \citenamefont {Vandersypen}}]{Zhang25}%
  \BibitemOpen
  \bibfield  {author} {\bibinfo {author} {\bibfnamefont {X.}~\bibnamefont {Zhang}}, \bibinfo {author} {\bibfnamefont {E.}~\bibnamefont {Morozova}}, \bibinfo {author} {\bibfnamefont {M.}~\bibnamefont {Rimbach-Russ}}, \bibinfo {author} {\bibfnamefont {D.}~\bibnamefont {Jirovec}}, \bibinfo {author} {\bibfnamefont {T.-K.}\ \bibnamefont {Hsiao}}, \bibinfo {author} {\bibfnamefont {P.~C.}\ \bibnamefont {Fariña}}, \bibinfo {author} {\bibfnamefont {C.-A.}\ \bibnamefont {Wang}}, \bibinfo {author} {\bibfnamefont {S.~D.}\ \bibnamefont {Oosterhout}}, \bibinfo {author} {\bibfnamefont {A.}~\bibnamefont {Sammak}}, \bibinfo {author} {\bibfnamefont {G.}~\bibnamefont {Scappucci}}, \bibinfo {author} {\bibfnamefont {M.}~\bibnamefont {Veldhorst}},\ and\ \bibinfo {author} {\bibfnamefont {L.~M.~K.}\ \bibnamefont {Vandersypen}},\ }\bibfield  {title} {\bibinfo {title} {Universal control of four singlet-triplet qubits},\ }\href {https://doi.org/10.1038/s41565-024-01817-9} {\bibfield  {journal} {\bibinfo  {journal} {Nature
  Nanotechnology}\ }\textbf {\bibinfo {volume} {20}},\ \bibinfo {pages} {209} (\bibinfo {year} {2025})}\BibitemShut {NoStop}%
\bibitem [{\citenamefont {Tsoukalas}\ \emph {et~al.}(2025)\citenamefont {Tsoukalas}, \citenamefont {von Lüpke}, \citenamefont {Orekhov}, \citenamefont {Hetényi}, \citenamefont {Seidler}, \citenamefont {Sommer}, \citenamefont {Kelly}, \citenamefont {Massai}, \citenamefont {Aldeghi}, \citenamefont {Pita-Vidal}, \citenamefont {Hendrickx}, \citenamefont {Bedell}, \citenamefont {Paredes}, \citenamefont {Schupp}, \citenamefont {Mergenthaler}, \citenamefont {Salis}, \citenamefont {Fuhrer},\ and\ \citenamefont {Harvey-Collard}}]{Tsoukalas2025}%
  \BibitemOpen
  \bibfield  {author} {\bibinfo {author} {\bibfnamefont {K.}~\bibnamefont {Tsoukalas}}, \bibinfo {author} {\bibfnamefont {U.}~\bibnamefont {von Lüpke}}, \bibinfo {author} {\bibfnamefont {A.}~\bibnamefont {Orekhov}}, \bibinfo {author} {\bibfnamefont {B.}~\bibnamefont {Hetényi}}, \bibinfo {author} {\bibfnamefont {I.}~\bibnamefont {Seidler}}, \bibinfo {author} {\bibfnamefont {L.}~\bibnamefont {Sommer}}, \bibinfo {author} {\bibfnamefont {E.~G.}\ \bibnamefont {Kelly}}, \bibinfo {author} {\bibfnamefont {L.}~\bibnamefont {Massai}}, \bibinfo {author} {\bibfnamefont {M.}~\bibnamefont {Aldeghi}}, \bibinfo {author} {\bibfnamefont {M.}~\bibnamefont {Pita-Vidal}}, \bibinfo {author} {\bibfnamefont {N.~W.}\ \bibnamefont {Hendrickx}}, \bibinfo {author} {\bibfnamefont {S.~W.}\ \bibnamefont {Bedell}}, \bibinfo {author} {\bibfnamefont {S.}~\bibnamefont {Paredes}}, \bibinfo {author} {\bibfnamefont {F.~J.}\ \bibnamefont {Schupp}}, \bibinfo {author} {\bibfnamefont {M.}~\bibnamefont {Mergenthaler}}, \bibinfo {author}
  {\bibfnamefont {G.}~\bibnamefont {Salis}}, \bibinfo {author} {\bibfnamefont {A.}~\bibnamefont {Fuhrer}},\ and\ \bibinfo {author} {\bibfnamefont {P.}~\bibnamefont {Harvey-Collard}},\ }\bibfield  {title} {\bibinfo {title} {A dressed singlet-triplet qubit in germanium},\ }\href {https://arxiv.org/abs/2501.14627} {\bibfield  {journal} {\bibinfo  {journal} {arXiv:2501.14627}\ } (\bibinfo {year} {2025})}\BibitemShut {NoStop}%
\bibitem [{\citenamefont {Bacon}\ \emph {et~al.}(2000)\citenamefont {Bacon}, \citenamefont {Kempe}, \citenamefont {Burkard},\ and\ \citenamefont {Whaley}}]{DiVincenzo00}%
  \BibitemOpen
  \bibfield  {author} {\bibinfo {author} {\bibfnamefont {D.}~\bibnamefont {Bacon}}, \bibinfo {author} {\bibfnamefont {J.}~\bibnamefont {Kempe}}, \bibinfo {author} {\bibfnamefont {G.}~\bibnamefont {Burkard}},\ and\ \bibinfo {author} {\bibfnamefont {K.~B.}\ \bibnamefont {Whaley}},\ }\bibfield  {title} {\bibinfo {title} {Universal quantum computation with the exchange interaction},\ }\href {https://doi.org/0.1038/35042541} {\bibfield  {journal} {\bibinfo  {journal} {Nature}\ }\textbf {\bibinfo {volume} {408}},\ \bibinfo {pages} {339} (\bibinfo {year} {2000})}\BibitemShut {NoStop}%
\bibitem [{\citenamefont {Laird}\ \emph {et~al.}(2010)\citenamefont {Laird}, \citenamefont {Taylor}, \citenamefont {DiVincenzo}, \citenamefont {Marcus}, \citenamefont {Hanson},\ and\ \citenamefont {Gossard}}]{Laird10}%
  \BibitemOpen
  \bibfield  {author} {\bibinfo {author} {\bibfnamefont {E.~A.}\ \bibnamefont {Laird}}, \bibinfo {author} {\bibfnamefont {J.~M.}\ \bibnamefont {Taylor}}, \bibinfo {author} {\bibfnamefont {D.~P.}\ \bibnamefont {DiVincenzo}}, \bibinfo {author} {\bibfnamefont {C.~M.}\ \bibnamefont {Marcus}}, \bibinfo {author} {\bibfnamefont {M.~P.}\ \bibnamefont {Hanson}},\ and\ \bibinfo {author} {\bibfnamefont {A.~C.}\ \bibnamefont {Gossard}},\ }\bibfield  {title} {\bibinfo {title} {Coherent spin manipulation in an exchange-only qubit},\ }\href {https://doi.org/10.1103/PhysRevB.82.075403} {\bibfield  {journal} {\bibinfo  {journal} {Physical Review B}\ }\textbf {\bibinfo {volume} {82}},\ \bibinfo {pages} {075403} (\bibinfo {year} {2010})}\BibitemShut {NoStop}%
\bibitem [{\citenamefont {Medford}\ \emph {et~al.}(2013)\citenamefont {Medford}, \citenamefont {Beil}, \citenamefont {Taylor}, \citenamefont {Rashba}, \citenamefont {Lu}, \citenamefont {Gossard},\ and\ \citenamefont {Marcus}}]{Medford13}%
  \BibitemOpen
  \bibfield  {author} {\bibinfo {author} {\bibfnamefont {J.}~\bibnamefont {Medford}}, \bibinfo {author} {\bibfnamefont {J.}~\bibnamefont {Beil}}, \bibinfo {author} {\bibfnamefont {J.~M.}\ \bibnamefont {Taylor}}, \bibinfo {author} {\bibfnamefont {E.~I.}\ \bibnamefont {Rashba}}, \bibinfo {author} {\bibfnamefont {H.}~\bibnamefont {Lu}}, \bibinfo {author} {\bibfnamefont {A.~C.}\ \bibnamefont {Gossard}},\ and\ \bibinfo {author} {\bibfnamefont {C.~M.}\ \bibnamefont {Marcus}},\ }\bibfield  {title} {\bibinfo {title} {Quantum-dot-based resonant exchange qubit},\ }\href {https://doi.org/10.1103/PhysRevLett.111.050501} {\bibfield  {journal} {\bibinfo  {journal} {Physical Review Letters}\ }\textbf {\bibinfo {volume} {111}},\ \bibinfo {pages} {050501} (\bibinfo {year} {2013})}\BibitemShut {NoStop}%
\bibitem [{\citenamefont {Eng}\ \emph {et~al.}(2015)\citenamefont {Eng}, \citenamefont {Ladd}, \citenamefont {Smith}, \citenamefont {Borselli}, \citenamefont {Kiselev}, \citenamefont {Fong}, \citenamefont {Holabird}, \citenamefont {Hazard}, \citenamefont {Huang}, \citenamefont {Deelman}, \citenamefont {Milosavljevic}, \citenamefont {Schmitz}, \citenamefont {Ross}, \citenamefont {Gyure},\ and\ \citenamefont {Hunter}}]{Eng15}%
  \BibitemOpen
  \bibfield  {author} {\bibinfo {author} {\bibfnamefont {K.}~\bibnamefont {Eng}}, \bibinfo {author} {\bibfnamefont {T.~D.}\ \bibnamefont {Ladd}}, \bibinfo {author} {\bibfnamefont {A.}~\bibnamefont {Smith}}, \bibinfo {author} {\bibfnamefont {M.~G.}\ \bibnamefont {Borselli}}, \bibinfo {author} {\bibfnamefont {A.~A.}\ \bibnamefont {Kiselev}}, \bibinfo {author} {\bibfnamefont {B.~H.}\ \bibnamefont {Fong}}, \bibinfo {author} {\bibfnamefont {K.~S.}\ \bibnamefont {Holabird}}, \bibinfo {author} {\bibfnamefont {T.~M.}\ \bibnamefont {Hazard}}, \bibinfo {author} {\bibfnamefont {B.}~\bibnamefont {Huang}}, \bibinfo {author} {\bibfnamefont {P.~W.}\ \bibnamefont {Deelman}}, \bibinfo {author} {\bibfnamefont {I.}~\bibnamefont {Milosavljevic}}, \bibinfo {author} {\bibfnamefont {A.~E.}\ \bibnamefont {Schmitz}}, \bibinfo {author} {\bibfnamefont {R.~S.}\ \bibnamefont {Ross}}, \bibinfo {author} {\bibfnamefont {M.~F.}\ \bibnamefont {Gyure}},\ and\ \bibinfo {author} {\bibfnamefont {A.~T.}\ \bibnamefont {Hunter}},\ }\bibfield
  {title} {\bibinfo {title} {Isotopically enhanced triple-quantum-dot qubit},\ }\href {https://doi.org/10.1126/sciadv.1500214} {\bibfield  {journal} {\bibinfo  {journal} {Science Advances}\ }\textbf {\bibinfo {volume} {1}},\ \bibinfo {pages} {e1500214} (\bibinfo {year} {2015})}\BibitemShut {NoStop}%
\bibitem [{\citenamefont {Weinstein}\ \emph {et~al.}(2023)\citenamefont {Weinstein}, \citenamefont {Reed}, \citenamefont {Jones}, \citenamefont {Andrews}, \citenamefont {Barnes}, \citenamefont {Blumoff}, \citenamefont {Euliss}, \citenamefont {Eng}, \citenamefont {Fong}, \citenamefont {Ha}, \citenamefont {Hulbert}, \citenamefont {Jackson}, \citenamefont {Jura}, \citenamefont {Keating}, \citenamefont {Kerckhoff}, \citenamefont {Kiselev}, \citenamefont {Matten}, \citenamefont {Sabbir}, \citenamefont {Smith}, \citenamefont {Wright}, \citenamefont {Rakher}, \citenamefont {Ladd},\ and\ \citenamefont {Borselli}}]{Weinstein23}%
  \BibitemOpen
  \bibfield  {author} {\bibinfo {author} {\bibfnamefont {A.~J.}\ \bibnamefont {Weinstein}}, \bibinfo {author} {\bibfnamefont {M.~D.}\ \bibnamefont {Reed}}, \bibinfo {author} {\bibfnamefont {A.~M.}\ \bibnamefont {Jones}}, \bibinfo {author} {\bibfnamefont {R.~W.}\ \bibnamefont {Andrews}}, \bibinfo {author} {\bibfnamefont {D.}~\bibnamefont {Barnes}}, \bibinfo {author} {\bibfnamefont {J.~Z.}\ \bibnamefont {Blumoff}}, \bibinfo {author} {\bibfnamefont {L.~E.}\ \bibnamefont {Euliss}}, \bibinfo {author} {\bibfnamefont {K.}~\bibnamefont {Eng}}, \bibinfo {author} {\bibfnamefont {B.~H.}\ \bibnamefont {Fong}}, \bibinfo {author} {\bibfnamefont {S.~D.}\ \bibnamefont {Ha}}, \bibinfo {author} {\bibfnamefont {D.~R.}\ \bibnamefont {Hulbert}}, \bibinfo {author} {\bibfnamefont {C.~A.~C.}\ \bibnamefont {Jackson}}, \bibinfo {author} {\bibfnamefont {M.}~\bibnamefont {Jura}}, \bibinfo {author} {\bibfnamefont {T.~E.}\ \bibnamefont {Keating}}, \bibinfo {author} {\bibfnamefont {J.}~\bibnamefont {Kerckhoff}}, \bibinfo {author}
  {\bibfnamefont {A.~A.}\ \bibnamefont {Kiselev}}, \bibinfo {author} {\bibfnamefont {J.}~\bibnamefont {Matten}}, \bibinfo {author} {\bibfnamefont {G.}~\bibnamefont {Sabbir}}, \bibinfo {author} {\bibfnamefont {A.}~\bibnamefont {Smith}}, \bibinfo {author} {\bibfnamefont {J.}~\bibnamefont {Wright}}, \bibinfo {author} {\bibfnamefont {M.~T.}\ \bibnamefont {Rakher}}, \bibinfo {author} {\bibfnamefont {T.~D.}\ \bibnamefont {Ladd}},\ and\ \bibinfo {author} {\bibfnamefont {M.~G.}\ \bibnamefont {Borselli}},\ }\bibfield  {title} {\bibinfo {title} {Universal logic with encoded spin qubits in silicon},\ }\href {https://doi.org/10.1038/s41586-023-05777-3} {\bibfield  {journal} {\bibinfo  {journal} {Nature}\ }\textbf {\bibinfo {volume} {615}},\ \bibinfo {pages} {817} (\bibinfo {year} {2023})}\BibitemShut {NoStop}%
\bibitem [{\citenamefont {Acuna}\ \emph {et~al.}(2024)\citenamefont {Acuna}, \citenamefont {Broz}, \citenamefont {Shyamsundar}, \citenamefont {Mei}, \citenamefont {Feeney}, \citenamefont {Smetanka}, \citenamefont {Davis}, \citenamefont {Lee}, \citenamefont {Choi}, \citenamefont {Boyd}, \citenamefont {Suh}, \citenamefont {Ha}, \citenamefont {Jennings}, \citenamefont {Pan}, \citenamefont {Sanchez}, \citenamefont {Reed},\ and\ \citenamefont {Petta}}]{Acuna24}%
  \BibitemOpen
  \bibfield  {author} {\bibinfo {author} {\bibfnamefont {E.}~\bibnamefont {Acuna}}, \bibinfo {author} {\bibfnamefont {J.~D.}\ \bibnamefont {Broz}}, \bibinfo {author} {\bibfnamefont {K.}~\bibnamefont {Shyamsundar}}, \bibinfo {author} {\bibfnamefont {A.~B.}\ \bibnamefont {Mei}}, \bibinfo {author} {\bibfnamefont {C.~P.}\ \bibnamefont {Feeney}}, \bibinfo {author} {\bibfnamefont {V.}~\bibnamefont {Smetanka}}, \bibinfo {author} {\bibfnamefont {T.}~\bibnamefont {Davis}}, \bibinfo {author} {\bibfnamefont {K.}~\bibnamefont {Lee}}, \bibinfo {author} {\bibfnamefont {M.~D.}\ \bibnamefont {Choi}}, \bibinfo {author} {\bibfnamefont {B.}~\bibnamefont {Boyd}}, \bibinfo {author} {\bibfnamefont {J.}~\bibnamefont {Suh}}, \bibinfo {author} {\bibfnamefont {W.}~\bibnamefont {Ha}}, \bibinfo {author} {\bibfnamefont {C.}~\bibnamefont {Jennings}}, \bibinfo {author} {\bibfnamefont {A.~S.}\ \bibnamefont {Pan}}, \bibinfo {author} {\bibfnamefont {D.~S.}\ \bibnamefont {Sanchez}}, \bibinfo {author} {\bibfnamefont {M.~D.}\ \bibnamefont
  {Reed}},\ and\ \bibinfo {author} {\bibfnamefont {J.~R.}\ \bibnamefont {Petta}},\ }\bibfield  {title} {\bibinfo {title} {Coherent control of a triangular exchange-only spin qubit},\ }\href {https://doi.org/10.1103/PhysRevApplied.22.044057} {\bibfield  {journal} {\bibinfo  {journal} {Physical Review Applied}\ }\textbf {\bibinfo {volume} {22}},\ \bibinfo {pages} {044057} (\bibinfo {year} {2024})}\BibitemShut {NoStop}%
\bibitem [{\citenamefont {Harvey-Collard}\ \emph {et~al.}(2019)\citenamefont {Harvey-Collard}, \citenamefont {Jacobson}, \citenamefont {Bureau-Oxton}, \citenamefont {Jock}, \citenamefont {Srinivasa}, \citenamefont {Mounce}, \citenamefont {Ward}, \citenamefont {Anderson}, \citenamefont {Manginell}, \citenamefont {Wendt}, \citenamefont {Pluym}, \citenamefont {Lilly}, \citenamefont {Luhman}, \citenamefont {Pioro-Ladrière},\ and\ \citenamefont {Carroll}}]{harvey-collard_spin-orbit_ST-Si}%
  \BibitemOpen
  \bibfield  {author} {\bibinfo {author} {\bibfnamefont {P.}~\bibnamefont {Harvey-Collard}}, \bibinfo {author} {\bibfnamefont {N.~T.}\ \bibnamefont {Jacobson}}, \bibinfo {author} {\bibfnamefont {C.}~\bibnamefont {Bureau-Oxton}}, \bibinfo {author} {\bibfnamefont {R.~M.}\ \bibnamefont {Jock}}, \bibinfo {author} {\bibfnamefont {V.}~\bibnamefont {Srinivasa}}, \bibinfo {author} {\bibfnamefont {A.~M.}\ \bibnamefont {Mounce}}, \bibinfo {author} {\bibfnamefont {D.~R.}\ \bibnamefont {Ward}}, \bibinfo {author} {\bibfnamefont {J.~M.}\ \bibnamefont {Anderson}}, \bibinfo {author} {\bibfnamefont {R.~P.}\ \bibnamefont {Manginell}}, \bibinfo {author} {\bibfnamefont {J.~R.}\ \bibnamefont {Wendt}}, \bibinfo {author} {\bibfnamefont {T.}~\bibnamefont {Pluym}}, \bibinfo {author} {\bibfnamefont {M.~P.}\ \bibnamefont {Lilly}}, \bibinfo {author} {\bibfnamefont {D.~R.}\ \bibnamefont {Luhman}}, \bibinfo {author} {\bibfnamefont {M.}~\bibnamefont {Pioro-Ladrière}},\ and\ \bibinfo {author} {\bibfnamefont {M.~S.}\ \bibnamefont
  {Carroll}},\ }\bibfield  {title} {\bibinfo {title} {Spin-orbit interactions for singlet-triplet qubits in silicon},\ }\href {https://doi.org/10.1103/PhysRevLett.122.217702} {\bibfield  {journal} {\bibinfo  {journal} {Physical Review Letters}\ }\textbf {\bibinfo {volume} {122}},\ \bibinfo {pages} {217702} (\bibinfo {year} {2019})}\BibitemShut {NoStop}%
\bibitem [{\citenamefont {Froning}\ \emph {et~al.}(2021{\natexlab{b}})\citenamefont {Froning}, \citenamefont {Rančić}, \citenamefont {Hetényi}, \citenamefont {Bosco}, \citenamefont {Rehmann}, \citenamefont {Li}, \citenamefont {Bakkers}, \citenamefont {Zwanenburg}, \citenamefont {Loss}, \citenamefont {Zumbühl},\ and\ \citenamefont {Braakman}}]{froning_strong_SO_2021}%
  \BibitemOpen
  \bibfield  {author} {\bibinfo {author} {\bibfnamefont {F.~N.~M.}\ \bibnamefont {Froning}}, \bibinfo {author} {\bibfnamefont {M.~J.}\ \bibnamefont {Rančić}}, \bibinfo {author} {\bibfnamefont {B.}~\bibnamefont {Hetényi}}, \bibinfo {author} {\bibfnamefont {S.}~\bibnamefont {Bosco}}, \bibinfo {author} {\bibfnamefont {M.~K.}\ \bibnamefont {Rehmann}}, \bibinfo {author} {\bibfnamefont {A.}~\bibnamefont {Li}}, \bibinfo {author} {\bibfnamefont {E.~P. A.~M.}\ \bibnamefont {Bakkers}}, \bibinfo {author} {\bibfnamefont {F.~A.}\ \bibnamefont {Zwanenburg}}, \bibinfo {author} {\bibfnamefont {D.}~\bibnamefont {Loss}}, \bibinfo {author} {\bibfnamefont {D.~M.}\ \bibnamefont {Zumbühl}},\ and\ \bibinfo {author} {\bibfnamefont {F.~R.}\ \bibnamefont {Braakman}},\ }\bibfield  {title} {\bibinfo {title} {Strong spin-orbit interaction and $g$-factor renormalization of hole spins in {Ge}/{Si} nanowire quantum dots},\ }\href {https://doi.org/10.1103/PhysRevResearch.3.013081} {\bibfield  {journal} {\bibinfo  {journal} {Physical
  Review Research}\ }\textbf {\bibinfo {volume} {3}},\ \bibinfo {pages} {013081} (\bibinfo {year} {2021}{\natexlab{b}})}\BibitemShut {NoStop}%
\bibitem [{\citenamefont {Geyer}\ \emph {et~al.}(2024)\citenamefont {Geyer}, \citenamefont {Hetényi}, \citenamefont {Bosco}, \citenamefont {Camenzind}, \citenamefont {Eggli}, \citenamefont {Fuhrer}, \citenamefont {Loss}, \citenamefont {Warburton}, \citenamefont {Zumbühl},\ and\ \citenamefont {Kuhlmann}}]{Geyer24}%
  \BibitemOpen
  \bibfield  {author} {\bibinfo {author} {\bibfnamefont {S.}~\bibnamefont {Geyer}}, \bibinfo {author} {\bibfnamefont {B.}~\bibnamefont {Hetényi}}, \bibinfo {author} {\bibfnamefont {S.}~\bibnamefont {Bosco}}, \bibinfo {author} {\bibfnamefont {L.~C.}\ \bibnamefont {Camenzind}}, \bibinfo {author} {\bibfnamefont {R.~S.}\ \bibnamefont {Eggli}}, \bibinfo {author} {\bibfnamefont {A.}~\bibnamefont {Fuhrer}}, \bibinfo {author} {\bibfnamefont {D.}~\bibnamefont {Loss}}, \bibinfo {author} {\bibfnamefont {R.~J.}\ \bibnamefont {Warburton}}, \bibinfo {author} {\bibfnamefont {D.~M.}\ \bibnamefont {Zumbühl}},\ and\ \bibinfo {author} {\bibfnamefont {A.~V.}\ \bibnamefont {Kuhlmann}},\ }\bibfield  {title} {\bibinfo {title} {Anisotropic exchange interaction of two hole-spin qubits},\ }\href {https://doi.org/10.1038/s41567-024-02481-5} {\bibfield  {journal} {\bibinfo  {journal} {Nature Physics}\ }\textbf {\bibinfo {volume} {20}},\ \bibinfo {pages} {1152} (\bibinfo {year} {2024})}\BibitemShut {NoStop}%
\bibitem [{\citenamefont {Bosco}\ and\ \citenamefont {Rimbach-Russ}(2024)}]{Bosco24}%
  \BibitemOpen
  \bibfield  {author} {\bibinfo {author} {\bibfnamefont {S.}~\bibnamefont {Bosco}}\ and\ \bibinfo {author} {\bibfnamefont {M.}~\bibnamefont {Rimbach-Russ}},\ }\bibfield  {title} {\bibinfo {title} {Exchange-only spin-orbit qubits in silicon and germanium},\ }\href {https://arxiv.org/abs/2410.05461} {\bibfield  {journal} {\bibinfo  {journal} {arXiv:2410.05461}\ } (\bibinfo {year} {2024})}\BibitemShut {NoStop}%
\bibitem [{\citenamefont {Stepanenko}\ \emph {et~al.}(2012)\citenamefont {Stepanenko}, \citenamefont {Rudner}, \citenamefont {Halperin},\ and\ \citenamefont {Loss}}]{Stepanenko12}%
  \BibitemOpen
  \bibfield  {author} {\bibinfo {author} {\bibfnamefont {D.}~\bibnamefont {Stepanenko}}, \bibinfo {author} {\bibfnamefont {M.}~\bibnamefont {Rudner}}, \bibinfo {author} {\bibfnamefont {B.~I.}\ \bibnamefont {Halperin}},\ and\ \bibinfo {author} {\bibfnamefont {D.}~\bibnamefont {Loss}},\ }\bibfield  {title} {\bibinfo {title} {Singlet-triplet splitting in double quantum dots due to spin-orbit and hyperfine interactions},\ }\href {https://doi.org/10.1103/PhysRevB.85.075416} {\bibfield  {journal} {\bibinfo  {journal} {Physical Review B}\ }\textbf {\bibinfo {volume} {85}},\ \bibinfo {pages} {075416} (\bibinfo {year} {2012})}\BibitemShut {NoStop}%
\bibitem [{\citenamefont {Het\'enyi}\ \emph {et~al.}(2020)\citenamefont {Het\'enyi}, \citenamefont {Kloeffel},\ and\ \citenamefont {Loss}}]{Hetenyi20}%
  \BibitemOpen
  \bibfield  {author} {\bibinfo {author} {\bibfnamefont {B.}~\bibnamefont {Het\'enyi}}, \bibinfo {author} {\bibfnamefont {C.}~\bibnamefont {Kloeffel}},\ and\ \bibinfo {author} {\bibfnamefont {D.}~\bibnamefont {Loss}},\ }\bibfield  {title} {\bibinfo {title} {Exchange interaction of hole-spin qubits in double quantum dots in highly anisotropic semiconductors},\ }\href {https://doi.org/10.1103/PhysRevResearch.2.033036} {\bibfield  {journal} {\bibinfo  {journal} {Physical Review Research}\ }\textbf {\bibinfo {volume} {2}},\ \bibinfo {pages} {033036} (\bibinfo {year} {2020})}\BibitemShut {NoStop}%
\bibitem [{\citenamefont {Mutter}\ and\ \citenamefont {Burkard}(2021)}]{mutter_all-electrical_ST-Ge}%
  \BibitemOpen
  \bibfield  {author} {\bibinfo {author} {\bibfnamefont {P.~M.}\ \bibnamefont {Mutter}}\ and\ \bibinfo {author} {\bibfnamefont {G.}~\bibnamefont {Burkard}},\ }\bibfield  {title} {\bibinfo {title} {All-electrical control of hole singlet-triplet spin qubits at low-leakage points},\ }\href {https://doi.org/10.1103/PhysRevB.104.195421} {\bibfield  {journal} {\bibinfo  {journal} {Physical Review B}\ }\textbf {\bibinfo {volume} {104}},\ \bibinfo {pages} {195421} (\bibinfo {year} {2021})}\BibitemShut {NoStop}%
\bibitem [{\citenamefont {Sen}\ \emph {et~al.}(2023)\citenamefont {Sen}, \citenamefont {Frank}, \citenamefont {Kolok}, \citenamefont {Danon},\ and\ \citenamefont {P\'alyi}}]{Sen23}%
  \BibitemOpen
  \bibfield  {author} {\bibinfo {author} {\bibfnamefont {A.}~\bibnamefont {Sen}}, \bibinfo {author} {\bibfnamefont {G.}~\bibnamefont {Frank}}, \bibinfo {author} {\bibfnamefont {B.}~\bibnamefont {Kolok}}, \bibinfo {author} {\bibfnamefont {J.}~\bibnamefont {Danon}},\ and\ \bibinfo {author} {\bibfnamefont {A.}~\bibnamefont {P\'alyi}},\ }\bibfield  {title} {\bibinfo {title} {Classification and magic magnetic field directions for spin-orbit-coupled double quantum dots},\ }\href {https://doi.org/10.1103/PhysRevB.108.245406} {\bibfield  {journal} {\bibinfo  {journal} {Physical Review B}\ }\textbf {\bibinfo {volume} {108}},\ \bibinfo {pages} {245406} (\bibinfo {year} {2023})}\BibitemShut {NoStop}%
\bibitem [{\citenamefont {{David Sherrill}}\ and\ \citenamefont {Schaefer}(1999)}]{Sherrill99}%
  \BibitemOpen
  \bibfield  {author} {\bibinfo {author} {\bibfnamefont {C.}~\bibnamefont {{David Sherrill}}}\ and\ \bibinfo {author} {\bibfnamefont {H.~F.}\ \bibnamefont {Schaefer}},\ }\bibfield  {title} {\bibinfo {title} {The configuration interaction method: Advances in highly correlated approaches}\ }(\bibinfo  {publisher} {Academic Press},\ \bibinfo {year} {1999})\ pp.\ \bibinfo {pages} {143--269}\BibitemShut {NoStop}%
\bibitem [{\citenamefont {Rontani}\ \emph {et~al.}(2006)\citenamefont {Rontani}, \citenamefont {Cavazzoni}, \citenamefont {Bellucci},\ and\ \citenamefont {Goldoni}}]{Rontani06}%
  \BibitemOpen
  \bibfield  {author} {\bibinfo {author} {\bibfnamefont {M.}~\bibnamefont {Rontani}}, \bibinfo {author} {\bibfnamefont {C.}~\bibnamefont {Cavazzoni}}, \bibinfo {author} {\bibfnamefont {D.}~\bibnamefont {Bellucci}},\ and\ \bibinfo {author} {\bibfnamefont {G.}~\bibnamefont {Goldoni}},\ }\bibfield  {title} {\bibinfo {title} {Full configuration interaction approach to the few-electron problem in artificial atoms},\ }\href {https://doi.org/10.1063/1.2179418} {\bibfield  {journal} {\bibinfo  {journal} {The Journal of Chemical Physics}\ }\textbf {\bibinfo {volume} {124}},\ \bibinfo {pages} {124102} (\bibinfo {year} {2006})}\BibitemShut {NoStop}%
\bibitem [{\citenamefont {Anderson}\ \emph {et~al.}(2022)\citenamefont {Anderson}, \citenamefont {Gyure}, \citenamefont {Quinn}, \citenamefont {Pan}, \citenamefont {Ross},\ and\ \citenamefont {Kiselev}}]{ElectroConfinedCI(EM)}%
  \BibitemOpen
  \bibfield  {author} {\bibinfo {author} {\bibfnamefont {C.~R.}\ \bibnamefont {Anderson}}, \bibinfo {author} {\bibfnamefont {M.~F.}\ \bibnamefont {Gyure}}, \bibinfo {author} {\bibfnamefont {S.}~\bibnamefont {Quinn}}, \bibinfo {author} {\bibfnamefont {A.}~\bibnamefont {Pan}}, \bibinfo {author} {\bibfnamefont {R.~S.}\ \bibnamefont {Ross}},\ and\ \bibinfo {author} {\bibfnamefont {A.~A.}\ \bibnamefont {Kiselev}},\ }\bibfield  {title} {\bibinfo {title} {{High-precision real-space simulation of electrostatically confined few-electron states}},\ }\href {https://doi.org/10.1063/5.0089350} {\bibfield  {journal} {\bibinfo  {journal} {AIP Advances}\ }\textbf {\bibinfo {volume} {12}},\ \bibinfo {pages} {065123} (\bibinfo {year} {2022})}\BibitemShut {NoStop}%
\bibitem [{\citenamefont {Climente}\ \emph {et~al.}(2007)\citenamefont {Climente}, \citenamefont {Bertoni}, \citenamefont {Goldoni}, \citenamefont {Rontani},\ and\ \citenamefont {Molinari}}]{Cimente07}%
  \BibitemOpen
  \bibfield  {author} {\bibinfo {author} {\bibfnamefont {J.~I.}\ \bibnamefont {Climente}}, \bibinfo {author} {\bibfnamefont {A.}~\bibnamefont {Bertoni}}, \bibinfo {author} {\bibfnamefont {G.}~\bibnamefont {Goldoni}}, \bibinfo {author} {\bibfnamefont {M.}~\bibnamefont {Rontani}},\ and\ \bibinfo {author} {\bibfnamefont {E.}~\bibnamefont {Molinari}},\ }\bibfield  {title} {\bibinfo {title} {Magnetic field dependence of triplet-singlet relaxation in quantum dots with spin-orbit coupling},\ }\href {https://doi.org/10.1103/PhysRevB.75.081303} {\bibfield  {journal} {\bibinfo  {journal} {Physical Review B}\ }\textbf {\bibinfo {volume} {75}},\ \bibinfo {pages} {081303} (\bibinfo {year} {2007})}\BibitemShut {NoStop}%
\bibitem [{\citenamefont {Baruffa}\ \emph {et~al.}(2010)\citenamefont {Baruffa}, \citenamefont {Stano},\ and\ \citenamefont {Fabian}}]{Baruffa10}%
  \BibitemOpen
  \bibfield  {author} {\bibinfo {author} {\bibfnamefont {F.}~\bibnamefont {Baruffa}}, \bibinfo {author} {\bibfnamefont {P.}~\bibnamefont {Stano}},\ and\ \bibinfo {author} {\bibfnamefont {J.}~\bibnamefont {Fabian}},\ }\bibfield  {title} {\bibinfo {title} {Theory of anisotropic exchange in laterally coupled quantum dots},\ }\href {https://doi.org/10.1103/PhysRevLett.104.126401} {\bibfield  {journal} {\bibinfo  {journal} {Physical Review Letters}\ }\textbf {\bibinfo {volume} {104}},\ \bibinfo {pages} {126401} (\bibinfo {year} {2010})}\BibitemShut {NoStop}%
\bibitem [{\citenamefont {Nielsen}\ \emph {et~al.}(2010)\citenamefont {Nielsen}, \citenamefont {Young}, \citenamefont {Muller},\ and\ \citenamefont {Carroll}}]{lownoisexchGCI}%
  \BibitemOpen
  \bibfield  {author} {\bibinfo {author} {\bibfnamefont {E.}~\bibnamefont {Nielsen}}, \bibinfo {author} {\bibfnamefont {R.~W.}\ \bibnamefont {Young}}, \bibinfo {author} {\bibfnamefont {R.~P.}\ \bibnamefont {Muller}},\ and\ \bibinfo {author} {\bibfnamefont {M.~S.}\ \bibnamefont {Carroll}},\ }\bibfield  {title} {\bibinfo {title} {Implications of simultaneous requirements for low-noise exchange gates in double quantum dots},\ }\href {https://doi.org/10.1103/PhysRevB.82.075319} {\bibfield  {journal} {\bibinfo  {journal} {Physical Review B}\ }\textbf {\bibinfo {volume} {82}},\ \bibinfo {pages} {075319} (\bibinfo {year} {2010})}\BibitemShut {NoStop}%
\bibitem [{\citenamefont {Nowak}\ and\ \citenamefont {Szafran}(2010)}]{timedepCI}%
  \BibitemOpen
  \bibfield  {author} {\bibinfo {author} {\bibfnamefont {M.~P.}\ \bibnamefont {Nowak}}\ and\ \bibinfo {author} {\bibfnamefont {B.}~\bibnamefont {Szafran}},\ }\bibfield  {title} {\bibinfo {title} {Time-dependent configuration-interaction simulations of spin swap in spin-orbit-coupled double quantum dots},\ }\href {https://doi.org/10.1103/PhysRevB.82.165316} {\bibfield  {journal} {\bibinfo  {journal} {Physical Review B}\ }\textbf {\bibinfo {volume} {82}},\ \bibinfo {pages} {165316} (\bibinfo {year} {2010})}\BibitemShut {NoStop}%
\bibitem [{\citenamefont {Barnes}\ \emph {et~al.}(2011)\citenamefont {Barnes}, \citenamefont {Kestner}, \citenamefont {Nguyen},\ and\ \citenamefont {Das~Sarma}}]{Barnes11}%
  \BibitemOpen
  \bibfield  {author} {\bibinfo {author} {\bibfnamefont {E.}~\bibnamefont {Barnes}}, \bibinfo {author} {\bibfnamefont {J.~P.}\ \bibnamefont {Kestner}}, \bibinfo {author} {\bibfnamefont {N.~T.~T.}\ \bibnamefont {Nguyen}},\ and\ \bibinfo {author} {\bibfnamefont {S.}~\bibnamefont {Das~Sarma}},\ }\bibfield  {title} {\bibinfo {title} {Screening of charged impurities with multielectron singlet-triplet spin qubits in quantum dots},\ }\href {https://doi.org/10.1103/PhysRevB.84.235309} {\bibfield  {journal} {\bibinfo  {journal} {Physical Review B}\ }\textbf {\bibinfo {volume} {84}},\ \bibinfo {pages} {235309} (\bibinfo {year} {2011})}\BibitemShut {NoStop}%
\bibitem [{\citenamefont {Nielsen}\ \emph {et~al.}(2012)\citenamefont {Nielsen}, \citenamefont {Muller},\ and\ \citenamefont {Carroll}}]{CtrlPhaseGCI}%
  \BibitemOpen
  \bibfield  {author} {\bibinfo {author} {\bibfnamefont {E.}~\bibnamefont {Nielsen}}, \bibinfo {author} {\bibfnamefont {R.~P.}\ \bibnamefont {Muller}},\ and\ \bibinfo {author} {\bibfnamefont {M.~S.}\ \bibnamefont {Carroll}},\ }\bibfield  {title} {\bibinfo {title} {Configuration interaction calculations of the controlled phase gate in double quantum dot qubits},\ }\href {https://doi.org/10.1103/PhysRevB.85.035319} {\bibfield  {journal} {\bibinfo  {journal} {Physical Review B}\ }\textbf {\bibinfo {volume} {85}},\ \bibinfo {pages} {035319} (\bibinfo {year} {2012})}\BibitemShut {NoStop}%
\bibitem [{\citenamefont {Raith}\ \emph {et~al.}(2012)\citenamefont {Raith}, \citenamefont {Stano},\ and\ \citenamefont {Fabian}}]{SpinRelaxCI}%
  \BibitemOpen
  \bibfield  {author} {\bibinfo {author} {\bibfnamefont {M.}~\bibnamefont {Raith}}, \bibinfo {author} {\bibfnamefont {P.}~\bibnamefont {Stano}},\ and\ \bibinfo {author} {\bibfnamefont {J.}~\bibnamefont {Fabian}},\ }\bibfield  {title} {\bibinfo {title} {Theory of spin relaxation in two-electron laterally coupled {Si}/{SiGe} quantum dots},\ }\href {https://doi.org/10.1103/PhysRevB.86.205321} {\bibfield  {journal} {\bibinfo  {journal} {Physical Review B}\ }\textbf {\bibinfo {volume} {86}},\ \bibinfo {pages} {205321} (\bibinfo {year} {2012})}\BibitemShut {NoStop}%
\bibitem [{\citenamefont {Nielsen}\ \emph {et~al.}(2013)\citenamefont {Nielsen}, \citenamefont {Barnes}, \citenamefont {Kestner},\ and\ \citenamefont {Das~Sarma}}]{Nielsen13}%
  \BibitemOpen
  \bibfield  {author} {\bibinfo {author} {\bibfnamefont {E.}~\bibnamefont {Nielsen}}, \bibinfo {author} {\bibfnamefont {E.}~\bibnamefont {Barnes}}, \bibinfo {author} {\bibfnamefont {J.~P.}\ \bibnamefont {Kestner}},\ and\ \bibinfo {author} {\bibfnamefont {S.}~\bibnamefont {Das~Sarma}},\ }\bibfield  {title} {\bibinfo {title} {Six-electron semiconductor double quantum dot qubits},\ }\href {https://doi.org/10.1103/PhysRevB.88.195131} {\bibfield  {journal} {\bibinfo  {journal} {Physical Review B}\ }\textbf {\bibinfo {volume} {88}},\ \bibinfo {pages} {195131} (\bibinfo {year} {2013})}\BibitemShut {NoStop}%
\bibitem [{\citenamefont {Joecker}\ \emph {et~al.}(2021)\citenamefont {Joecker}, \citenamefont {Baczewski}, \citenamefont {Gamble}, \citenamefont {Pla}, \citenamefont {Saraiva},\ and\ \citenamefont {Morello}}]{ExCplDonCI(EM)}%
  \BibitemOpen
  \bibfield  {author} {\bibinfo {author} {\bibfnamefont {B.}~\bibnamefont {Joecker}}, \bibinfo {author} {\bibfnamefont {A.~D.}\ \bibnamefont {Baczewski}}, \bibinfo {author} {\bibfnamefont {J.~K.}\ \bibnamefont {Gamble}}, \bibinfo {author} {\bibfnamefont {J.~J.}\ \bibnamefont {Pla}}, \bibinfo {author} {\bibfnamefont {A.}~\bibnamefont {Saraiva}},\ and\ \bibinfo {author} {\bibfnamefont {A.}~\bibnamefont {Morello}},\ }\bibfield  {title} {\bibinfo {title} {Full configuration interaction simulations of exchange-coupled donors in silicon using multi-valley effective mass theory},\ }\href {https://doi.org/10.1088/1367-2630/ac0abf} {\bibfield  {journal} {\bibinfo  {journal} {New Journal of Physics}\ }\textbf {\bibinfo {volume} {23}},\ \bibinfo {pages} {073007} (\bibinfo {year} {2021})}\BibitemShut {NoStop}%
\bibitem [{\citenamefont {Deng}\ and\ \citenamefont {Barnes}(2020)}]{Deng20}%
  \BibitemOpen
  \bibfield  {author} {\bibinfo {author} {\bibfnamefont {K.}~\bibnamefont {Deng}}\ and\ \bibinfo {author} {\bibfnamefont {E.}~\bibnamefont {Barnes}},\ }\bibfield  {title} {\bibinfo {title} {Interplay of exchange and superexchange in triple quantum dots},\ }\href {https://doi.org/10.1103/PhysRevB.102.035427} {\bibfield  {journal} {\bibinfo  {journal} {Physical Review B}\ }\textbf {\bibinfo {volume} {102}},\ \bibinfo {pages} {035427} (\bibinfo {year} {2020})}\BibitemShut {NoStop}%
\bibitem [{\citenamefont {Chan}\ \emph {et~al.}(2021)\citenamefont {Chan}, \citenamefont {Sahasrabudhe}, \citenamefont {Huang}, \citenamefont {Wang}, \citenamefont {Yang}, \citenamefont {Veldhorst}, \citenamefont {Hwang}, \citenamefont {Mohiyaddin}, \citenamefont {Hudson}, \citenamefont {Itoh}, \citenamefont {Saraiva}, \citenamefont {Morello}, \citenamefont {Laucht}, \citenamefont {Rahman},\ and\ \citenamefont {Dzurak}}]{Chan2021}%
  \BibitemOpen
  \bibfield  {author} {\bibinfo {author} {\bibfnamefont {K.~W.}\ \bibnamefont {Chan}}, \bibinfo {author} {\bibfnamefont {H.}~\bibnamefont {Sahasrabudhe}}, \bibinfo {author} {\bibfnamefont {W.}~\bibnamefont {Huang}}, \bibinfo {author} {\bibfnamefont {Y.}~\bibnamefont {Wang}}, \bibinfo {author} {\bibfnamefont {H.~C.}\ \bibnamefont {Yang}}, \bibinfo {author} {\bibfnamefont {M.}~\bibnamefont {Veldhorst}}, \bibinfo {author} {\bibfnamefont {J.~C.~C.}\ \bibnamefont {Hwang}}, \bibinfo {author} {\bibfnamefont {F.~A.}\ \bibnamefont {Mohiyaddin}}, \bibinfo {author} {\bibfnamefont {F.~E.}\ \bibnamefont {Hudson}}, \bibinfo {author} {\bibfnamefont {K.~M.}\ \bibnamefont {Itoh}}, \bibinfo {author} {\bibfnamefont {A.}~\bibnamefont {Saraiva}}, \bibinfo {author} {\bibfnamefont {A.}~\bibnamefont {Morello}}, \bibinfo {author} {\bibfnamefont {A.}~\bibnamefont {Laucht}}, \bibinfo {author} {\bibfnamefont {R.}~\bibnamefont {Rahman}},\ and\ \bibinfo {author} {\bibfnamefont {A.~S.}\ \bibnamefont {Dzurak}},\ }\bibfield  {title}
  {\bibinfo {title} {Exchange coupling in a linear chain of three quantum-dot spin qubits in silicon},\ }\href {https://doi.org/10.1021/acs.nanolett.0c04771} {\bibfield  {journal} {\bibinfo  {journal} {Nano Letters}\ }\textbf {\bibinfo {volume} {21}},\ \bibinfo {pages} {1517} (\bibinfo {year} {2021})}\BibitemShut {NoStop}%
\bibitem [{\citenamefont {Secchi}\ \emph {et~al.}(2021{\natexlab{a}})\citenamefont {Secchi}, \citenamefont {Bellentani}, \citenamefont {Bertoni},\ and\ \citenamefont {Troiani}}]{IntHolesCI(KP)}%
  \BibitemOpen
  \bibfield  {author} {\bibinfo {author} {\bibfnamefont {A.}~\bibnamefont {Secchi}}, \bibinfo {author} {\bibfnamefont {L.}~\bibnamefont {Bellentani}}, \bibinfo {author} {\bibfnamefont {A.}~\bibnamefont {Bertoni}},\ and\ \bibinfo {author} {\bibfnamefont {F.}~\bibnamefont {Troiani}},\ }\bibfield  {title} {\bibinfo {title} {Interacting holes in {Si} and {Ge} double quantum dots: From a multiband approach to an effective-spin picture},\ }\href {https://doi.org/10.1103/PhysRevB.104.035302} {\bibfield  {journal} {\bibinfo  {journal} {Physical Review B}\ }\textbf {\bibinfo {volume} {104}},\ \bibinfo {pages} {035302} (\bibinfo {year} {2021}{\natexlab{a}})}\BibitemShut {NoStop}%
\bibitem [{\citenamefont {Chan}\ and\ \citenamefont {Wang}(2022{\natexlab{a}})}]{MicroSweetSpotCI}%
  \BibitemOpen
  \bibfield  {author} {\bibinfo {author} {\bibfnamefont {G.~X.}\ \bibnamefont {Chan}}\ and\ \bibinfo {author} {\bibfnamefont {X.}~\bibnamefont {Wang}},\ }\bibfield  {title} {\bibinfo {title} {Microscopic theory of a magnetic-field-tuned sweet spot of exchange interactions in multielectron quantum-dot systems},\ }\href {https://doi.org/10.1103/PhysRevB.105.245409} {\bibfield  {journal} {\bibinfo  {journal} {Physical Review B}\ }\textbf {\bibinfo {volume} {105}},\ \bibinfo {pages} {245409} (\bibinfo {year} {2022}{\natexlab{a}})}\BibitemShut {NoStop}%
\bibitem [{\citenamefont {Chan}\ and\ \citenamefont {Wang}(2022{\natexlab{b}})}]{4eDDQDCI}%
  \BibitemOpen
  \bibfield  {author} {\bibinfo {author} {\bibfnamefont {G.~X.}\ \bibnamefont {Chan}}\ and\ \bibinfo {author} {\bibfnamefont {X.}~\bibnamefont {Wang}},\ }\bibfield  {title} {\bibinfo {title} {Robust entangling gate for capacitively coupled few-electron singlet-triplet qubits},\ }\href {https://doi.org/10.1103/PhysRevB.106.075417} {\bibfield  {journal} {\bibinfo  {journal} {Physical Review B}\ }\textbf {\bibinfo {volume} {106}},\ \bibinfo {pages} {075417} (\bibinfo {year} {2022}{\natexlab{b}})}\BibitemShut {NoStop}%
\bibitem [{\citenamefont {Dodson}\ \emph {et~al.}(2022)\citenamefont {Dodson}, \citenamefont {Ercan}, \citenamefont {Corrigan}, \citenamefont {Losert}, \citenamefont {Holman}, \citenamefont {McJunkin}, \citenamefont {Edge}, \citenamefont {Friesen}, \citenamefont {Coppersmith},\ and\ \citenamefont {Eriksson}}]{Dodson22}%
  \BibitemOpen
  \bibfield  {author} {\bibinfo {author} {\bibfnamefont {J.~P.}\ \bibnamefont {Dodson}}, \bibinfo {author} {\bibfnamefont {H.~E.}\ \bibnamefont {Ercan}}, \bibinfo {author} {\bibfnamefont {J.}~\bibnamefont {Corrigan}}, \bibinfo {author} {\bibfnamefont {M.~P.}\ \bibnamefont {Losert}}, \bibinfo {author} {\bibfnamefont {N.}~\bibnamefont {Holman}}, \bibinfo {author} {\bibfnamefont {T.}~\bibnamefont {McJunkin}}, \bibinfo {author} {\bibfnamefont {L.~F.}\ \bibnamefont {Edge}}, \bibinfo {author} {\bibfnamefont {M.}~\bibnamefont {Friesen}}, \bibinfo {author} {\bibfnamefont {S.~N.}\ \bibnamefont {Coppersmith}},\ and\ \bibinfo {author} {\bibfnamefont {M.~A.}\ \bibnamefont {Eriksson}},\ }\bibfield  {title} {\bibinfo {title} {How valley-orbit states in silicon quantum dots probe quantum well interfaces},\ }\href {https://doi.org/10.1103/PhysRevLett.128.146802} {\bibfield  {journal} {\bibinfo  {journal} {Physical Review Letters}\ }\textbf {\bibinfo {volume} {128}},\ \bibinfo {pages} {146802} (\bibinfo {year}
  {2022})}\BibitemShut {NoStop}%
\bibitem [{\citenamefont {Shehata}\ \emph {et~al.}(2023)\citenamefont {Shehata}, \citenamefont {Simion}, \citenamefont {Li}, \citenamefont {Mohiyaddin}, \citenamefont {Wan}, \citenamefont {Mongillo}, \citenamefont {Govoreanu}, \citenamefont {Radu}, \citenamefont {De~Greve},\ and\ \citenamefont {Van~Dorpe}}]{Shehata23}%
  \BibitemOpen
  \bibfield  {author} {\bibinfo {author} {\bibfnamefont {M.~M. E.~K.}\ \bibnamefont {Shehata}}, \bibinfo {author} {\bibfnamefont {G.}~\bibnamefont {Simion}}, \bibinfo {author} {\bibfnamefont {R.}~\bibnamefont {Li}}, \bibinfo {author} {\bibfnamefont {F.~A.}\ \bibnamefont {Mohiyaddin}}, \bibinfo {author} {\bibfnamefont {D.}~\bibnamefont {Wan}}, \bibinfo {author} {\bibfnamefont {M.}~\bibnamefont {Mongillo}}, \bibinfo {author} {\bibfnamefont {B.}~\bibnamefont {Govoreanu}}, \bibinfo {author} {\bibfnamefont {I.}~\bibnamefont {Radu}}, \bibinfo {author} {\bibfnamefont {K.}~\bibnamefont {De~Greve}},\ and\ \bibinfo {author} {\bibfnamefont {P.}~\bibnamefont {Van~Dorpe}},\ }\bibfield  {title} {\bibinfo {title} {Modeling semiconductor spin qubits and their charge noise environment for quantum gate fidelity estimation},\ }\href {https://doi.org/10.1103/PhysRevB.108.045305} {\bibfield  {journal} {\bibinfo  {journal} {Physical Review B}\ }\textbf {\bibinfo {volume} {108}},\ \bibinfo {pages} {045305} (\bibinfo {year}
  {2023})}\BibitemShut {NoStop}%
\bibitem [{\citenamefont {Bryant}(1987)}]{Bryant87}%
  \BibitemOpen
  \bibfield  {author} {\bibinfo {author} {\bibfnamefont {G.~W.}\ \bibnamefont {Bryant}},\ }\bibfield  {title} {\bibinfo {title} {Electronic structure of ultrasmall quantum-well boxes},\ }\href {https://doi.org/10.1103/PhysRevLett.59.1140} {\bibfield  {journal} {\bibinfo  {journal} {Physical Review Letters}\ }\textbf {\bibinfo {volume} {59}},\ \bibinfo {pages} {1140} (\bibinfo {year} {1987})}\BibitemShut {NoStop}%
\bibitem [{\citenamefont {Reimann}\ \emph {et~al.}(2000)\citenamefont {Reimann}, \citenamefont {Koskinen},\ and\ \citenamefont {Manninen}}]{Reimann00}%
  \BibitemOpen
  \bibfield  {author} {\bibinfo {author} {\bibfnamefont {S.~M.}\ \bibnamefont {Reimann}}, \bibinfo {author} {\bibfnamefont {M.}~\bibnamefont {Koskinen}},\ and\ \bibinfo {author} {\bibfnamefont {M.}~\bibnamefont {Manninen}},\ }\bibfield  {title} {\bibinfo {title} {Formation of wigner molecules in small quantum dots},\ }\href {https://doi.org/10.1103/PhysRevB.62.8108} {\bibfield  {journal} {\bibinfo  {journal} {Physical Review B}\ }\textbf {\bibinfo {volume} {62}},\ \bibinfo {pages} {8108} (\bibinfo {year} {2000})}\BibitemShut {NoStop}%
\bibitem [{\citenamefont {Ellenberger}\ \emph {et~al.}(2006)\citenamefont {Ellenberger}, \citenamefont {Ihn}, \citenamefont {Yannouleas}, \citenamefont {Landman}, \citenamefont {Ensslin}, \citenamefont {Driscoll},\ and\ \citenamefont {Gossard}}]{Ellenberg06}%
  \BibitemOpen
  \bibfield  {author} {\bibinfo {author} {\bibfnamefont {C.}~\bibnamefont {Ellenberger}}, \bibinfo {author} {\bibfnamefont {T.}~\bibnamefont {Ihn}}, \bibinfo {author} {\bibfnamefont {C.}~\bibnamefont {Yannouleas}}, \bibinfo {author} {\bibfnamefont {U.}~\bibnamefont {Landman}}, \bibinfo {author} {\bibfnamefont {K.}~\bibnamefont {Ensslin}}, \bibinfo {author} {\bibfnamefont {D.}~\bibnamefont {Driscoll}},\ and\ \bibinfo {author} {\bibfnamefont {A.~C.}\ \bibnamefont {Gossard}},\ }\bibfield  {title} {\bibinfo {title} {Excitation spectrum of two correlated electrons in a lateral quantum dot with negligible zeeman splitting},\ }\href {https://doi.org/10.1103/PhysRevLett.96.126806} {\bibfield  {journal} {\bibinfo  {journal} {Physical Review Letters}\ }\textbf {\bibinfo {volume} {96}},\ \bibinfo {pages} {126806} (\bibinfo {year} {2006})}\BibitemShut {NoStop}%
\bibitem [{\citenamefont {Kalliakos}\ \emph {et~al.}(2008)\citenamefont {Kalliakos}, \citenamefont {Rontani}, \citenamefont {Pellegrini}, \citenamefont {Garc{\'i}a}, \citenamefont {Pinczuk}, \citenamefont {Goldoni}, \citenamefont {Molinari}, \citenamefont {Pfeiffer},\ and\ \citenamefont {West}}]{Kalliakos08}%
  \BibitemOpen
  \bibfield  {author} {\bibinfo {author} {\bibfnamefont {S.}~\bibnamefont {Kalliakos}}, \bibinfo {author} {\bibfnamefont {M.}~\bibnamefont {Rontani}}, \bibinfo {author} {\bibfnamefont {V.}~\bibnamefont {Pellegrini}}, \bibinfo {author} {\bibfnamefont {C.~P.}\ \bibnamefont {Garc{\'i}a}}, \bibinfo {author} {\bibfnamefont {A.}~\bibnamefont {Pinczuk}}, \bibinfo {author} {\bibfnamefont {G.}~\bibnamefont {Goldoni}}, \bibinfo {author} {\bibfnamefont {E.}~\bibnamefont {Molinari}}, \bibinfo {author} {\bibfnamefont {L.~N.}\ \bibnamefont {Pfeiffer}},\ and\ \bibinfo {author} {\bibfnamefont {K.~W.}\ \bibnamefont {West}},\ }\bibfield  {title} {\bibinfo {title} {A molecular state of correlated electrons in a quantum dot},\ }\href {https://doi.org/10.1038/nphys944} {\bibfield  {journal} {\bibinfo  {journal} {Nature Physics}\ }\textbf {\bibinfo {volume} {4}},\ \bibinfo {pages} {467} (\bibinfo {year} {2008})}\BibitemShut {NoStop}%
\bibitem [{\citenamefont {Singha}\ \emph {et~al.}(2010)\citenamefont {Singha}, \citenamefont {Pellegrini}, \citenamefont {Pinczuk}, \citenamefont {Pfeiffer}, \citenamefont {West},\ and\ \citenamefont {Rontani}}]{Singha10}%
  \BibitemOpen
  \bibfield  {author} {\bibinfo {author} {\bibfnamefont {A.}~\bibnamefont {Singha}}, \bibinfo {author} {\bibfnamefont {V.}~\bibnamefont {Pellegrini}}, \bibinfo {author} {\bibfnamefont {A.}~\bibnamefont {Pinczuk}}, \bibinfo {author} {\bibfnamefont {L.~N.}\ \bibnamefont {Pfeiffer}}, \bibinfo {author} {\bibfnamefont {K.~W.}\ \bibnamefont {West}},\ and\ \bibinfo {author} {\bibfnamefont {M.}~\bibnamefont {Rontani}},\ }\bibfield  {title} {\bibinfo {title} {Correlated electrons in optically tunable quantum dots: Building an electron dimer molecule},\ }\href {https://doi.org/10.1103/PhysRevLett.104.246802} {\bibfield  {journal} {\bibinfo  {journal} {Physical Review Letters}\ }\textbf {\bibinfo {volume} {104}},\ \bibinfo {pages} {246802} (\bibinfo {year} {2010})}\BibitemShut {NoStop}%
\bibitem [{\citenamefont {Abadillo-Uriel}\ \emph {et~al.}(2021)\citenamefont {Abadillo-Uriel}, \citenamefont {Martinez}, \citenamefont {Filippone},\ and\ \citenamefont {Niquet}}]{Abadillo21}%
  \BibitemOpen
  \bibfield  {author} {\bibinfo {author} {\bibfnamefont {J.~C.}\ \bibnamefont {Abadillo-Uriel}}, \bibinfo {author} {\bibfnamefont {B.}~\bibnamefont {Martinez}}, \bibinfo {author} {\bibfnamefont {M.}~\bibnamefont {Filippone}},\ and\ \bibinfo {author} {\bibfnamefont {Y.-M.}\ \bibnamefont {Niquet}},\ }\bibfield  {title} {\bibinfo {title} {Two-body wigner molecularization in asymmetric quantum dot spin qubits},\ }\href {https://doi.org/10.1103/PhysRevB.104.195305} {\bibfield  {journal} {\bibinfo  {journal} {Physical Review B}\ }\textbf {\bibinfo {volume} {104}},\ \bibinfo {pages} {195305} (\bibinfo {year} {2021})}\BibitemShut {NoStop}%
\bibitem [{\citenamefont {Corrigan}\ \emph {et~al.}(2021)\citenamefont {Corrigan}, \citenamefont {Dodson}, \citenamefont {Ercan}, \citenamefont {Abadillo-Uriel}, \citenamefont {Thorgrimsson}, \citenamefont {Knapp}, \citenamefont {Holman}, \citenamefont {McJunkin}, \citenamefont {Neyens}, \citenamefont {MacQuarrie}, \citenamefont {Foote}, \citenamefont {Edge}, \citenamefont {Friesen}, \citenamefont {Coppersmith},\ and\ \citenamefont {Eriksson}}]{Corrigan21}%
  \BibitemOpen
  \bibfield  {author} {\bibinfo {author} {\bibfnamefont {J.}~\bibnamefont {Corrigan}}, \bibinfo {author} {\bibfnamefont {J.~P.}\ \bibnamefont {Dodson}}, \bibinfo {author} {\bibfnamefont {H.~E.}\ \bibnamefont {Ercan}}, \bibinfo {author} {\bibfnamefont {J.~C.}\ \bibnamefont {Abadillo-Uriel}}, \bibinfo {author} {\bibfnamefont {B.}~\bibnamefont {Thorgrimsson}}, \bibinfo {author} {\bibfnamefont {T.~J.}\ \bibnamefont {Knapp}}, \bibinfo {author} {\bibfnamefont {N.}~\bibnamefont {Holman}}, \bibinfo {author} {\bibfnamefont {T.}~\bibnamefont {McJunkin}}, \bibinfo {author} {\bibfnamefont {S.~F.}\ \bibnamefont {Neyens}}, \bibinfo {author} {\bibfnamefont {E.~R.}\ \bibnamefont {MacQuarrie}}, \bibinfo {author} {\bibfnamefont {R.~H.}\ \bibnamefont {Foote}}, \bibinfo {author} {\bibfnamefont {L.~F.}\ \bibnamefont {Edge}}, \bibinfo {author} {\bibfnamefont {M.}~\bibnamefont {Friesen}}, \bibinfo {author} {\bibfnamefont {S.~N.}\ \bibnamefont {Coppersmith}},\ and\ \bibinfo {author} {\bibfnamefont {M.~A.}\ \bibnamefont {Eriksson}},\
  }\bibfield  {title} {\bibinfo {title} {Coherent control and spectroscopy of a semiconductor quantum dot wigner molecule},\ }\href {https://doi.org/10.1103/PhysRevLett.127.127701} {\bibfield  {journal} {\bibinfo  {journal} {Physical Review Letters}\ }\textbf {\bibinfo {volume} {127}},\ \bibinfo {pages} {127701} (\bibinfo {year} {2021})}\BibitemShut {NoStop}%
\bibitem [{\citenamefont {Ercan}\ \emph {et~al.}(2021)\citenamefont {Ercan}, \citenamefont {Coppersmith},\ and\ \citenamefont {Friesen}}]{e-eCI(TB)}%
  \BibitemOpen
  \bibfield  {author} {\bibinfo {author} {\bibfnamefont {H.~E.}\ \bibnamefont {Ercan}}, \bibinfo {author} {\bibfnamefont {S.~N.}\ \bibnamefont {Coppersmith}},\ and\ \bibinfo {author} {\bibfnamefont {M.}~\bibnamefont {Friesen}},\ }\bibfield  {title} {\bibinfo {title} {Strong electron-electron interactions in {Si}/{SiGe} quantum dots},\ }\href {https://doi.org/10.1103/PhysRevB.104.235302} {\bibfield  {journal} {\bibinfo  {journal} {Physical Review B}\ }\textbf {\bibinfo {volume} {104}},\ \bibinfo {pages} {235302} (\bibinfo {year} {2021})}\BibitemShut {NoStop}%
\bibitem [{\citenamefont {Yannouleas}\ and\ \citenamefont {Landman}(2022{\natexlab{a}})}]{ValleytronicCI(EM)}%
  \BibitemOpen
  \bibfield  {author} {\bibinfo {author} {\bibfnamefont {C.}~\bibnamefont {Yannouleas}}\ and\ \bibinfo {author} {\bibfnamefont {U.}~\bibnamefont {Landman}},\ }\bibfield  {title} {\bibinfo {title} {Valleytronic full configuration-interaction approach: Application to the excitation spectra of {Si} double-dot qubits},\ }\href {https://doi.org/10.1103/PhysRevB.106.195306} {\bibfield  {journal} {\bibinfo  {journal} {Physical Review B}\ }\textbf {\bibinfo {volume} {106}},\ \bibinfo {pages} {195306} (\bibinfo {year} {2022}{\natexlab{a}})}\BibitemShut {NoStop}%
\bibitem [{\citenamefont {Yannouleas}\ and\ \citenamefont {Landman}(2022{\natexlab{b}})}]{Yannouleas22}%
  \BibitemOpen
  \bibfield  {author} {\bibinfo {author} {\bibfnamefont {C.}~\bibnamefont {Yannouleas}}\ and\ \bibinfo {author} {\bibfnamefont {U.}~\bibnamefont {Landman}},\ }\bibfield  {title} {\bibinfo {title} {Molecular formations and spectra due to electron correlations in three-electron hybrid double-well qubits},\ }\href {https://doi.org/10.1103/PhysRevB.105.205302} {\bibfield  {journal} {\bibinfo  {journal} {Physical Review B}\ }\textbf {\bibinfo {volume} {105}},\ \bibinfo {pages} {205302} (\bibinfo {year} {2022}{\natexlab{b}})}\BibitemShut {NoStop}%
\bibitem [{\citenamefont {Luttinger}(1956)}]{Luttinger56}%
  \BibitemOpen
  \bibfield  {author} {\bibinfo {author} {\bibfnamefont {J.~M.}\ \bibnamefont {Luttinger}},\ }\bibfield  {title} {\bibinfo {title} {Quantum theory of cyclotron resonance in semiconductors: General theory},\ }\href {https://doi.org/10.1103/PhysRev.102.1030} {\bibfield  {journal} {\bibinfo  {journal} {Physical Review}\ }\textbf {\bibinfo {volume} {102}},\ \bibinfo {pages} {1030} (\bibinfo {year} {1956})}\BibitemShut {NoStop}%
\bibitem [{\citenamefont {Lew Yan~Voon}\ and\ \citenamefont {Willatzen}(2009)}]{KP09}%
  \BibitemOpen
  \bibfield  {author} {\bibinfo {author} {\bibfnamefont {L.~C.}\ \bibnamefont {Lew Yan~Voon}}\ and\ \bibinfo {author} {\bibfnamefont {M.}~\bibnamefont {Willatzen}},\ }\href {https://doi.org/10.1007/978-3-540-92872-0} {\emph {\bibinfo {title} {The k p Method}}}\ (\bibinfo  {publisher} {Springer},\ \bibinfo {address} {Berlin},\ \bibinfo {year} {2009})\BibitemShut {NoStop}%
\bibitem [{Note1()}]{Note1}%
  \BibitemOpen
  \bibinfo {note} {The mesh is inhomogeneous along $z$ outside the well}\BibitemShut {NoStop}%
\bibitem [{\citenamefont {Sleijpen}\ and\ \citenamefont {Van~der Vorst}(2000)}]{Sleijpen00}%
  \BibitemOpen
  \bibfield  {author} {\bibinfo {author} {\bibfnamefont {G.}~\bibnamefont {Sleijpen}}\ and\ \bibinfo {author} {\bibfnamefont {H.}~\bibnamefont {Van~der Vorst}},\ }\bibfield  {title} {\bibinfo {title} {A {Jacobi}--{Davidson} iteration method for linear eigenvalue problems},\ }\href {https://doi.org/10.1137/S0036144599363084} {\bibfield  {journal} {\bibinfo  {journal} {SIAM Review}\ }\textbf {\bibinfo {volume} {42}},\ \bibinfo {pages} {267} (\bibinfo {year} {2000})}\BibitemShut {NoStop}%
\bibitem [{\citenamefont {Bai}\ \emph {et~al.}(2000)\citenamefont {Bai}, \citenamefont {Demmel}, \citenamefont {Dongarra}, \citenamefont {Ruhe},\ and\ \citenamefont {van~der Vorst}}]{Templates00}%
  \BibitemOpen
  \bibinfo {editor} {\bibfnamefont {Z.}~\bibnamefont {Bai}}, \bibinfo {editor} {\bibfnamefont {J.}~\bibnamefont {Demmel}}, \bibinfo {editor} {\bibfnamefont {J.}~\bibnamefont {Dongarra}}, \bibinfo {editor} {\bibfnamefont {A.}~\bibnamefont {Ruhe}},\ and\ \bibinfo {editor} {\bibfnamefont {H.}~\bibnamefont {van~der Vorst}},\ eds.,\ \href {https://doi.org/10.1137/1.9780898719581} {\emph {\bibinfo {title} {Templates for the Solution of Algebraic Eigenvalue Problems: A Practical Guide}}}\ (\bibinfo  {publisher} {SIAM},\ \bibinfo {address} {Philadelphia},\ \bibinfo {year} {2000})\BibitemShut {NoStop}%
\bibitem [{Note2()}]{Note2}%
  \BibitemOpen
  \bibinfo {note} {This is the case as long as screening by remote hole gases is not included in, e.g., the Thomas-Fermi approximation. Even in that case, the response of the hole gases can be linearized with respect to small variations of the potentials, which turns Poisson's equation into a linear Helmholtz equation.}\BibitemShut {Stop}%
\bibitem [{\citenamefont {Vinsome}\ and\ \citenamefont {Richardson}(1971)}]{Vinsome71}%
  \BibitemOpen
  \bibfield  {author} {\bibinfo {author} {\bibfnamefont {P.~K.~W.}\ \bibnamefont {Vinsome}}\ and\ \bibinfo {author} {\bibfnamefont {D.}~\bibnamefont {Richardson}},\ }\bibfield  {title} {\bibinfo {title} {The dielectric function in zincblende semiconductors},\ }\href {https://doi.org/10.1088/0022-3719/4/16/030} {\bibfield  {journal} {\bibinfo  {journal} {Journal of Physics C: Solid State Physics}\ }\textbf {\bibinfo {volume} {4}},\ \bibinfo {pages} {2650} (\bibinfo {year} {1971})}\BibitemShut {NoStop}%
\bibitem [{\citenamefont {Richardson}\ and\ \citenamefont {Vinsome}(1971)}]{Richardson71}%
  \BibitemOpen
  \bibfield  {author} {\bibinfo {author} {\bibfnamefont {D.}~\bibnamefont {Richardson}}\ and\ \bibinfo {author} {\bibfnamefont {P.~K.~W.}\ \bibnamefont {Vinsome}},\ }\bibfield  {title} {\bibinfo {title} {Dielectric function in semiconductors},\ }\href {https://doi.org/https://doi.org/10.1016/0375-9601(71)90035-1} {\bibfield  {journal} {\bibinfo  {journal} {Physics Letters A}\ }\textbf {\bibinfo {volume} {36}},\ \bibinfo {pages} {3} (\bibinfo {year} {1971})}\BibitemShut {NoStop}%
\bibitem [{\citenamefont {Secchi}\ \emph {et~al.}(2021{\natexlab{b}})\citenamefont {Secchi}, \citenamefont {Bellentani}, \citenamefont {Bertoni},\ and\ \citenamefont {Troiani}}]{BandCoulCI(KP)}%
  \BibitemOpen
  \bibfield  {author} {\bibinfo {author} {\bibfnamefont {A.}~\bibnamefont {Secchi}}, \bibinfo {author} {\bibfnamefont {L.}~\bibnamefont {Bellentani}}, \bibinfo {author} {\bibfnamefont {A.}~\bibnamefont {Bertoni}},\ and\ \bibinfo {author} {\bibfnamefont {F.}~\bibnamefont {Troiani}},\ }\bibfield  {title} {\bibinfo {title} {Inter- and intraband coulomb interactions between holes in silicon nanostructures},\ }\href {https://doi.org/10.1103/PhysRevB.104.205409} {\bibfield  {journal} {\bibinfo  {journal} {Physical Review B}\ }\textbf {\bibinfo {volume} {104}},\ \bibinfo {pages} {205409} (\bibinfo {year} {2021}{\natexlab{b}})}\BibitemShut {NoStop}%
\bibitem [{Note3()}]{Note3}%
  \BibitemOpen
  \bibinfo {note} {Moreover, the low-energy states of this system are almost pure HH states, and $u_{+3/2}^*(\protect \bm {r},\sigma ) u_{-3/2}(\protect \bm {r},\sigma )=0$ since the two Bloch functions have different physical spins.}\BibitemShut {Stop}%
\bibitem [{\citenamefont {Castelano}\ \emph {et~al.}(2018)\citenamefont {Castelano}, \citenamefont {de~Lima}, \citenamefont {Madureira}, \citenamefont {Degani},\ and\ \citenamefont {Maialle}}]{Castelano2018}%
  \BibitemOpen
  \bibfield  {author} {\bibinfo {author} {\bibfnamefont {L.~K.}\ \bibnamefont {Castelano}}, \bibinfo {author} {\bibfnamefont {E.~F.}\ \bibnamefont {de~Lima}}, \bibinfo {author} {\bibfnamefont {J.~R.}\ \bibnamefont {Madureira}}, \bibinfo {author} {\bibfnamefont {M.~H.}\ \bibnamefont {Degani}},\ and\ \bibinfo {author} {\bibfnamefont {M.~Z.}\ \bibnamefont {Maialle}},\ }\bibfield  {title} {\bibinfo {title} {Optimal control of universal quantum gates in a double quantum dot},\ }\href {https://doi.org/10.1103/PhysRevB.97.235301} {\bibfield  {journal} {\bibinfo  {journal} {Physical Review B}\ }\textbf {\bibinfo {volume} {97}},\ \bibinfo {pages} {235301} (\bibinfo {year} {2018})}\BibitemShut {NoStop}%
\bibitem [{Note4()}]{Note4}%
  \BibitemOpen
  \bibinfo {note} {All single-particle states $\protect \ket {\psi _n}$ and energies $E_n=\protect \bra {\psi _n}H_0\protect \ket {\psi _n}$ are converged so that $|H_0\protect \ket {\psi _n}-E_n\protect \ket {\psi _n}|<10^{-10}$ Hartree.}\BibitemShut {Stop}%
\bibitem [{Note5()}]{Note5}%
  \BibitemOpen
  \bibinfo {note} {They can be slightly recoupled to $S(1,1)$ by the magnetic field (that breaks symmetries), but make negligible contributions.}\BibitemShut {Stop}%
\bibitem [{Note6()}]{Note6}%
  \BibitemOpen
  \bibinfo {note} {It is more precisely the difference between the charging energy in the $(0,2)/(2,0)$ and $(1,1)$ states.}\BibitemShut {Stop}%
\bibitem [{Note7()}]{Note7}%
  \BibitemOpen
  \bibinfo {note} {This net charging energy is computed as $U=U_{02}-U_{11}$, where $U_{02}=2.32$\protect \,meV is the charging energy in the $(0,2)$ state (beyond the anti-crossing) and $U_{11}=0.346$\protect \,meV is the charging in the $(1,1)$ state (at the symmetric operation point). The former is the difference $U_{02}=E[S(0,2)]-E[S^*(0,2)]$ between the energies $E[S(0,2)]$ and $E[S^*(0,2)]=2E_0$ of the interacting and non-interacting singlets, respectively, with $E_0$ the single-particle ground-state energy. Likewise, $U_{11}=E[S(1,1)]-E[S^*(1,1)]=E[S(1,1)]-2E_b$, with $E_b$ the bonding ground-state energy at zero detuning.}\BibitemShut {Stop}%
\bibitem [{\citenamefont {Calder\'on}\ \emph {et~al.}(2006)\citenamefont {Calder\'on}, \citenamefont {Koiller},\ and\ \citenamefont {Das~Sarma}}]{Calderon06}%
  \BibitemOpen
  \bibfield  {author} {\bibinfo {author} {\bibfnamefont {M.~J.}\ \bibnamefont {Calder\'on}}, \bibinfo {author} {\bibfnamefont {B.}~\bibnamefont {Koiller}},\ and\ \bibinfo {author} {\bibfnamefont {S.}~\bibnamefont {Das~Sarma}},\ }\bibfield  {title} {\bibinfo {title} {Exchange coupling in semiconductor nanostructures: Validity and limitations of the {Heitler}-{London} approach},\ }\href {https://doi.org/10.1103/PhysRevB.74.045310} {\bibfield  {journal} {\bibinfo  {journal} {Physical Review B}\ }\textbf {\bibinfo {volume} {74}},\ \bibinfo {pages} {045310} (\bibinfo {year} {2006})}\BibitemShut {NoStop}%
\bibitem [{\citenamefont {Pedersen}\ \emph {et~al.}(2007)\citenamefont {Pedersen}, \citenamefont {Flindt}, \citenamefont {Mortensen},\ and\ \citenamefont {Jauho}}]{Pedersen07}%
  \BibitemOpen
  \bibfield  {author} {\bibinfo {author} {\bibfnamefont {J.}~\bibnamefont {Pedersen}}, \bibinfo {author} {\bibfnamefont {C.}~\bibnamefont {Flindt}}, \bibinfo {author} {\bibfnamefont {N.~A.}\ \bibnamefont {Mortensen}},\ and\ \bibinfo {author} {\bibfnamefont {A.-P.}\ \bibnamefont {Jauho}},\ }\bibfield  {title} {\bibinfo {title} {Failure of standard approximations of the exchange coupling in nanostructures},\ }\href {https://doi.org/10.1103/PhysRevB.76.125323} {\bibfield  {journal} {\bibinfo  {journal} {Physical Review B}\ }\textbf {\bibinfo {volume} {76}},\ \bibinfo {pages} {125323} (\bibinfo {year} {2007})}\BibitemShut {NoStop}%
\bibitem [{\citenamefont {Lieb}\ and\ \citenamefont {Mattis}(1962)}]{Lieb62}%
  \BibitemOpen
  \bibfield  {author} {\bibinfo {author} {\bibfnamefont {E.}~\bibnamefont {Lieb}}\ and\ \bibinfo {author} {\bibfnamefont {D.}~\bibnamefont {Mattis}},\ }\bibfield  {title} {\bibinfo {title} {Theory of ferromagnetism and the ordering of electronic energy levels},\ }\href {https://doi.org/10.1103/PhysRev.125.164} {\bibfield  {journal} {\bibinfo  {journal} {Physical Review}\ }\textbf {\bibinfo {volume} {125}},\ \bibinfo {pages} {164} (\bibinfo {year} {1962})}\BibitemShut {NoStop}%
\bibitem [{\citenamefont {Hedin}(1965)}]{Hedin65}%
  \BibitemOpen
  \bibfield  {author} {\bibinfo {author} {\bibfnamefont {L.}~\bibnamefont {Hedin}},\ }\bibfield  {title} {\bibinfo {title} {New method for calculating the one-particle green's function with application to the electron-gas problem},\ }\href {https://doi.org/10.1103/PhysRev.139.A796} {\bibfield  {journal} {\bibinfo  {journal} {Physical Review}\ }\textbf {\bibinfo {volume} {139}},\ \bibinfo {pages} {A796} (\bibinfo {year} {1965})}\BibitemShut {NoStop}%
\bibitem [{\citenamefont {White}\ and\ \citenamefont {Ramon}(2018)}]{White18}%
  \BibitemOpen
  \bibfield  {author} {\bibinfo {author} {\bibfnamefont {Z.}~\bibnamefont {White}}\ and\ \bibinfo {author} {\bibfnamefont {G.}~\bibnamefont {Ramon}},\ }\bibfield  {title} {\bibinfo {title} {Extended orbital modeling of spin qubits in double quantum dots},\ }\href {https://doi.org/10.1103/PhysRevB.97.045306} {\bibfield  {journal} {\bibinfo  {journal} {Physical Review B}\ }\textbf {\bibinfo {volume} {97}},\ \bibinfo {pages} {045306} (\bibinfo {year} {2018})}\BibitemShut {NoStop}%
\bibitem [{\citenamefont {Merkt}\ \emph {et~al.}(1991)\citenamefont {Merkt}, \citenamefont {Huser},\ and\ \citenamefont {Wagner}}]{Merkt91}%
  \BibitemOpen
  \bibfield  {author} {\bibinfo {author} {\bibfnamefont {U.}~\bibnamefont {Merkt}}, \bibinfo {author} {\bibfnamefont {J.}~\bibnamefont {Huser}},\ and\ \bibinfo {author} {\bibfnamefont {M.}~\bibnamefont {Wagner}},\ }\bibfield  {title} {\bibinfo {title} {Energy spectra of two electrons in a harmonic quantum dot},\ }\href {https://doi.org/10.1103/PhysRevB.43.7320} {\bibfield  {journal} {\bibinfo  {journal} {Physical Review B}\ }\textbf {\bibinfo {volume} {43}},\ \bibinfo {pages} {7320} (\bibinfo {year} {1991})}\BibitemShut {NoStop}%
\bibitem [{\citenamefont {Abragam}\ and\ \citenamefont {Bleaney}(1971)}]{Abragam1970}%
  \BibitemOpen
  \bibfield  {author} {\bibinfo {author} {\bibfnamefont {A.}~\bibnamefont {Abragam}}\ and\ \bibinfo {author} {\bibfnamefont {B.}~\bibnamefont {Bleaney}},\ }\href@noop {} {\emph {\bibinfo {title} {Electron Paramagnetic Resonance of Transition Ions}}},\ Resonance Paramagnetique Electronique Des Ions de Transition\ (\bibinfo  {publisher} {{Presses Universitaires de France}},\ \bibinfo {address} {{France}},\ \bibinfo {year} {1971})\BibitemShut {NoStop}%
\bibitem [{\citenamefont {Chibotaru}\ \emph {et~al.}(2008)\citenamefont {Chibotaru}, \citenamefont {Ceulemans},\ and\ \citenamefont {Bolvin}}]{Chiboratu08}%
  \BibitemOpen
  \bibfield  {author} {\bibinfo {author} {\bibfnamefont {L.~F.}\ \bibnamefont {Chibotaru}}, \bibinfo {author} {\bibfnamefont {A.}~\bibnamefont {Ceulemans}},\ and\ \bibinfo {author} {\bibfnamefont {H.}~\bibnamefont {Bolvin}},\ }\bibfield  {title} {\bibinfo {title} {Unique definition of the zeeman-splitting $g$ tensor of a kramers doublet},\ }\href {https://doi.org/10.1103/PhysRevLett.101.033003} {\bibfield  {journal} {\bibinfo  {journal} {Physical Review Letters}\ }\textbf {\bibinfo {volume} {101}},\ \bibinfo {pages} {033003} (\bibinfo {year} {2008})}\BibitemShut {NoStop}%
\bibitem [{Note8()}]{Note8}%
  \BibitemOpen
  \bibinfo {note} {The spin-orbit frame is not uniquely defined either, as a same unitary transform on both dots leaves $T$ invariant. A particular spin-orbit frame can always be built starting from given $g_L$, $g_R$ matrices and tunneling block $T$ parametrized as $T=t(\cos \theta _\protect \mathrm {so}I_2-i\sin \theta _\protect \mathrm {so}\protect \bm {n}_\protect \mathrm {so}\cdot \protect \bm {\sigma })$, with $\protect \bm {n}_\protect \mathrm {so}$ and $\theta _\protect \mathrm {so}$ the so-called spin-orbit vector and spin-orbit angle. Indeed, the transformations $U_L=\exp (-i\theta _\protect \mathrm {so}\protect \bm {n}_\protect \mathrm {so}\cdot \protect \bm {\sigma }/2)$ in the $\{\protect \ket {L\uparrow },\protect \ket {L\downarrow }\}$ subspace and $U_R=\exp (i\theta _\protect \mathrm {so}\protect \bm {n}_\protect \mathrm {so}\cdot \protect \bm {\sigma }/2)$ in the $\{\protect \ket {R\uparrow },\protect \ket {R\downarrow }\}$ subspace diagonalize the $T$ block $\protect \tilde {T}=U_L^\dagger
  TU_R=tI_2$. The $g$-matrices then become $\protect \tilde {g}_L=R(\protect \bm {n}_\protect \mathrm {so},-\theta _\protect \mathrm {so})g_L$ and $\protect \tilde {g}_R=R(\protect \bm {n}_\protect \mathrm {so},\theta _\protect \mathrm {so})g_R$ with $R(\protect \bm {u},\theta )$ the matrix of the real-space rotation of angle $\theta $ around the unit vector $\protect \bm {u}$ \cite {Geyer24,Saezmollejo24}.}\BibitemShut {Stop}%
\bibitem [{\citenamefont {Yu}\ \emph {et~al.}(2023)\citenamefont {Yu}, \citenamefont {Zihlmann}, \citenamefont {Abadillo-Uriel}, \citenamefont {Michal}, \citenamefont {Rambal}, \citenamefont {Niebojewski}, \citenamefont {Bedecarrats}, \citenamefont {Vinet}, \citenamefont {Dumur}, \citenamefont {Filippone}, \citenamefont {Bertrand}, \citenamefont {De~Franceschi}, \citenamefont {Niquet},\ and\ \citenamefont {Maurand}}]{yu2022strong}%
  \BibitemOpen
  \bibfield  {author} {\bibinfo {author} {\bibfnamefont {C.~X.}\ \bibnamefont {Yu}}, \bibinfo {author} {\bibfnamefont {S.}~\bibnamefont {Zihlmann}}, \bibinfo {author} {\bibfnamefont {J.~C.}\ \bibnamefont {Abadillo-Uriel}}, \bibinfo {author} {\bibfnamefont {V.~P.}\ \bibnamefont {Michal}}, \bibinfo {author} {\bibfnamefont {N.}~\bibnamefont {Rambal}}, \bibinfo {author} {\bibfnamefont {H.}~\bibnamefont {Niebojewski}}, \bibinfo {author} {\bibfnamefont {T.}~\bibnamefont {Bedecarrats}}, \bibinfo {author} {\bibfnamefont {M.}~\bibnamefont {Vinet}}, \bibinfo {author} {\bibfnamefont {{\'E}.}~\bibnamefont {Dumur}}, \bibinfo {author} {\bibfnamefont {M.}~\bibnamefont {Filippone}}, \bibinfo {author} {\bibfnamefont {B.}~\bibnamefont {Bertrand}}, \bibinfo {author} {\bibfnamefont {S.}~\bibnamefont {De~Franceschi}}, \bibinfo {author} {\bibfnamefont {Y.-M.}\ \bibnamefont {Niquet}},\ and\ \bibinfo {author} {\bibfnamefont {R.}~\bibnamefont {Maurand}},\ }\bibfield  {title} {\bibinfo {title} {Strong coupling between a photon and a
  hole spin in silicon},\ }\href {https://doi.org/10.1038/s41565-023-01332-3} {\bibfield  {journal} {\bibinfo  {journal} {Nature Nanotechnology}\ }\textbf {\bibinfo {volume} {18}},\ \bibinfo {pages} {741} (\bibinfo {year} {2023})}\BibitemShut {NoStop}%
\bibitem [{\citenamefont {Kelly}\ \emph {et~al.}(2025)\citenamefont {Kelly}, \citenamefont {Massai}, \citenamefont {Hetényi}, \citenamefont {Pita-Vidal}, \citenamefont {Orekhov}, \citenamefont {Carlsson}, \citenamefont {Seidler}, \citenamefont {Tsoukalas}, \citenamefont {Sommer}, \citenamefont {Aldeghi}, \citenamefont {Bedell}, \citenamefont {Paredes}, \citenamefont {Schupp}, \citenamefont {Mergenthaler}, \citenamefont {Fuhrer}, \citenamefont {Salis},\ and\ \citenamefont {Harvey-Collard}}]{Kelly2025}%
  \BibitemOpen
  \bibfield  {author} {\bibinfo {author} {\bibfnamefont {E.~G.}\ \bibnamefont {Kelly}}, \bibinfo {author} {\bibfnamefont {L.}~\bibnamefont {Massai}}, \bibinfo {author} {\bibfnamefont {B.}~\bibnamefont {Hetényi}}, \bibinfo {author} {\bibfnamefont {M.}~\bibnamefont {Pita-Vidal}}, \bibinfo {author} {\bibfnamefont {A.}~\bibnamefont {Orekhov}}, \bibinfo {author} {\bibfnamefont {C.}~\bibnamefont {Carlsson}}, \bibinfo {author} {\bibfnamefont {I.}~\bibnamefont {Seidler}}, \bibinfo {author} {\bibfnamefont {K.}~\bibnamefont {Tsoukalas}}, \bibinfo {author} {\bibfnamefont {L.}~\bibnamefont {Sommer}}, \bibinfo {author} {\bibfnamefont {M.}~\bibnamefont {Aldeghi}}, \bibinfo {author} {\bibfnamefont {S.~W.}\ \bibnamefont {Bedell}}, \bibinfo {author} {\bibfnamefont {S.}~\bibnamefont {Paredes}}, \bibinfo {author} {\bibfnamefont {F.~J.}\ \bibnamefont {Schupp}}, \bibinfo {author} {\bibfnamefont {M.}~\bibnamefont {Mergenthaler}}, \bibinfo {author} {\bibfnamefont {A.}~\bibnamefont {Fuhrer}}, \bibinfo {author} {\bibfnamefont
  {G.}~\bibnamefont {Salis}},\ and\ \bibinfo {author} {\bibfnamefont {P.}~\bibnamefont {Harvey-Collard}},\ }\bibfield  {title} {\bibinfo {title} {Identifying and mitigating errors in hole spin qubit readout},\ }\href {https://arxiv.org/abs/2504.06898} {\bibfield  {journal} {\bibinfo  {journal} {arXiv: 2504.06898}\ } (\bibinfo {year} {2025})}\BibitemShut {NoStop}%
\bibitem [{\citenamefont {Danon}\ and\ \citenamefont {Nazarov}(2009)}]{danon_pauli_2009}%
  \BibitemOpen
  \bibfield  {author} {\bibinfo {author} {\bibfnamefont {J.}~\bibnamefont {Danon}}\ and\ \bibinfo {author} {\bibfnamefont {Y.~V.}\ \bibnamefont {Nazarov}},\ }\bibfield  {title} {\bibinfo {title} {Pauli spin blockade in the presence of strong spin-orbit coupling},\ }\href {https://doi.org/10.1103/PhysRevB.80.041301} {\bibfield  {journal} {\bibinfo  {journal} {Physical Review B}\ }\textbf {\bibinfo {volume} {80}},\ \bibinfo {pages} {041301} (\bibinfo {year} {2009})}\BibitemShut {NoStop}%
\bibitem [{Note9()}]{Note9}%
  \BibitemOpen
  \bibinfo {note} {For the squeezed dots, $\delta V_\protect \mathrm {d}=\delta V_\protect \mathrm {L}-\delta V_\protect \mathrm {R}$ with $\delta V_\protect \mathrm {R}=-1.035\delta V_\protect \mathrm {L}$.}\BibitemShut {Stop}%
\bibitem [{\citenamefont {Aleiner}\ and\ \citenamefont {Fal'ko}(2001)}]{Aleiner01}%
  \BibitemOpen
  \bibfield  {author} {\bibinfo {author} {\bibfnamefont {I.~L.}\ \bibnamefont {Aleiner}}\ and\ \bibinfo {author} {\bibfnamefont {V.~I.}\ \bibnamefont {Fal'ko}},\ }\bibfield  {title} {\bibinfo {title} {Spin-orbit coupling effects on quantum transport in lateral semiconductor dots},\ }\href {https://doi.org/10.1103/PhysRevLett.87.256801} {\bibfield  {journal} {\bibinfo  {journal} {Physical Review Letters}\ }\textbf {\bibinfo {volume} {87}},\ \bibinfo {pages} {256801} (\bibinfo {year} {2001})}\BibitemShut {NoStop}%
\bibitem [{\citenamefont {Levitov}\ and\ \citenamefont {Rashba}(2003)}]{Levitov03}%
  \BibitemOpen
  \bibfield  {author} {\bibinfo {author} {\bibfnamefont {L.~S.}\ \bibnamefont {Levitov}}\ and\ \bibinfo {author} {\bibfnamefont {E.~I.}\ \bibnamefont {Rashba}},\ }\bibfield  {title} {\bibinfo {title} {Dynamical spin-electric coupling in a quantum dot},\ }\href {https://doi.org/10.1103/PhysRevB.67.115324} {\bibfield  {journal} {\bibinfo  {journal} {Physical Review B}\ }\textbf {\bibinfo {volume} {67}},\ \bibinfo {pages} {115324} (\bibinfo {year} {2003})}\BibitemShut {NoStop}%
\bibitem [{Note10()}]{Note10}%
  \BibitemOpen
  \bibinfo {note} {These expressions apply for positive $g_{xx}$'s and negative $g_{yy}$'s.}\BibitemShut {Stop}%
\bibitem [{Note11()}]{Note11}%
  \BibitemOpen
  \bibinfo {note} {We emphasize, though, that the cost of the method we have chosen to solve the time-dependent Schrodinger equation scales as $M^3$ (with $M$ the dimension of the basis set). Other methods (Trotter, ...) may show a better ($\propto M^2$) scaling. Anyhow, the gain in the dressed basis set remains very significant.}\BibitemShut {Stop}%
\bibitem [{\citenamefont {Flindt}\ \emph {et~al.}(2006)\citenamefont {Flindt}, \citenamefont {S\o{}rensen},\ and\ \citenamefont {Flensberg}}]{Flindt06}%
  \BibitemOpen
  \bibfield  {author} {\bibinfo {author} {\bibfnamefont {C.}~\bibnamefont {Flindt}}, \bibinfo {author} {\bibfnamefont {A.~S.}\ \bibnamefont {S\o{}rensen}},\ and\ \bibinfo {author} {\bibfnamefont {K.}~\bibnamefont {Flensberg}},\ }\bibfield  {title} {\bibinfo {title} {Spin-orbit mediated control of spin qubits},\ }\href {https://doi.org/10.1103/PhysRevLett.97.240501} {\bibfield  {journal} {\bibinfo  {journal} {Physical Review Letters}\ }\textbf {\bibinfo {volume} {97}},\ \bibinfo {pages} {240501} (\bibinfo {year} {2006})}\BibitemShut {NoStop}%
\bibitem [{\citenamefont {Trif}\ \emph {et~al.}(2007)\citenamefont {Trif}, \citenamefont {Golovach},\ and\ \citenamefont {Loss}}]{Trif07}%
  \BibitemOpen
  \bibfield  {author} {\bibinfo {author} {\bibfnamefont {M.}~\bibnamefont {Trif}}, \bibinfo {author} {\bibfnamefont {V.~N.}\ \bibnamefont {Golovach}},\ and\ \bibinfo {author} {\bibfnamefont {D.}~\bibnamefont {Loss}},\ }\bibfield  {title} {\bibinfo {title} {Spin-spin coupling in electrostatically coupled quantum dots},\ }\href {https://doi.org/10.1103/PhysRevB.75.085307} {\bibfield  {journal} {\bibinfo  {journal} {Physical Review B}\ }\textbf {\bibinfo {volume} {75}},\ \bibinfo {pages} {085307} (\bibinfo {year} {2007})}\BibitemShut {NoStop}%
\bibitem [{\citenamefont {Trifunovic}\ \emph {et~al.}(2012)\citenamefont {Trifunovic}, \citenamefont {Dial}, \citenamefont {Trif}, \citenamefont {Wootton}, \citenamefont {Abebe}, \citenamefont {Yacoby},\ and\ \citenamefont {Loss}}]{Trifunovic12}%
  \BibitemOpen
  \bibfield  {author} {\bibinfo {author} {\bibfnamefont {L.}~\bibnamefont {Trifunovic}}, \bibinfo {author} {\bibfnamefont {O.}~\bibnamefont {Dial}}, \bibinfo {author} {\bibfnamefont {M.}~\bibnamefont {Trif}}, \bibinfo {author} {\bibfnamefont {J.~R.}\ \bibnamefont {Wootton}}, \bibinfo {author} {\bibfnamefont {R.}~\bibnamefont {Abebe}}, \bibinfo {author} {\bibfnamefont {A.}~\bibnamefont {Yacoby}},\ and\ \bibinfo {author} {\bibfnamefont {D.}~\bibnamefont {Loss}},\ }\bibfield  {title} {\bibinfo {title} {Long-distance spin-spin coupling via floating gates},\ }\href {https://doi.org/10.1103/PhysRevX.2.011006} {\bibfield  {journal} {\bibinfo  {journal} {Physical Review X}\ }\textbf {\bibinfo {volume} {2}},\ \bibinfo {pages} {011006} (\bibinfo {year} {2012})}\BibitemShut {NoStop}%
\end{thebibliography}%

\end{document}